\documentclass[a4paper,fleqn,useAMS]{mnras}

\usepackage{graphicx}	
\usepackage{amsmath}	
\usepackage{amssymb}	
\usepackage{multicol}        
\usepackage{pdflscape}	

\usepackage[T1]{fontenc}
\usepackage{ae,aecompl}

\usepackage{times,txfonts}
\usepackage{amsfonts,amsopn,amssymb,capt-of,stmaryrd,xspace}

\newcommand{\ec}{$\eta$~Car\xspace}
\newcommand{\SimpleX}{\textsc{simplex}\xspace}

\title[He~I absorption variability in $\eta$ Car]{To $v_\infty$ and beyond! The He~I absorption variability across the 2014.6 periastron passage of $\eta$ Carinae\thanks{Based in part on observations made with the NASA/ESA Hubble Space Telescope, obtained at the Space Telescope Science Institute, which is operated by the Association of Universities for Research in Astronomy, Inc., under NASA contract NAS 5-26555. These observations are associated with program \#13395.}}

\author[N. D. Richardson et al.]{Noel~D.~Richardson$^{1,2}$\thanks{E-mail:noel.richardson@utoledo.edu},
Thomas I. Madura$^{3,4}$,
Lucas St-Jean$^{2}$,
Anthony F. J. Moffat$^{2}$,
\newauthor Theodore R. Gull$^{3}$,
Christopher M.~P.~Russell$^{3}$,
Augusto Damineli$^5$,
Mairan Teodoro$^{3,4}$,
\newauthor Michael F. Corcoran$^{6}$,
Frederick M. Walter$^7$,
Nicola Clementel$^{8}$,
Jos\'e H. Groh$^{9}$,
\newauthor Kenji Hamaguchi$^{6}$,
and D. John Hillier$^{10}$ \\
\\
$^{1}$ Ritter Observatory, Department of Physics and Astronomy, The University of Toledo, Toledo, OH 43606-3390, USA\\
$^{2}$D\'epartement de physique and Centre de Recherche en Astrophysique du Qu\'ebec (CRAQ), Universit\'e de Montr\'eal, C.P. 6128,\\  Succ.~Centre-Ville, Montr\'eal, Qu\'ebec, H3C 3J7, Canada\\
$^{3}$Astrophysics Science Division, Code 667, NASA Goddard Space Flight Center, Greenbelt, MD 20771, USA\\
$^{4}$Universities Space Research Association, 7178 Columbia Gateway Drive, Columbia, MD 20146 USA \\
$^{5}$IAG-USP, Rua do Mat\~ao, 1226, 05508-090, S\~ao Paulo, SP Brazil\\
$^{6}$CRESST and X-ray Astrophysics Laboratory, NASA/Goddard Space Flight Center, Greenbelt, MD 20771, USA\\
$^7$Department of Physics and Astronomy, Stony Brook University, Stony Brook, NY, 11794-3800, USA\\
$^{8}$South African Astronomical Observatory, P.O. Box 9, 7935 Observatory, South Africa\\
$^{9}$School of Physics, Trinity College Dublin, The University of Dublin, Dublin 2, Ireland \\
$^{10}$Department of Physics and Astronomy and Pittsburgh Particle physics, Astrophysics, and Cosmology Center (PITT PACC),\\University
of Pittsburgh, 3941 O'Hara Street, Pittsburgh, PA 15260, USA\\
}

\begin{document}
\bibliographystyle{mn2e}

\date{}

\pagerange{\pageref{firstpage}--\pageref{lastpage}} \pubyear{2015}

\maketitle

\label{firstpage}

\begin{abstract}
We have monitored the massive binary star $\eta$ Carinae with the CTIO/SMARTS 1.5~m telescope and CHIRON spectrograph from the previous apastron passage of the system through the recent 2014.6 periastron passage. Our monitoring has resulted in a large, homogeneous data set with an unprecedented time-sampling, spectral resolving power, and signal-to-noise. This allowed us to investigate temporal variability previously unexplored in the system and discover a kinematic structure in the P Cygni absorption troughs of neutral helium wind lines. The features observed occurred prior to the periastron passage and are seen as we look through the trailing arm of the wind-wind collision shock cone. We show that the bulk of the variability is repeatable across the last five periastron passages, and that the absorption occurs in the inner 230 AU of the system. In addition, we found an additional, high-velocity absorption component super-imposed on the P Cygni absorption troughs that has been previously un-observed in these lines, but which bears resemblance to the observations of the He~I $\lambda$10830 \AA\ feature across previous cycles. Through a comparison of the current smoothed particle hydrodynamical simulations, we show that the observed variations are likely caused by instabilities in the wind-wind collision region in our line-of-sight, coupled with stochastic variability related to clumping in the winds.
\end{abstract}

\begin{keywords}
stars: early-type
-- binaries: close
-- stars: individual ($\eta$ Car)
-- stars: winds, outflows
-- stars: mass loss
-- stars: variables: S Doradus
\end{keywords}

\section{Introduction}

Massive stars love company, and it seems that more than 90\% of massive stars are formed in binaries (Sana et al.~2014), with a majority of them interacting during their lives (Sana et al.~2012). These interactions could take the form of mergers, Roche lobe overflow and accretion, or wind interactions through wind-wind collisions (WWCs) when both stars are evolved and have strong winds.
WWCs are best known in Wolf-Rayet$+$O binaries, which occur quite frequently and with dramatic effects. In such systems, we observe hot, dense stellar winds, high terminal wind speeds, and phase-dependent shock cones through studies of spectral line excesses, X-ray light curves and spectroscopy, non-thermal radio emission, and dust production in some cases (e.g. Moffat 1998). One classical example of these colliding-wind WR$+$O binaries is the WC7pd$+$O5.5fc system WR 140 (Fahed et al. 2011; Monnier et al. 2011), with an orbital period of 7.94 yr and a high eccentricity of 0.88. The denser WR wind pushes back the O star wind to form a shock cone with a half opening angle $\sim50^\circ$, which wraps around the system in successive orbits. WWCs are occasionally observed in O$+$O binaries, but with correspondingly lower excess emission fluxes.

$\eta$~Car is a massive, evolved binary residing in the nearby ($D \sim2.3$~kpc; Smith 2006) Carina nebula and star-forming region. It has been well-studied, being the subject of a Hubble Treasury program for spatially resolved spectroscopy, while also being observed with ground-based telescopes. Recent studies have allowed state-of-the-art observations and theory to meet, providing constraints on the orbital, stellar, and wind parameters of the system, and our understanding of massive star WWCs. The secondary star has eluded observers since the discovery of the binary nature of the system and the 5.54-yr periodicities (Damineli et al.~1997, 2008). The WWC is dramatic because of the lower terminal wind speed, $v_{\infty,1}$ ($\sim 500$~km~s$^{-1}$; Hillier et al.~2001) and higher mass-loss rate of the primary ($\dot{M}\approx 8.5 \times 10^{-4}$~$\mathrm{M}_{\odot}$~yr$^{-1}$; Groh et al.~2012a) compared to typical WR$+$O systems. The secondary star has wind parameters indicative of either a WR or extreme O star, with the strongest constraints coming from the X-ray analysis (Pittard \& Corcoran 2002; Parkin et al.~2009), which points to $\dot{M}_{2} \approx 10^{-5}$~$\mathrm{M}_{\odot}$~yr$^{-1}$ and $v_{\infty, 2}\sim 3000$~km~s$^{-1}$.

\ec has a high eccentricity ($\sim 0.9$) and an orbital period of 5.54 yr, and many observational phenomena are modulated on this time-scale, such as the X-ray emission (Corcoran 2005, Corcoran et al.~2010), multi-wavelength photometry (Fern\'andez-Laj\'us et al.~2003, 2010; Whitelock et al.~2004), He~I narrow emission (Damineli et al.~1997), the H$\alpha$ and other wind line profile morphologies (Richardson et al.~2010, 2015), He~II emission from near the colliding winds (Steiner \& Damineli 2004; Teodoro et al.~2012, 2016; Mehner et al.~2015; Davidson et al.~2015), and spatially-extended forbidden line emission (Gull et al. 2009, 2011; Teodoro et al. 2013). The 2003 periastron passage was well-documented thanks to a large Treasury program with the {\it Hubble Space Telescope} ({\it HST}) and Space Telescope Imaging Spectrograph (STIS)\footnote{http://etacar.umn.edu}. The more recent 2009 periastron passage was observed in the optical only with ground-based facilities due to a failure of the STIS that was repaired during Servicing Mission 4, after the 2009 periastron passage. However, this event brought about one of the most intense ground-based spectroscopic data-sets compiled, both for the He~II emission (Teodoro et al.~2012), and the optical spectrum and wind transitions (Richardson et al.~2015), including the strong H$\alpha$ profile (Richardson et al.~2010) that is often saturated in ground-based observations. The intense time-series collected will allow for many future theoretical studies about the detailed system parameters.

Three-dimensional (3D) hydrodynamical and radiative transfer simulations of \ec's binary colliding winds have provided crucial insights on the time-dependent geometry and ionization state of the WWCs, helping constrain the binary orientation, wind parameters, and locations where various observed emission and absorption features arise (Okazaki et al. 2008; Parkin et al. 2009, 2011; Groh et al.~2012a,b; Madura et al. 2012, 2013, 2015; Clementel et al. 2014, 2015a,b). Using a 3D dynamical model of the broad, extended [Fe III] emission observed in \ec by the {\it HST}/STIS (Gull et al. 2009), Madura et al. (2012) confirmed the orbital inclination and argument of periapsis derived by Okazaki et al. (2008) and Parkin et al. (2009) using X-ray data. More importantly, Madura et al. (2012) broke the degeneracy inherent to models based solely on X-rays or other spatially-unresolved data and constrained the 3D orientation of the \ec binary. They found that the binary has an argument of periapsis $\omega \approx 240^{\circ} - 285^{\circ}$, with the orbital axis closely aligned with the Homunculus nebula's polar symmetry axis at an inclination $i \approx 130^{\circ} - 145^{\circ}$ and position angle on the sky $\mathrm{PA} \approx 302^{\circ} - 327^{\circ}$, implying that the binary companion star at apastron is on the observer's side of the system and that the companion orbits clockwise on the sky.

Most recently, Clementel et al. (2015a,b) performed 3D radiative transfer simulations of \ec's innermost ($\lesssim$~155~AU) interacting winds at apastron and periastron, applying the \SimpleX algorithm to the 3D smoothed particle hydrodynamics (SPH) simulations of Madura et al. (2013). Clementel et al. (2015a,b) focus specifically on the ionization structure of He in the stellar winds and WWC regions, producing 3D ionization maps showing the regions where He is singly- and doubly-ionized due to photoionization by the hot companion star\footnote{$T\sim 40,000$~K from analysis of the forbidden line emission; e.g.,~Mehner et al.~2010a} and collisional ionization in the X-ray generating WWC shocks, respectively. The results of Clementel et al. (2015a,b) help constrain the regions where the observed He~I emission and absorption lines can arise in \ec and provide further support for a binary orientation in which the companion star at apastron is on the observer's side of the system, with the companion at periastron deeply embedded within the optically-thick primary wind on the side opposite that of the observer.

Groh et al.~(2010) observed the He~I lines of $\eta$ Car as it approached its 2009.0 periastron passage, concentrating on the NIR transition of He~I $\lambda$10830. A high-velocity absorption component at velocities of $v_r\sim-900$~km~s$^{-1}$ or lower was observed between phases 0.976 and 1.023, which was postulated to originate in the shocked gas from the WWC zone as the trailing arm of the shock cone passes through our line-of-sight. From the spatial scale of these absorption variations and a comparison to 3D models of the colliding winds, these observations pointed to a WWC at scales of 15--45~AU from the central binary. Other optical He~I lines examined from the {\it HST} Treasury program during the 2003 spectroscopic event (Nielsen et al. 2007; Groh et al. 2010) show a noticeable high-velocity absorption in the two triplet line profiles of He~I $\lambda$3888 and $\lambda$5876 extending to $\sim -900$~km~s$^{-1}$ at phase 0.995. Singlet lines such as He~I $\lambda$6680 show weaker absorption up to $\sim -800$~km~s$^{-1}$. 
However, no visible-wavelength He~I lines showed obvious highly-blueshifted absorptions ($< -900$ km s$^{-1}$). This is explainable by the difference in lower levels of the transitions. The He~I $\lambda$10830 absorptions originate from the metastable 2S$^2$ level, but the visible lines are absorbed from levels readily depopulated by permitted transitions.

The interpretation of He I emission in \ec is still somewhat debated in the literature. Some of the He I emission and absorption can certainly be accounted for by the primary wind (e.g.~Hillier et al.~2001, Groh et al.~2012a). This is supported by the presence of extremely similar He~I profiles in the LBV HDE 316285 (Hillier et al.~1998, 2001), which has no known binary companion. The spectral modeling of both $\eta$~Car and HDE 316285 (Hillier et al. 1998, 2001) can reproduce both stars and show similar properties. Many authors have argued that the He I is related to, but not necessarily coming from, the companion (see e.g. Mehner et al.~2012, 2015). This would allow for several variations we observe to be accounted for in the $\eta$~Car system. For example, Richardson et al.~(2015) developed a simple cartoon model for the variability of the He~I lines that allows for some fraction of the emission to be formed in an ionised hemisphere of the primary wind facing the secondary star. When coupled with the expected radial velocity variations of the primary, the model does fairly well at explaining the variations, but future 3D SPH models with \SimpleX might better account for the variability. 

The recent 2014.6 spectroscopic event provided observers a fresh look at the variations related to the periastron passage of the system, despite an unfortunate timing for ground-based observatories. Davidson et al.~(2015) reported a potential cycle-to-cycle change in the He~II $\lambda$4686 emission based upon {\it HST}/STIS measurements, but Teodoro et al.~(2016) have compiled a large, extensive dataset with spatial mappings from {\it HST}/STIS used to correct for differences between ground-based and space-based observations. The dataset of Teodoro et al.~(2016) shows very few changes in the global behaviour of He~II when compared to the previous four periastron passages observed, but some minor variability is present. Davidson et al.~also point out the appearance of N~II absorptions in the blue spectrum, but did not discuss these changes in detail. The {\it HST}/STIS spatial mappings are being analysed by Gull et al.~(in prep.) and show a distinctive changing ionisation balance when comparing the forbidden [Fe~III] and [Fe~II] emission during the periastron passage. Mehner et al.~(2015) have further analysed several ground- and space-based observations of $\eta$~Car across its 2014 event, including important observations of the reflected polar spectrum, which show fewer changes in the broad wind emission lines than those seen in direct line-of-sight.

Further complicating the story of $\eta$~Car's 2014 spectroscopic event are \emph{Swift} X-ray monitoring observations of the X-ray minimum (Corcoran et al., in prep.). Comparison of the recent \emph{Swift} measurements to \emph{RXTE} observations of $\eta$~Car's 1998.0, 2003.5, and  2009.0 events shows that the start of the X-ray minimum is remarkably reproducible, to within one day over the last $\sim 18$~years, helping define the 2023~day period (Corcoran et al., in prep.). However, the recovery from the X-ray minimum is not as well behaved, with the 2009.0 recovery occurring $\sim 30$~days earlier compared to the 1998.0 and 2003.5 recoveries, and the 2014.6 recovery time being intermediate to those of the short 2009.0 recovery time compared to the longer 1998.0 and 2003.5 recoveries (Corcoran et al., in prep.).

In an attempt to understand better $\eta$~Car's spectroscopic events, we obtained frequent ground-based, optical, high-resolution ($R\sim90,000$) spectroscopy of the system as the periastron passage approached in 2014 (Section 2), allowing for comparisons to the previous 2009 periastron passage. Among the many resulting observations, we noticed a strong, additional absorption on the He~I lines as the system approached the X-ray minimum and higher energy forbidden and permitted lines reached a minimum, all of which happened just prior to the recent periastron passage. We discuss this result in Section 3. In Section 4, we discuss how the absorptions are related to the binary orbit and their repeatability. Section 5 presents a very detailed account of the formation of these spectral features and comparisons with our expectations from hydrodynamical simulations at different orbital epochs through the periastron passage. We conclude the study in Section 6 where we highlight our main findings of this study.

\section{Observations}

During the 2009 periastron passage of $\eta$ Carinae, we began monitoring the system with the fiber-fed echelle spectrograph (Richardson et al.~2010) on the CTIO 1.5 m telescope operated by the SMARTS Consortium. In 2011, this spectrograph was decommissioned in favor of a new, high-efficiency spectrograph, CHIRON (Tokovinin et al.~2013) that covers the optical from $\sim4500$ \AA\ to $\sim8500$ \AA. Both of these spectrographs were fed starlight through a multi-mode fibre that has a size of 2.7\arcsec on the sky. We began monitoring the system with the ``slit'' mode that has a resolving power of $\sim 90,000$ in early 2012, which we continued whenever possible leading up to the current 2014 spectroscopic event and recovery. In total we collected 207 spectra between the calendar dates 2012 March 03 and 2015 July 27. Some spectra were obtained with CHIRON in the ``slicer'' mode, which delivers spectra with a resolving power of $\sim80,000$. All spectra were corrected for bias and flat field effects and wavelength calibrated through the CHIRON pipeline. This leaves a very strong blaze function present on each order which is difficult to remove in the presence of strong emission lines with widths similar to the echelle order range in some cases. Therefore, we observed HR 4468 (B9.5Vn) and $\mu$ Col (O9.5V) to fit the continuum blaze function empirically on orders without spectral lines. The remaining orders (primarily around H$\alpha$ and H$\beta$) were interpolated to match adjacent orders. The resulting spectra were then normalized and combined into standard one-dimensional spectra, and the resulting order-overlaps in the blue showed us that the blaze removal was accurate to $<0.5\%$. A global normalization was then applied to obtain a unit continuum, which we adjusted in the regions adjacent to our spectral lines of interest for analysis.

 \begin{figure*}
\includegraphics[height=8in, angle=0]{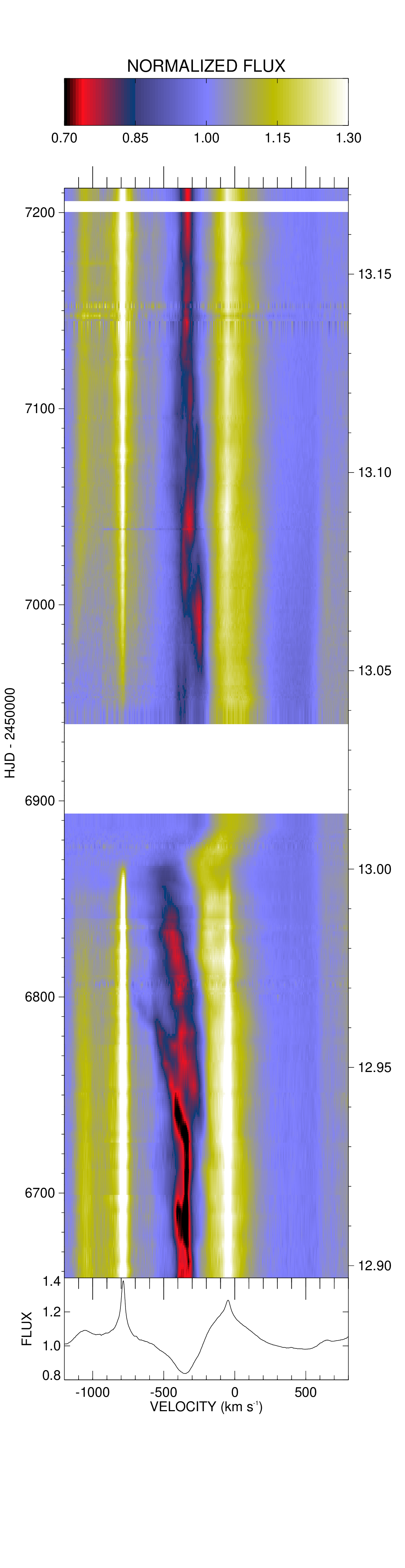}
\includegraphics[height=8in, angle=0]{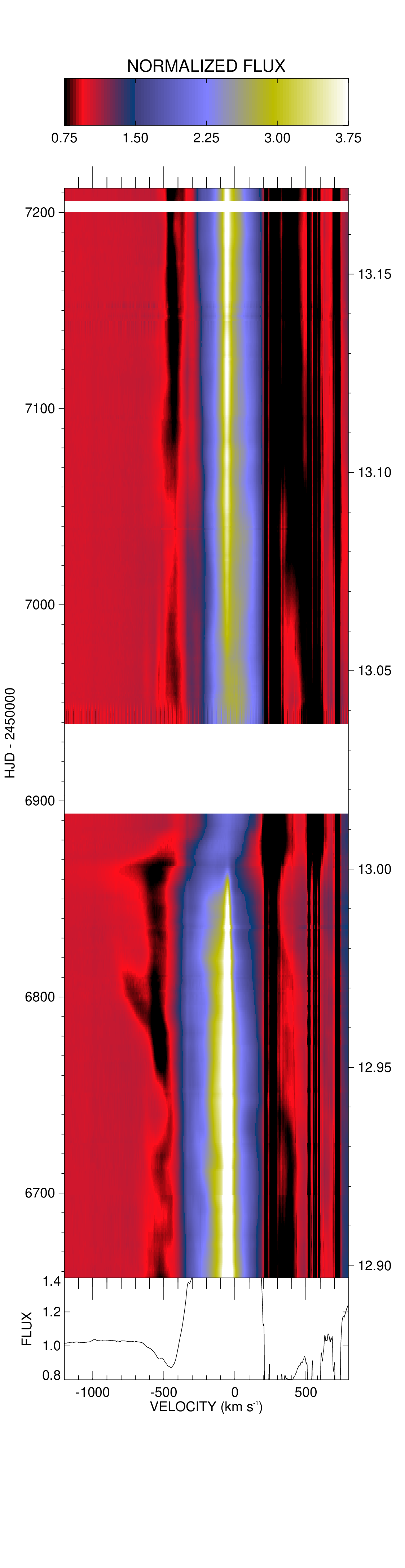}
\includegraphics[height=8in, angle=0]{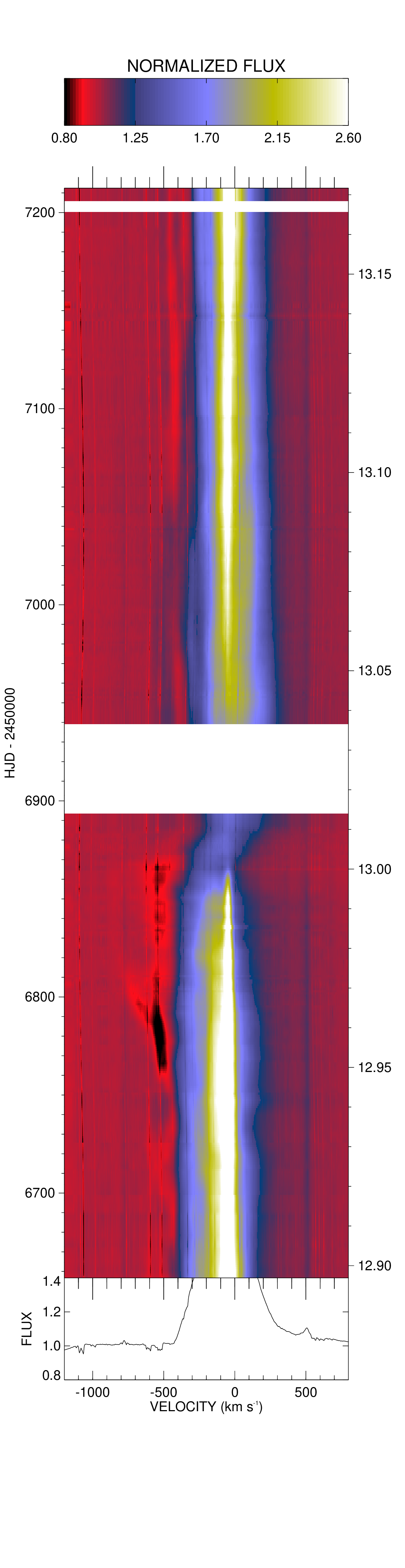}

\caption{\label{figdynam} Dynamical representations of He~I $\lambda$4713, He~I $\lambda$5876, and He~I $\lambda$7065 are shown (left to right). Time increases in the vertical direction, with average profiles calculated from a simple combination of all spectra shown in the bottom panels. Colour bars are given on the top for each representative figure. Interpolation was done for any gap less than 6~d, leaving only two gaps in the data stream, one following the periastron passage, and a short time-period around HJD~2,457,200.
All individual line profiles are shown in Appendix B1. }

\end{figure*}

In order to compare our observations to the binary clock, we adopt the orbital cycle terminology of Teodoro et al.~(2016), corresponding to a period of 2022.7~d, and a zero-point of HJD 2,456,874.4. This zero point corresponds approximately to the time of periastron passage, which was derived by a detailed modeling of the orbital modulation of the He~II $\lambda$4686 emission. For cycle 13, this zero point corresponds to HJD 2,456,865.2. The spectroscopic cycles are numbered since the first observed low/high state transition in 1948 (Gaviola 1953) as discovered by Damineli (1996), with the recent 2014 minimum corresponding to the beginning of cycle 13, leading to the ephemeris
$$\phi = {{ (HJD -2,452,819.2) }\over 2022.7} +11.$$

High spatial resolution, moderate dispersion observations were obtained with {\it HST}/STIS across the 2003.0 and 2014.6 periastron events through several programs including the {\it Hubble} Treasury Program (Davidson 2002, 2003) and a series of mapping programs (Gull et al. 2011; Teodoro et al. 2013).  All four programs used the 52\arcsec$\times$0.1\arcsec\ aperture, taking advantage of the 0.1\arcsec\ near-diffraction limit of {\it HST}. Spectra of interest were recorded centred at $5734$ \AA\ and included the He I $\lambda 5876$ \AA\ and Na~D lines. The observations recorded across the 2003.0 periastron event were single exposures centred on \ec at position angles defined by {\it HST} solar panel considerations. Nebular emissions, scattered starlight, including continuum, wind line emission and absorption noticeably extended well beyond the stellar core (Gull et al. 2009, 2011). For this reason, spectral extractions of the spatially-resolved spectral image were limited to the three rows (52\arcsec$\times$0.1\arcsec offset by 0.05") centered on \ec.

Spatial mapping, as demonstrated by Gull et al. (2011), revealed forbidden line emission extending out to at least 0.7\arcsec\ from the central stars that was determined to originate from fossil wind structures formed during previous periastron passages 5.5, 11, and 16.5 years earlier (Teodoro et al. 2013). At critical phases of the 5.54-yr cycle, 2\arcsec$\times$2\arcsec\ mappings were obtained (Gull et al. 2011; Teodoro et al. 2013). Given the typical seeing being about 2\arcsec\ and the size of the fiber optic feeding into the CHIRON spectrograph being 2.7\arcsec\ in diameter, comparison spectra of a 0.15\arcsec$\times$0.15\arcsec\ area centred on \ec and the integrated 2\arcsec$\times$2\arcsec\ area will be presented later in this paper.

\section{Results}

The average profile of each of the He~I lines investigated here is shown in the bottom panel of Fig.~1. In these figures, it can be seen that the profile generally has a P Cygni shape and that the bulk of the absorption is at velocities between $-400$ and $-500$ km s$^{-1}$, i.e. near the terminal speed of the primary LBV's wind ($420$ km s$^{-1}$).
We observed some additional P Cygni absorption components separate from this primary absorption in several He~I lines in the optical spectrum of $\eta$~Car prior to its 2014 event, mainly between approximately HJD 2,456,750 and 2,456,820, and between HJD 2,456,860 and 2,456,870. These additional absorptions appear separate from the normal P Cygni absorption, and usually at velocities much greater than $\sim$$420$ km s$^{-1}$. This was easiest to observe in He~I $\lambda\lambda$4713, 5876, 7065 (atomic data in Table 1). The other He~I profiles observed by CHIRON are He~I $\lambda\lambda$4921, 5015, 6678. Of these profiles, He~I $\lambda\lambda$4921, 5015 are blended with stronger Fe II transitions at the same wavelength, and the P Cygni absorption component of He~I $\lambda$6678 is blended with [Ni II] emission. Therefore, we will not consider these lines in this paper. Telluric line contamination made the analysis of He~I $\lambda$7065 difficult, but still allowed us to confirm trends in other lines. The blue edge of the He~I $\lambda$4713 profile is marginally blended with [Fe III] $\lambda$4702. In comparison to all of the other He I lines, the He~I $\lambda$5876 profile is clean on the blue edge, with no other strong emission lines present until $\sim5835$ \AA\ ([Fe II]; Zethson et al.~2012). In the emission portion of the profile, there is a weak [Fe II] emission that peaks at $v_r \sim -350$~km~s$^{-1}$ in our rest frame, and multiple Na D$_1$ and D$_2$ components on the red portion of the profile. The red side is also affected by telluric absorption, but it mainly affects the Na D transitions, and does not interfere with the He~I line.

\begin{table}
 \centering
 \caption{He~I atomic transitions investigated}
\begin{tabular}{lll}
\hline
Transition &  Wavelength(Vac)  &   $f$   \\
\hline
1s 2p $^3$P\textsuperscript{o} --  1s 3s $^3$S  & 7067.2\,\AA & 0.0695   \\
1s 2p $^3$P\textsuperscript{o} --  1s 3d $^3$D  & 5877.3\,\AA  & 0.610   \\
1s 2p $^3$P\textsuperscript{o} --  1s 4s $^3$S  &  4714.5\,\AA & 0.0106  \\
\hline
\end{tabular}
\end{table}  

In Fig. 1, we show three dynamical representations of the CHIRON data, one for each transition of He~I $\lambda\lambda$4713, 5876, 7065. The first column shows the He I $\lambda$4713 spectra leading up to and including the minimum, followed by its progression towards recovery (phases 12.89--13.17). While the data were interpolated to display this, no interpolation happened with a gap of more than 6~d. The key exception is the large data gap that exists between HJD 2,456,887.5 and 2,456,944.9 (phases 13.0110--13.0349). While our coverage near the periastron passage is exquisite, we were not able to have coverage as long into the event as Richardson et al.~(2015) had for the 2009 event due to constraints on the telescope and the season in which periastron occurred in 2014. Nevertheless, these datasets are compatible as they integrate light over both the same kinematical region, as well as use the same entrance fibre for the spectrographs.

The absorption variability is clearly seen in Fig.~1 and the kinematic signature of an accelerating component is clear between HJD 2,456,750 (phase 12.938) and 2,456,820 (phase 12.973) in He~I $\lambda$5876 and He~I $\lambda$7065, while it is seen for an extended time in He~I $\lambda$4713. The top portions of the plots show the data obtained as the system recovered from the periastron passage. Less variation is seen in these data, but there is a strengthening in the P Cygni absorption starting around HJD 2,457,080 (phase 13.10).

We attempted to measure the new pre-periastron absorption component observed between HJD 2,456,750 and 2,456,820 in a variety of ways. First, we attempted a numerical derivative method that was used by Richardson et al.~(2011) to examine the variability in the P Cygni absorption of the H$\alpha$ line in P Cygni. Unfortunately, this is not possible with these data as the absorption is very weak and blends with the P Cygni absorption from the stellar wind. We attempted to perform multiple Gaussian fits to the P Cygni profile, but with the mixture of the He~I wind emission, [Fe II] $\lambda$5871 emission, narrow He~I emission from the Weigelt knots, and the Na~D components, we were unable to produce adequate results. The only remaining method was an interactive approach to determine an ``edge" velocity to the absorption. We performed this interaction 10 times, with consistent results ($\sigma \sim 5$~km~s$^{-1}$ dispersion from an average). These are shown in Fig.~2. We note that these are not the typical edge velocities as in the cases of UV resonance lines, but rather a point of a minimum flux in the local portion of the profile, as seen in Fig.~2. This represents a velocity related to the largest optical depth of the feature in the system's outflow.

\begin{figure}
\includegraphics[width=57mm, angle=90]{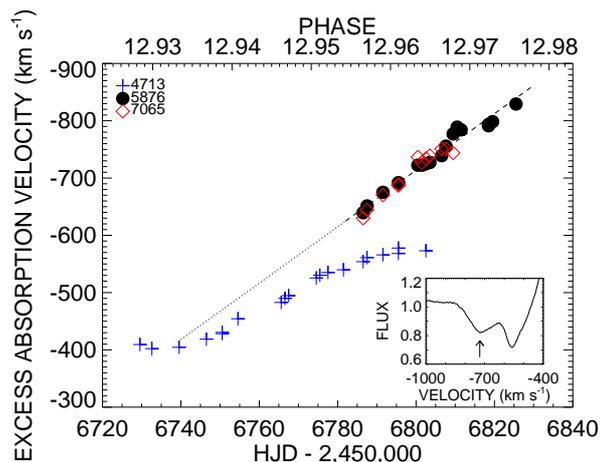}
\caption{\label{figacc} The kinematics of the excess absorption are shown for three He~I lines. The inset spectrum shows the absorption of a representative He~ I $\lambda$5876 profile and indicates the measured velocity with an arrow.}
\end{figure}

We fit a linear trend to the measurements of the peculiar high-velocity absorption component of He~I $\lambda$5876 and He~I $\lambda$7065 data (which were internally consistent) that resulted in a measured acceleration of $-4.93\pm0.19$~km~s$^{-1}$~d$^{-1}$. We overplot this linear trend on the data shown in Fig.~2. The overarching feature lasted for about 40--50 days, or $\sim 0.02$ in phase, centred near $\phi = 12.965$. We note that the feature measured for He~I $\lambda$4713 is likely different than the feature measured for He~I $\lambda\lambda$ 5876, 7065. This is discussed in more detail later. This feature was strong for about 40--50 days for He~I $\lambda\lambda$5876, 7065, with the accelerating feature measured in the He~I $\lambda$4713 profile originating in a different absorption component. We note that all three of these absorptions have the same lower energy state, but the optical depths in the He~I $\lambda$4713 and He~I $\lambda\lambda$5876, 7065 transitions differ by nearly a factor of 60. Near the end of this time frame, the feature evolved into a barely
discernible absorption where spectra with S/N greater than 200 are
required to detect it. The offset is likely due to these absorptions corresponding to different sources of the excess absorptions (see Fig.~6 and corresponding discussion.

\begin{figure*}
\includegraphics[width=125mm, angle=90]{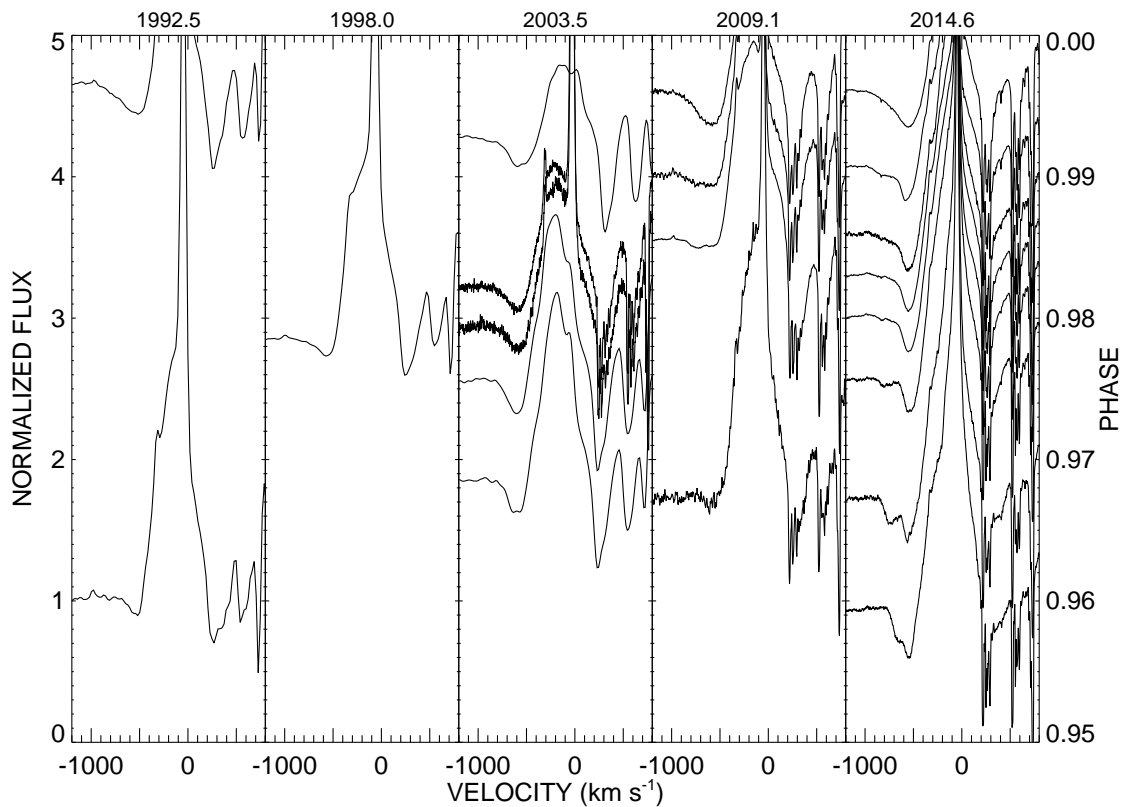}
\caption{\label{fig2009_2014} A direct comparison of He I $\lambda$5876 spectra obtained prior to the last five events shows that the P Cygni absorption seems fairly repeatable, with the exception of the secondary, excess absorption observed this cycle. The data sources for the events are 1992.5: Damineli et al.~(1998), 1998.0: Damineli et al.~(2000), 2003.5: {\it HST} Treasury program and FEROS results from Teodoro et al.~(2012), 2009.1: the CTIO 1.5~m and fiber-fed echelle (Richardson et al.~2015) and FEROS (Teodoro et al.~2012), and 2014.6: the CTIO 1.5~m and CHIRON (this work).}
\end{figure*}

The principal question regarding these absorption variations is their repeatability. The best time-series for comparison at similar orbital phases is available through the {\it HST} Treasury program\footnote{http://etacar.umn.edu/}, which had regularly-scheduled spectroscopic observations with STIS during the 2003 spectroscopic event. In Fig.~3, we show a direct comparison of CHIRON observations from the 2014 event with ground-based spectroscopy and {\it HST} observations from the previous four events. Unfortunately, our new CHIRON spectra sample a seeing-limited 2.7$\arcsec$ fibre, while {\it HST} observations are taken with a $52 \arcsec \times 0.1\arcsec$ slit with spatial resolution of 0.13$\arcsec$.

We compared the He I $\lambda$5876 profiles taken at the same phases. These results show that qualitatively the absorption behaviour is similar and repeatable, with some weaker absorptions seen in the 2009 event. The faint high-velocity ($\sim -750$~km~s$^{-1}$) absorption component centred near phase 12.965 (but lasting $\sim 0.02$ in phase) appears absent in the four previous cycles. This component seems to be a new phenomenon this cycle, as it was not seen at a comparable phase in 2009 (Richardson et al.~2015, Fig.~3). Is this feature due to some short-lived instability or is it representative of other long-term changes in the system (see e.g. Mehner et al.~2010b, 2012)? We note the extreme similarities at phases very close to periastron (0.99) in 1992, 2009, and 2014, which are indistinguishable, along with strong similarities for the spectra seen at phase 0.98 in 1998, 2003, and 2014. Of particular importance is the very high velocity absorption ($\lesssim -900$~km~s$^{-1}$) observed in He~I $\lambda$5876 near phase 12.995 (see line profiles in Appendix B1). This is nearly identical to the $\sim-900$~km~s$^{-1}$ absorption observed in the same line at the same phase during the 2003.5 event (Nielsen et al.~2007), and to the high velocity blue-shifted absorption observed in He~I $\lambda$10830 by Groh et al.~(2010) during the 2009 event between phases 11.976 and 12.023. This high velocity absorption component near phase 12.995 thus appears to be periodic and repeatable, and very likely a result of the trailing arm of the WWC region crossing our line-of-sight just before periastron (see Section~6).

The bulk of the P Cygni absorption profiles appear to qualitatively repeat from cycle to cycle, but there are some notable changes, especially around phase 0.96. As we will discuss later in the paper, this is a portion of the orbit where the trailing arm of the shock cone passes through our line-of-sight. With a small change (few percent) of mass-loss rate or wind speed of either star, the opening angle of the shock cone can easily change by a few degrees. Considering the system's geometry relative to our line-of-sight, these very small changes can cause us to look through the shock cone during one orbit but not during other orbits. Therefore, the excess absorption components seen at velocities much higher than the terminal wind speed of the primary that were observed during some events but not others can be associated with the various (Rayleigh-Taylor, non-linear thin shell, etc.) instabilities occurring in the shock region.

Mehner et al.~(2015) described the overall variability of He I $\lambda$4713 as observed with {\it HST}/STIS from 1998 through the present epochs. They found that the P Cygni absorption component may be strengthening with time, but seems to strengthen around orbital phases 0.90--0.98, and again several months following periastron passage (phases near 0.1--0.2). They argue that this can be explained through a qualitative comparison with the expected ionization-zone shapes. They also show a long-term rise in the emission component which may be correlated with an increase in the continuum emission. Our comparison of data from 2003, 2009, and 2014 does not show very large changes (Fig.~3), and the profiles from 2009 and 2014 at the same orbital epoch look identical in strength when viewed at the same resolution. There are some differences in ground-based and space-based spectra that could allow for some changes, but qualitatively these profiles seem very similar except for the fainter high-velocity absorption component near phase 12.965 seeming to appear and/or be stronger this cycle.

To further check for repeatability, we measured the He~I $\lambda$5876 profile via two methods. First, we did a direct comparison of the equivalent width to that of Richardson et al.~(2015). This integrates the profile from $-1000$ to $+2000$~km~s$^{-1}$, which also includes the Na I D emission and absorption profiles entirely. This is shown in the top panel of Fig.~4. The resulting trends show that the events are statistically indistinguishable, implying that there cannot be drastic changes in the mass-loss rates of the stars. Second, we measured the point in the P Cygni absorption profile corresponding to ``minimum flux'' both by a direct measure (velocity at the pixel with the fewest recorded photons) and through a gaussian fit to the $\sim11$ pixels surrounding that point. We show these measurements in the middle panel of Fig.~4, along with the measurements from Richardson et al.~(2015). Again, the points are nearly indistinguishable for the two periastron passages. Nielsen et al.~(2007) investigated whether these absorptions move with the primary star's wind, and could thus be related to the spectroscopic orbit, where the $\gamma$ velocity is offset by the primary terminal wind speed ($\sim 420$~km~s$^{-1}$). We show an overplotted orbit with similar orbital parameters in Fig.~4, and discuss this in Section~4.

\begin{figure}
\begin{center}
\includegraphics[width=80mm, angle=0]{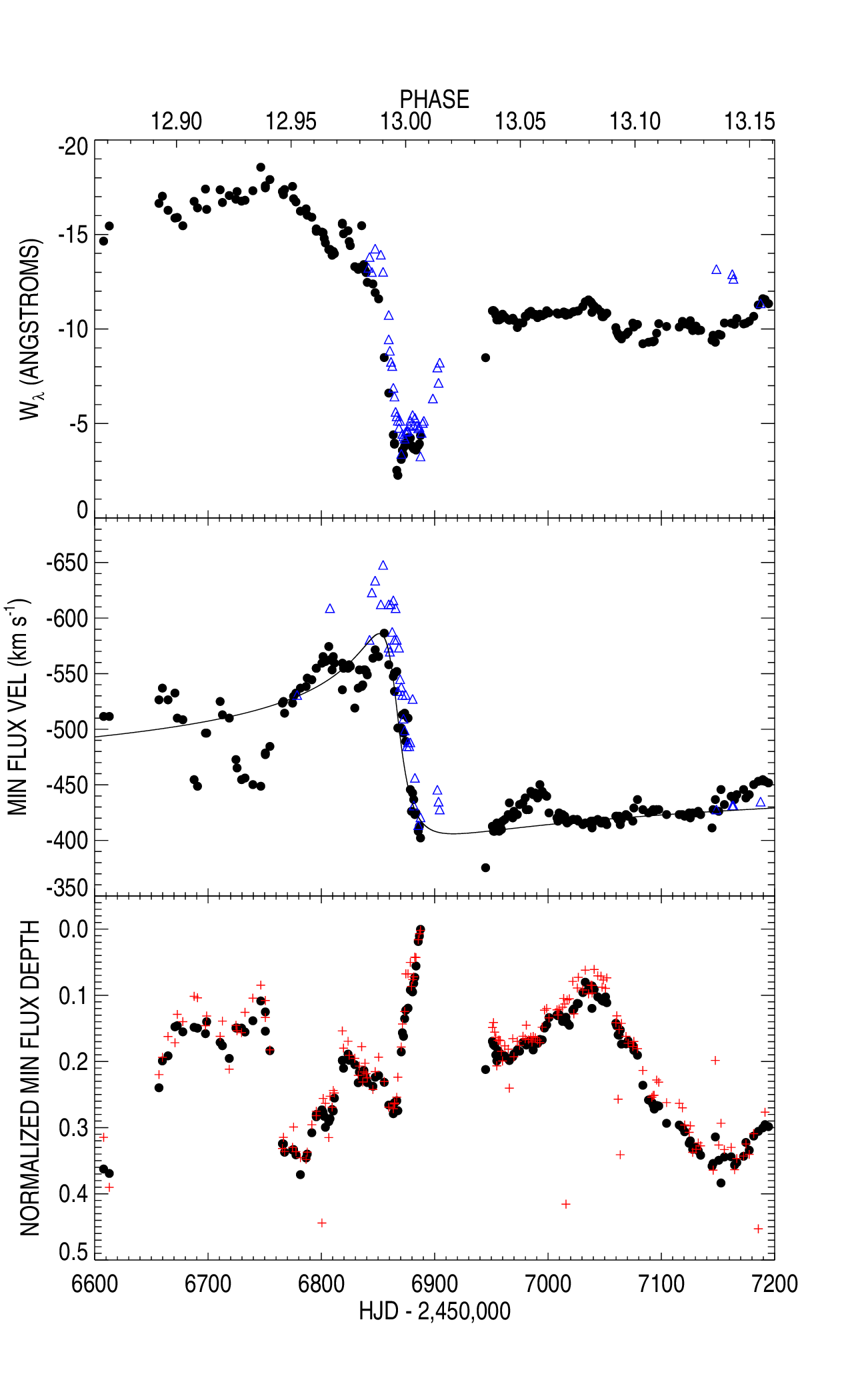}
\end{center}
\caption{\label{ew_minflux} Measurements of He~I $\lambda$5876. The top panel shows the equivalent width. The second panel shows the measured velocity where the minimum flux occurs in the P Cygni absorption component (black circle = measured, red $+$ is a gaussian fit) with an overplotted orbital velocity curve (Section 4.1), and the third panel shows the depth of the P Cygni absorption in continuum normalized units. In the top two panels, we overplot as blue triangles measurements from Richardson et al.~(2015), whereas the measurements of the absorption depth were too complicated for the lower signal-to-noise in the data from Richardson et al.~(2015). All measurements are tabulated in Table A1. We note that the top panel is dominated by emission line changes, while the bottom two panels are uniquely associated with the P Cygni absorption components.}
\end{figure}

\section{The He~I Absorptions and Their Relation to the Central Binary}

The program to map the spatial distribution of emission from \ec's fossil WWCs using {\it HST}/STIS (Gull et al.~2011; Teodoro et al. 2013; Gull et al., submitted) had a mapping visit at a time coincident to when we observed relatively high velocity ($\sim -700$~km~s$^{-1}$) He~I absorption (HJD 2,456,817; $\phi=12.977$) not seen during earlier events. The {\it HST}/STIS observations mapped the distribution of both absorption and emission of all three He~I transitions discussed to an angular resolution of $\sim 0.1 \arcsec$. The {\it HST}/STIS data show that the absorption profile at $\phi = 12.977$ is entirely concentrated on the central source, telling us that the absorption must arise in the inner $\lesssim 230$~AU of the system, close to the unresolved central binary and colliding winds region (Fig.~5). This is consistent with the results of earlier {\it HST}/STIS observations, which show that the broad He~I features are unresolved and almost certainly originate less than $\sim$ 100--200 AU from the primary star (Nielsen et al. 2007; Humphreys, Davidson, \& Koppelman 2008). The inferred location of high-velocity absorption seen in He~I $\lambda$10830 during \ec's 2009 periastron event (Groh et al.~2010) also agrees with this result.

\begin{figure}
\includegraphics[width=80mm, angle=0]{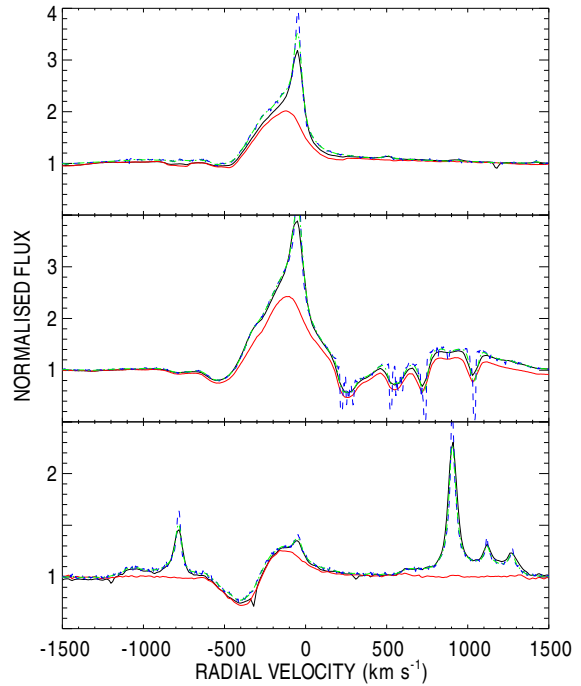}
\caption{\label{figchiron_stis} Comparison between ground-based CHIRON and {\it HST}/STIS spectroscopy. The spatially resolved spectra obtained with {\it HST}/STIS (GO program 13395; PI Gull), using the 52\arcsec$\times$0.1\arcsec\ aperture, are compared to ground-based, seeing limited CHIRON data. Red: the central source of the system (0.15\arcsec) extracted from the spectral map obtained at phase 12.977. Black: summation of the STIS mapping data over the 2\arcsec$\times$2\arcsec\ total STIS mapped area. Blue-dashed line: simultaneous CHIRON spectrum; green-dot-dahed line: CHIRON data smoothed to the resolving power of STIS. Differences between the integrated STIS and smoothed CHIRON spectra are nearly indistinguishable. }
\end{figure}

The He~I absorption during the five months prior to periastron is very dynamic. In Fig.~6, we show a dynamical representation of the difference between the observed profiles and the average profile of He~I $\lambda$4713 and He~I $\lambda$5876. From this view, we clearly see four, possibly five, distinct absorption components. We labeled these for He~I $\lambda$4713, and placed markers at the same locations for the He~I $\lambda$5876 panel. We note that some absorption features for He~I $\lambda$4713 in Fig.~6 may extend to velocities more negative than $\sim-750$~km~s$^{-1}$ since emission from [Fe III] $\lambda$4702 is marginally blended with He~I $\lambda$4713 and prevents us from seeing the maximum absorption velocity present. However, He~I $\lambda\lambda$5876 and 7065 do not suffer from this problem and may be used to help determine the max blue-shifted velocity of the absorption. The resemblance of the two panels is striking, although the stronger emission component of He~I $\lambda$5876 makes the portion of the diagram between $-200$ and $0$ km s$^{1}$ more difficult to interpret.

The first and most obvious absorption component (\#1 in Fig.~6) begins before HJD 2,456,700 ($\phi = 12.9138$), centred at $\sim -340$~km~s$^{-1}$, and remains roughly at that velocity until about HJD 2,456,725 ($\phi = 12.9261$), at which point the absorption component curves and extends with a slight slope to $\sim -550$~km~s$^{-1}$ near HJD 2,456,790 ($\phi = 12.9583$).
The second component (\#2 in Fig.~6) is centred at $\sim -275$~km~s$^{-1}$ and begins near HJD 2,456,740 ($\phi = 12.9336$), with fainter absorption extending to near HJD 2,456,770 ($\phi = 12.9484$).
A third component that is possibly related to the second component (\#3 and the `?' in Fig.~6) is centred at $-450$~km~s$^{-1}$ and extends from
HJD 2,456,810 ($\phi = 12.9682$) to HJD 2,456,840 ($\phi = 12.9830$). We suspect that this is oversubtracted when compared to the mean profile in Fig.~6, as it seems to extend between the time of \#2 and \#3 in Fig.~1, where we plot the observed profiles.
The final component (\#4 in Fig.~6) is fairly weak in He~I $\lambda 4713$ but easily discernible for He~I $\lambda5876$. It has a shallow slope, extending from $\sim -425$~km~s$^{-1}$ on HJD 2,456,770 ($\phi = 12.9484$) to $\sim -750$~km~s$^{-1}$ on HJD 2,456,810 ($\phi = 12.9682$), and represents the acceleration measured in Fig.~2 for the He~I $\lambda\lambda$5876, 7065 profiles, while the measurements for He~I $\lambda$4713 originate in component \#1.

\begin{figure*}
\includegraphics[width=75mm, angle=0]{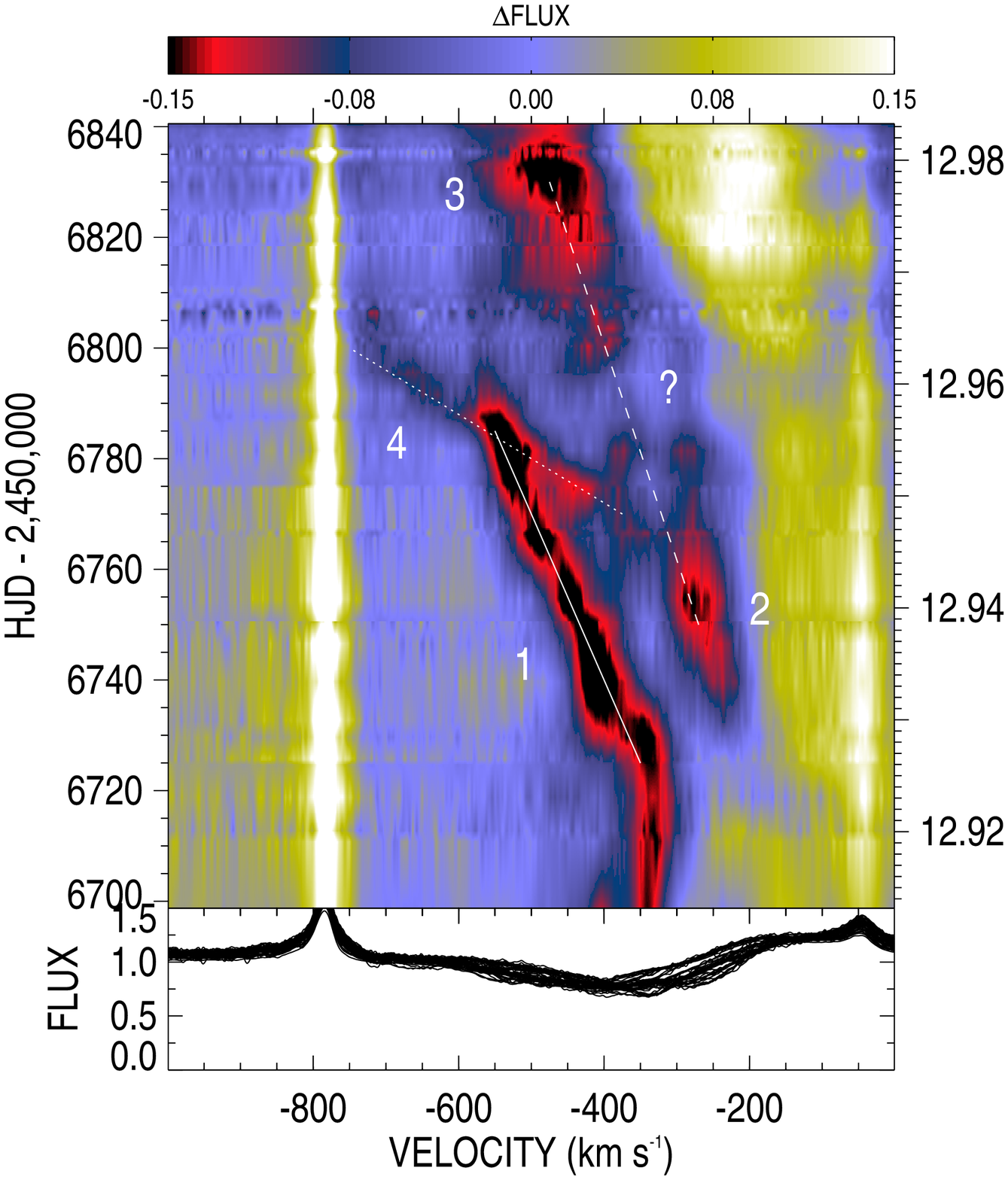}
\includegraphics[width=75mm, angle=0]{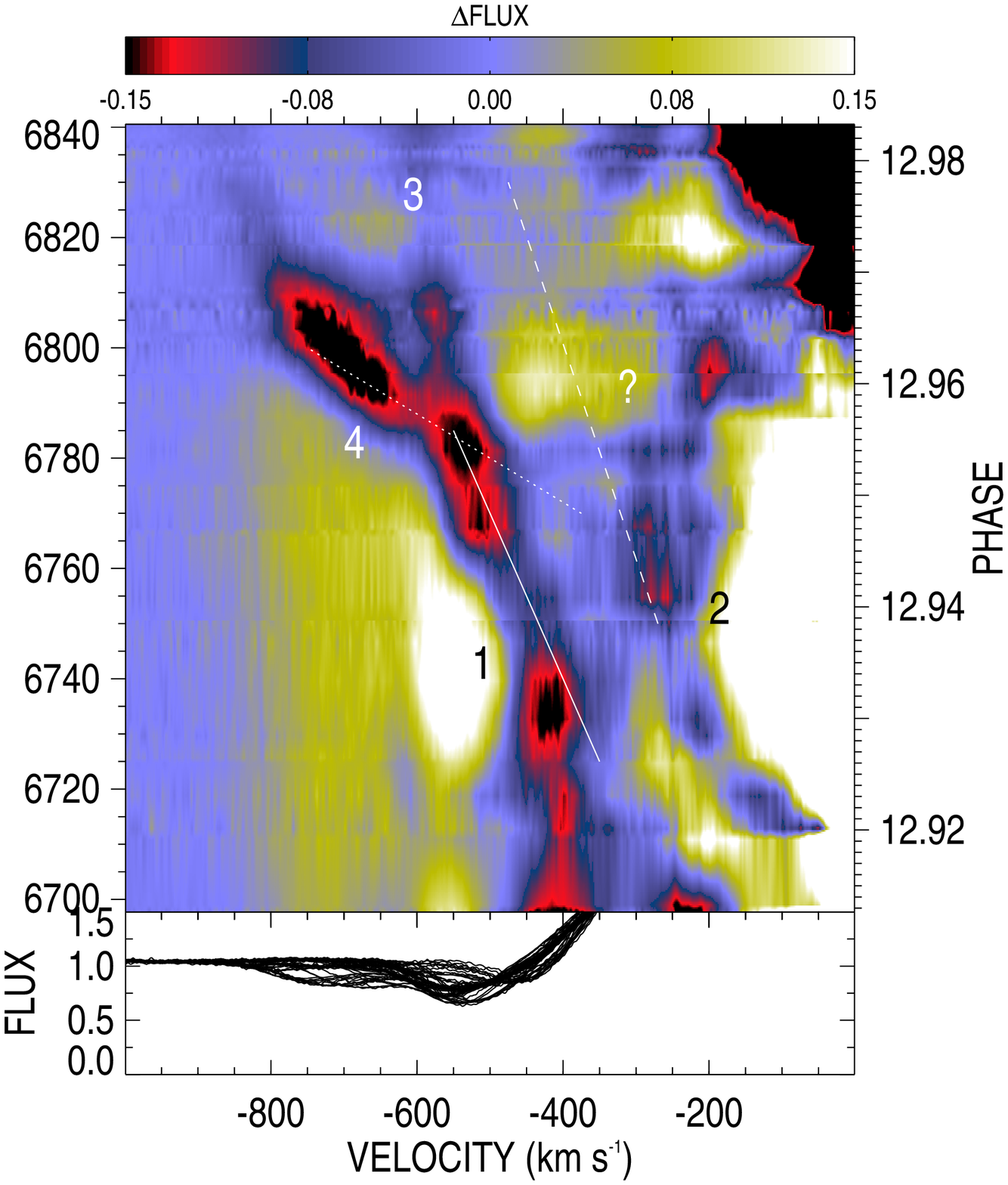}

\caption{\label{dynamzoom} A dynamical representation of the difference between the observed He~I $\lambda$4713 (left) and $\lambda$5876 (right) profiles and an average profile. The individual profiles are plotted on the bottom panel, with a representative colour bar on top. Individual labeled features are discussed in the text, and the lines are positioned with the same slope for both panels. }
\end{figure*}

Some of these absorption components, especially those with velocities more positive than $-450$~km~s$^{-1}$, are not easily visible in the dynamic spectra of He~I $\lambda\lambda$5876, 7065 (columns two and three of Fig.~1) due to the much stronger emission in these lines. However, the weak, lowest-velocity absorption component (\#4 in Fig.~6) that extends to $\sim -800$~km~s$^{-1}$ at HJD 2,456,810 is present in all three He~I lines. Moreover, in the He~I $\lambda$5876 profiles (middle column of Fig.~1) there is absorption that extends to $\lesssim -900$~km~s$^{-1}$ just before periastron, between about HJD 2,456,859 ($\phi = 12.9924$) and HJD 2,456,868 ($\phi = 12.9968$, see also Appendix B). Interestingly, at about HJD 2,456,870 ($\phi = 12.9980$), the most negative velocity absorption in the He~I $\lambda$5876 line extends to only $\sim -700$~km~s$^{-1}$, and continues to decrease in absolute value through periastron, where the absorption is centred at $\sim -450$~km~s$^{-1}$. The He~I $\lambda$5876 absorption remains centred about $\sim -400$ to $-450$~km~s$^{-1}$ and fairly weak until about HJD 2,457,070 ($\phi = 13.1000$), when it starts to strengthen (i.e. deepen) significantly. Meanwhile, the He~I $\lambda\lambda$4713, 7065 absorptions remain absent/extremely weak all through periastron passage and do not return strongly until the observations resume at HJD 2,456,944 ($\phi = 13.0388$). When they do, there are clearly two absorption components present in the He~I $\lambda$4713 profile centred at $\sim -250$~km~s$^{-1}$ and $\sim -400$~km~s$^{-1}$. These components remain until about HJD 2,456,970 ($\phi = 13.0511$), when the velocity component at $-250$~km~s$^{-1}$ starts to deepen and dominate the He~I$\lambda$4713 absorption. This continues until about HJD 2,457,001 ($\phi = 13.0670$), when another strong velocity component at $\sim -350$~km~s$^{-1}$ appears. These two absorption components eventually merge into a single strong broad component centred at $\sim -350$~km~s$^{-1}$ at $\phi = 13.0818$. This remains the dominant absorption component in He~I $\lambda$4713 for the remainder of the observations. However, in He~I $\lambda$5876, the absorption that strengthens near $\phi = 13.1000$ does so at a lower velocity of $\sim -400$ to $-450$~km~s$^{-1}$. For the remainder of the observations, the absorption appears strongest in He~I $\lambda$5876, centred near $-450$~km~s$^{-1}$. He~I $\lambda$7065 shows very similar absorption at these times, but is more difficult to measure due to the telluric contamination present in this region of the spectrum.

Nielsen et al.~(2007) initially assumed that the P~Cygni absorption mimics orbital motion of the primary. We further tested this hypothesis with the measurements of He~I $\lambda$5876, and in particular the velocity of the deepest absorption. This absorption can be seen to vary greatly in its strength (Fig~4., bottom panel), and we also see in the dynamical plots of Fig.~1 that the motion is sometimes not as simple as orbital motion, where between HJD 2,456,670 and 2,456,750 the absorption seems to have a sinusoidal appearance to it. In the middle panel of Fig.~4, we overplot a potential orbit based upon the periastron passage calculated by Teodoro et al.~(2015), an eccentricity of 0.88, a semi-amplitude of 90~km~s$^{-1}$, $\omega_{\rm primary}=243^\circ$, a $\gamma$ velocity of $-460$~km~s$^{-1}$, and orbital period of 2022.7~d. This overplotted orbit may appear to represent the overall variability fairly well, but it cannot explain deviations in the velocities. The overplotted orbit is also very misleading given the complexity of the \ec binary and does not take into account important photoionization effects due to the presence of the hot companion star (Clementel et al.~2014, 2015a,b), nor the presence of high velocity absorbing material in line-of-sight that is located within the WWC regions (Groh et al.~2010; Madura et al.~2013). As noted by Damineli et al.~(2008b), and supported by the work of Groh et al.~(2010) and Clementel et al.~(2015a,b), while some of the He~I absorption in \ec may originate in the pre-shock wind of the primary star at velocities near the wind terminal speed of $\sim420$ km~s$^{-1}$, a significant and likely dominant amount of absorption comes from dense post-shock wind material located in the WWC region. One simply \emph{cannot} associate the He~I absorption with solely one of the stars, regardless of the assumed orbital orientation, since around periastron passage material from the WWC will cross between the observer and the primary continuum source, leading to multiple absorption components at a range of different velocities.

\section{Origin of the He~I Absorption Components}

\subsection{The Primary Continuum Source and Background Theory on the Colliding Winds}

Since the He~I lines are recombination lines, they are produced in regions of He$^{+}$ rather than regions of neutral He. The broad optical He~I wind lines are thought to be excited by the UV radiation of the companion star and arise somewhere in/near the WWC region between the stars (Nielsen et al. 2007; Damineli et al. 2008a). Because we are analyzing absorption lines in which the absorption is always blue-shifted, the regions responsible for the different absorption components must be located between the primary continuum source and the observer.

Due to its very dense wind, the primary star in \ec has an extended continuum emitting region at visible and near-IR wavelengths (Hillier et al. 2001, 2006; Groh et al. 2012a,b). At the optical wavelengths of the observed He~I lines in this paper, the continuum emitting region of the primary has a radius of $\sim$ 1.5--2~AU (based on the models of Hillier et al. 2001, 2006; Groh et al. 2012a, and interferometric measurements in the $K$-band by van~Boekel et al. 2003; Weigelt et al. 2007). Therefore, in our discussion below, we assume a primary radius of 1.5~AU for the size of the continuum emitting region at 4713\AA\ and 5876\AA, and a radius of 2~AU at 7065\AA. While rather large, the continuum emitting region is still much smaller than the extent of the inner WWC zone (which extends $\sim \pm200$~AU), and comparable in size to the thickness of the densest part of the WWC zone ($\sim$1~AU; Madura et al. 2013; see Figs.~7--14).

When the two stellar winds in \ec collide, they produce a pair of hydrodynamical shocks (i.e. discontinuities in fluid flow) separated by a contact discontinuity (CD, a surface of pressure equilibrium between material of different composition and entropy), creating a WWC region. The shape of the WWC region is determined mainly by the wind momentum ratio $\eta \equiv (\dot{M}v_{\infty})_{2}/ (\dot{M}v_{\infty})_{1}$, where $\dot{M}$ and $v_{\infty}$ are the stellar mass loss rates and wind terminal speeds, respectively, for the two stars (e.g., Stevens et al. 1992). The aberration angle and degree of downstream curvature of the WWC zone are determined by the ratio of the orbital speed to the pre-shock wind speed (Pittard 2009). In the absence of significant orbital motion, the WWC zone has a simple paraboloidal shape that curves towards the star with the weaker wind (Canto et al.~1996). In highly elliptical systems like \ec, as the stars approach periastron, orbital speeds increase approaching the primary wind speed, highly distorting the WWC region, giving it a distinct spiral shape after periastron (Okazaki et al. 2008; Parkin et al. 2011; Madura et al. 2012, 2013, 2015). Orbital motion also alters our viewing angle to different portions of the WWC zone, and hence the strength and radial velocity of observed emission and absorption lines (Damineli et al. 2008). Different portions of the WWC regions can also be illuminated by or shadowed from different types of stellar radiation during the orbit, resulting in a `lighthouse effect' (Madura \& Groh 2012). Different lines can then be observed in different regions as the stellar `beams' of radiation sweep through the system across the orbital cycle.

Additionally, at the two shock fronts, the enormous kinetic energy of the incoming winds is converted into heat, causing a huge increase in the temperature of the post-shock gas, up to $\sim$10$^{8}$~K. The physical properties of the gas in the WWC regions then depend on the efficiency of radiative cooling. If cooling proceeds slowly compared to the wind flow time, the post-shock gas is adiabatic, and its density can increase by a factor of four relative to the pre-shock wind density. The adiabatic regime is typically encountered where pre-shock wind densities are low. If cooling operates quickly, then the post-shock gas is radiative, and its density can increase by several orders of magnitude relative to the pre-shock wind density. This frequently occurs in short-period systems, and/or high mass loss rates. In certain cases, such as \ec, both can occur simultaneously, and sometimes, depending on the orbital phase, the shocks can switch from one cooling regime to the other (Madura et al.~2013).

Throughout \ec's 5.54-yr orbit, the post-shock gas on the primary's side of the CD is radiative, while for most of the orbit, the post-shock gas on the companion's side of the CD is adiabatic (Parkin et al. 2011; Madura et al. 2013). However, due to the extremely high orbital eccentricity and subsequent decrease in pre-shock secondary wind velocity, during periastron passage, the post-shock wind on the companion's side of the CD can briefly switch from the adiabatic to the radiative regime and back, possibly explaining the changes observed in the timing of the recovery of the X-ray light curve out of periastron (Corcoran et al. 2010; Madura et al. 2013; Corcoran et al., in prep.).

\subsection{Interpretation of the Observations Based on 3D Hydrodynamical Simulations}

With these details in mind, we can use 3D hydrodynamical simulations (e.g. Madura et al. 2013) to help understand qualitatively where the different observed He~I absorption components arise in \ec. For simplicity, we assume that orbital phase is nearly the same as the spectroscopic phase we adopted. Based on the recent analysis in Teodoro et al.~(2016), the two are not believed to differ by more than approximately one week. 

\begin{figure*}
\includegraphics[width=17cm]{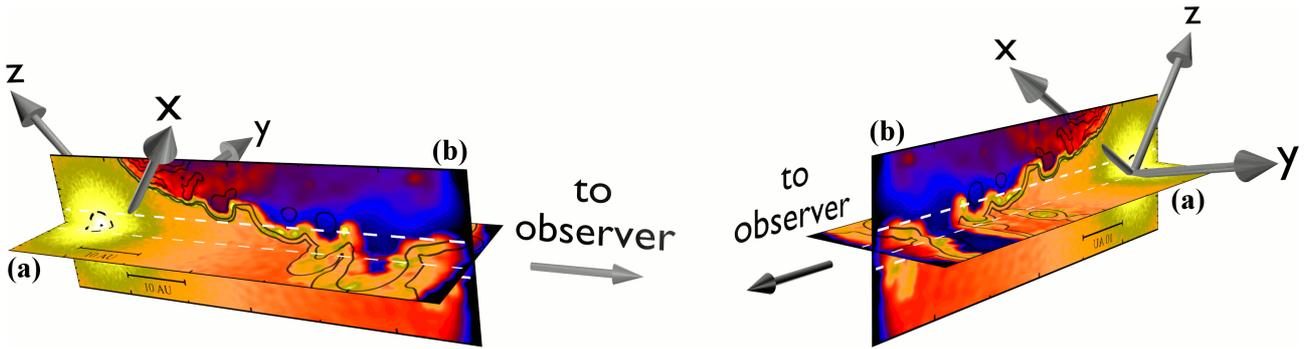}
\caption{\label{fig8} Diagram illustrating the orientation of the 3D simulation planes shown in Figs.~8--14 relative to the orientation of the \ec binary orbit and the observer. The planes shown correspond to the log density planes in the top row of Fig.~8 ($\phi = 0.9140$) and the top axes relate to the binary orbit which lies in the $x-y$ plane. See text of Section~5.2 for details.}
\end{figure*}

\begin{figure*}
\centering
\includegraphics[width=16cm]{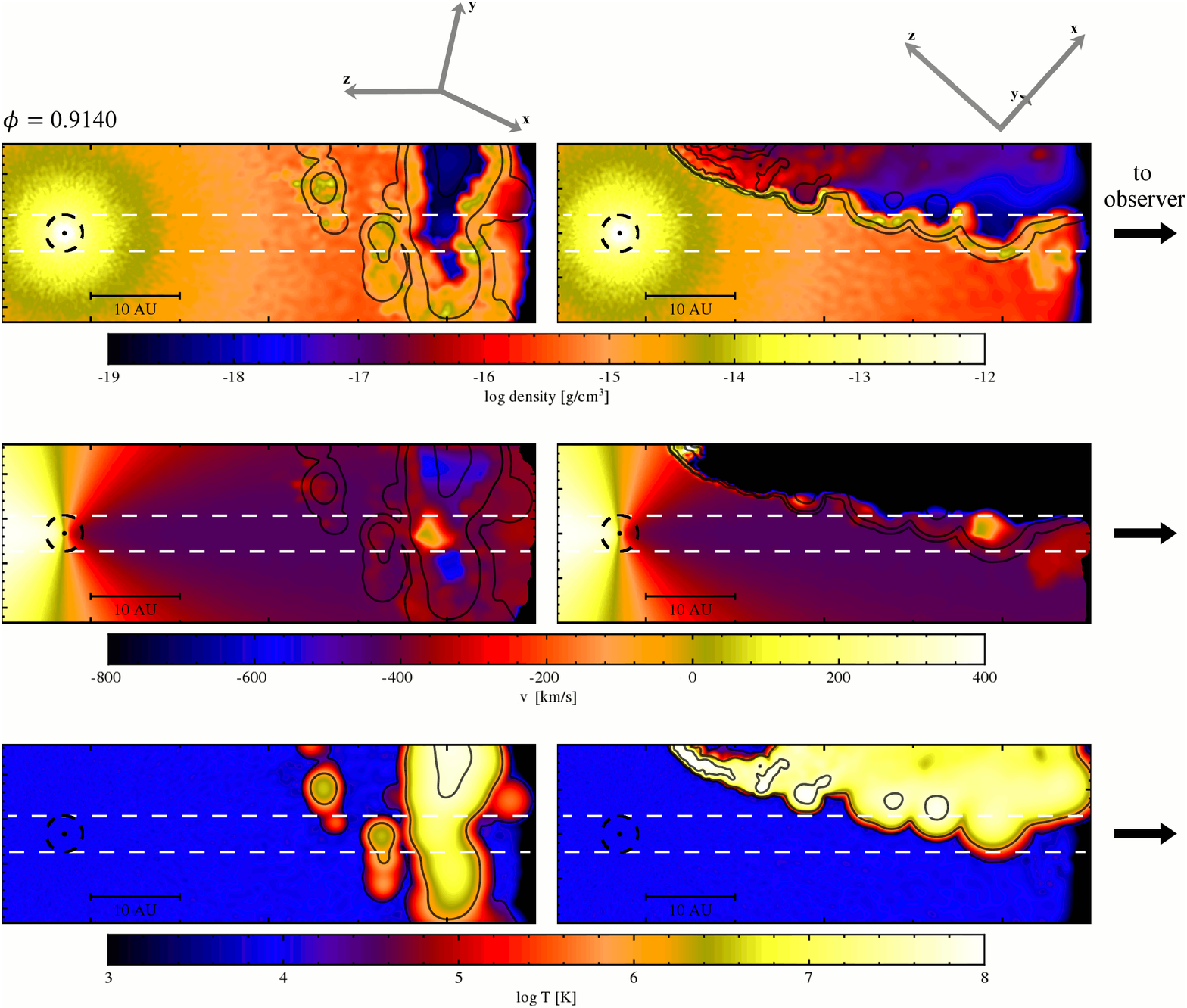}
\caption{\label{phase0p9140} Slices showing log density, radial velocity, and log temperature (rows, top to bottom) in two perpendicular planes containing the observer's line-of-sight at $\phi = 0.9140$ from a 3D SPH simulation of \ec's binary colliding winds. Fig.~7 illustrates the orientation of these planes [(a), left column; (b), right column] relative to the coordinate system that defines the binary orbit and the direction to the observer. Each plane passes through the centre of the primary star, and the orbit is assumed to have an inclination $i = 138^{\circ}$ and argument of periapsis $\omega_{\rm p} = 252^{\circ}$. The projected $x, y,$ and $z$ axes are shown at the top of each column. Panels show a zoomed region spanning $\pm$10~AU about centre of the primary star in the vertical direction and $-$10~AU to $+$50~AU in the horizontal direction. The length scale is indicated in each panel. In all panels the observer views the system from the right, indicated by the black arrows. The line-of-sight to the primary star is marked by two parallel white dashed lines, separated by the approximate diameter of the primary continuum-emitting region (4~AU). The continuum emitting region of the primary is indicated by a black dashed circle 4~AU in diameter for the He~I wavelengths of interest. Three temperature contours at levels of $1.5 \times 10^{4}$, $10^{6}$, and $5 \times 10^{7}$~K are overlaid on each panel to help in identifying the regions of cooler material likely responsible for the He~I absorption. In the first column, the stars orbit counterclockwise.}
\end{figure*}

Figs.~8--14 show log density, radial velocity, and log temperature in two perpendicular planes containing the observer's line-of-sight, at several orbital phases from a 3D SPH simulation of \ec's binary colliding winds. Fig.~7 illustrates the orientation of these planes relative to the coordinate system that defines the binary orbit and the direction to the observer. The labeled \emph{x,y,z} axes in Fig.~7 have their origin at the binary system centre of mass, where the $xy$ plane defines the binary orbital plane and the $x$ and $y$ axes lie along the semimajor and semiminor axes of the orbit, respectively. The $z$ axis defines the orbital angular momentum axis and is perpendicular to the $xy$ orbital plane (for reference, see Figs.~3 and 12 of Madura et al. 2012 and Fig.~10 of Madura et al. 2013). Each plane in Figs.~8--14 passes through the centre of the primary star (which is offset from the system centre of mass), and the orbit is assumed to have an inclination $i = 138^{\circ}$ and argument of periapsis $\omega_{\rm p} = 252^{\circ}$, consistent with the currently best-known binary orientation parameters (Okazaki et al. 2008; Parkin et al. 2009; Groh et al. 2010, 2012a,b; Madura \& Groh 2012; Madura et al. 2012, 2013; Teodoro et al. 2016).

The planes in the left column of Figs.~8--14 correspond to the plane labeled (a) in Fig.~7, while the right column in Figs.~8--14 corresponds to plane (b) in Fig.~7. For added clarity, the left and right columns of Figs.~8--14 include a display of the projected $x, y,$ and $z$ axes at the top of each column. The line-of-sight to the primary star is marked by two parallel white dashed lines, separated by the approximate diameter of the primary continuum-emitting region (4~AU). The observer is located to the right of each panel. The primary star is marked by a small filled black circle surrounded by a dashed black circle of diameter 4~AU. The exact location of the secondary star is not shown since it lies outside of the panels in the figures.

The SPH simulation used for Figs.~8--14 is identical to the `Case~A' simulations in Madura et al.~(2013), but uses a spherical computational domain with radius $r \approx 50$~AU. We refer the reader to Madura et al.~(2013) for details on the SPH code, numerical methods, and simulation parameters. The orbital phases chosen for Figs.~8--14represent those at which specific He~I absorption features are observed leading into periastron passage. The figures show a zoomed region spanning $\pm$10~AU about the centre of the primary star in the vertical direction and $-$10~AU to $+$50~AU in the horizontal direction, which should be fairly representative of the dynamical region responsible for the bulk of the time-variable He~I absorption. Color in the panels of the middle row of Figs.~9--15 shows the radial velocity of the gas with respect to the observer for the assumed orbital orientation. We note negative velocities indicate blue-shifted material moving toward the observer, while positive velocities represent redshifted material. 

We note that with the exception of $\phi = 0.9951$ and periastron ($\phi = 1.000$), the 3D SPH simulation shows the presence of large amounts of extremely hot ($>$ $10^{6}$~K) post-shock secondary wind (bottom row of Figs.~8--14), which is responsible for the observed time-variable X-ray emission in the system (Corcoran et al.~2010). At such high temperatures, helium is fully ionized to He$^{2+}$. Because the inclination of the orbit is $\approx 138^{\circ}$, the observer's line-of-sight to the primary continuum source lies very close to the WWC region and passes through the post-shock secondary wind and the pre- and post-shock primary wind. There is little or no pre-shock secondary wind in the line-of-sight that could lead to any measurable He~I absorption. Thus, we may safely exclude the pre- and post-shock secondary wind as the source of any of the observed He~I absorption components at phases between $\phi \approx 0.91$ and $0.99$.

\subsubsection{Absorptions at $\phi =$ 0.9140 and 0.9386}

Figs.~8 and 9 show that at $\phi = 0.9140$ and 0.9386, respectively, the observer's line-of-sight to the primary continuum source passes through the turbulent WWC region and pre-shock primary wind. Most of the material in the WWC region is too hot ($\gtrsim10^{6}~K$) to produce any significant He~I absorption. The density and temperature plots show that there is, however, a significant column of cooler pre-shock primary wind in the line-of-sight. This material has radial velocities between $\sim$$-360$~km~s$^{-1}$ to $-440$~km~s$^{-1}$ (see middle row of Figs.~8 and 9). This is consistent with the He~I absorption velocities observed at these phases (Figs.~1 and 6). This suggests that at these phases, the observed absorption is mainly due to the dense pre-shock primary wind in the line-of-sight. 

\begin{figure*}
\includegraphics[width=16cm]{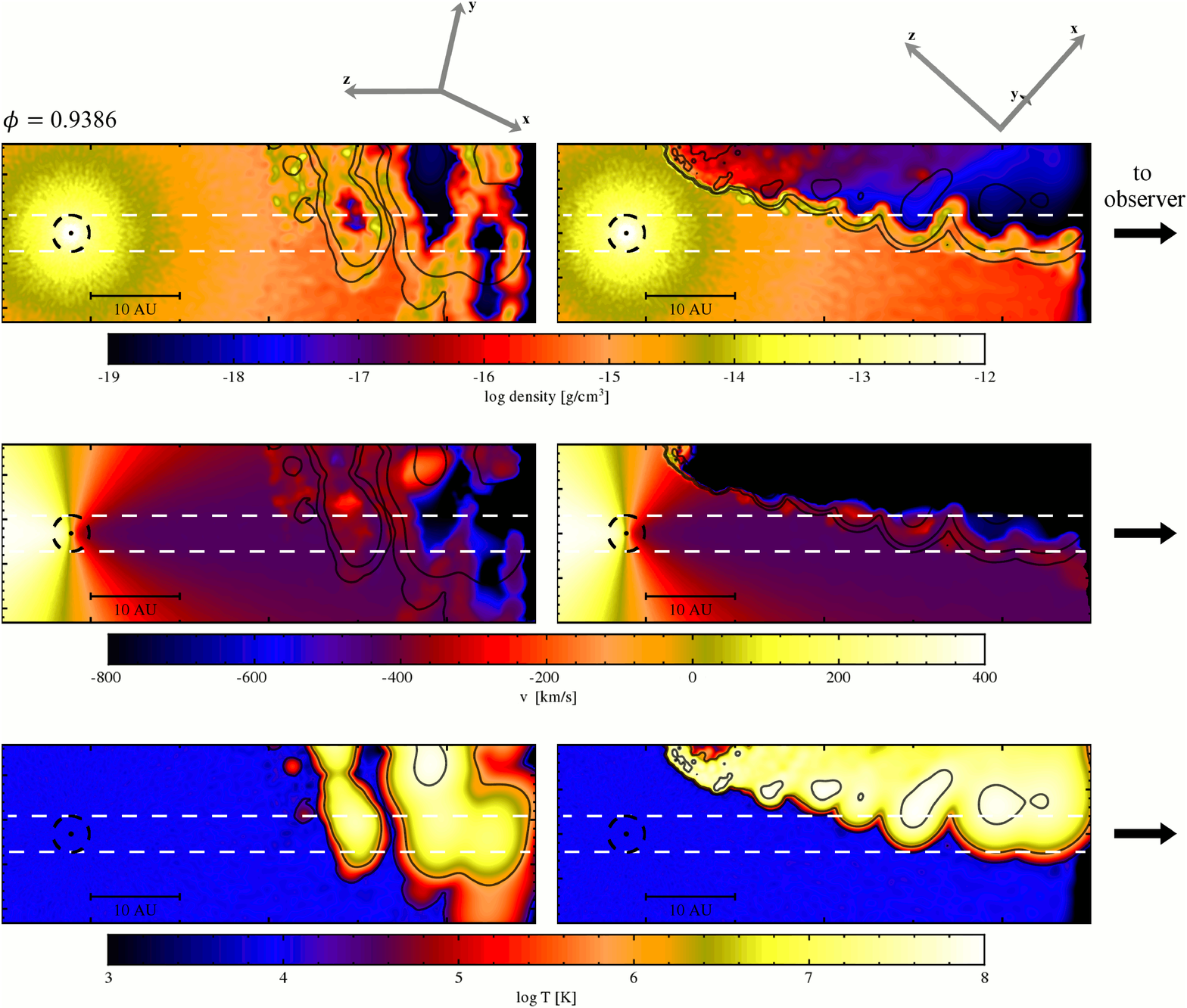}
\caption{\label{phase0p9386} Same as Fig.~\ref{phase0p9140}, but at $\phi = 0.9386$.}
\end{figure*}

\subsubsection{Absorptions at $\phi =$ 0.9569 and Features \#1 and \#2}

As the binary system approaches periastron and the shape of the WWC region begins to distort in the direction of orbital motion due to the increasing orbital speeds of the stars, the WWC region becomes even more turbulent. Fig.~10 shows that the line-of-sight at $\phi = 0.9569$ passes through a $\sim$5~AU thick region of density-enhanced, cool post-shock primary wind in the WWC region prior to passing through the pre-shock primary wind. This post-shock primary wind is composed of blobs of density-enhanced cool material moving at radial velocities on the order of $\sim$$-350$~km~s$^{-1}$ to $-300$~km~s$^{-1}$ (middle row of Fig.~10), which is noticeably less in magnitude than the primary's wind terminal velocity. We hypothesize that it is these dense blobs of slower-moving compressed post-shock primary wind that are responsible for absorption feature \#2 in Fig.~6. Absorption feature \#2 appears at roughly the same time in our simulations as in the observations (although it starts to appear $\sim 0.012$ earlier in phase in the observations) and at nearly the same radial velocity as the dense material in the post-shock primary wind. We note that the line profile of He~I $\lambda$4713 in Fig.~B1 shows a second absorption feature at phase 12.9570 centred at $\sim$$-350$~km~s$^{-1}$, in perfect agreement with the simulations.

Detailed examination of the velocity plots in Figs.~8--10 also shows that as the system approaches periastron, there is a small, but noticeable increase in the magnitude of the radial velocity of the pre-shock primary wind in the line-of-sight, a direct result of the stellar orbital motion (for details, see Appendix~A of Madura et al.~2013). The radial velocity of the pre-shock primary wind in the line-of-sight now approaches $\sim$$-480$~km~s$^{-1}$ to $-500$~km~s$^{-1}$. This increase in radial velocity of the pre-shock primary wind between $\phi = 0.9140$ and $0.9569$, helps explain feature \#1 in Fig.~6. However, as can be seen in Figs.~8--10, the post-shock primary wind material in the WWC zone is composed of material at a range of velocities, some of which are nearly identical to the pre-shock primary wind velocity. Thus, disentangling exactly to what extent feature \#1 is due solely to orbital motion of the primary rather than absorbing material in the WWC region remains extremely difficult. This moreover illustrates why it is so easy to compute (incorrectly) a large semi-amplitude, and hence a large mass function, for \ec when analyzing the P~Cygni absorption in the He~I observations. The increasing orbital speeds of the stars likely also increases the magnitude of the radial velocity of the absorbing post-shock primary wind material in the WWC, which could similarly explain the connection between absorption features \#2 and \#3 in Fig.~6, which appear to be connected and exhibit a slope similar to absorption feature \#1.

\begin{figure*}
\includegraphics[width=16cm]{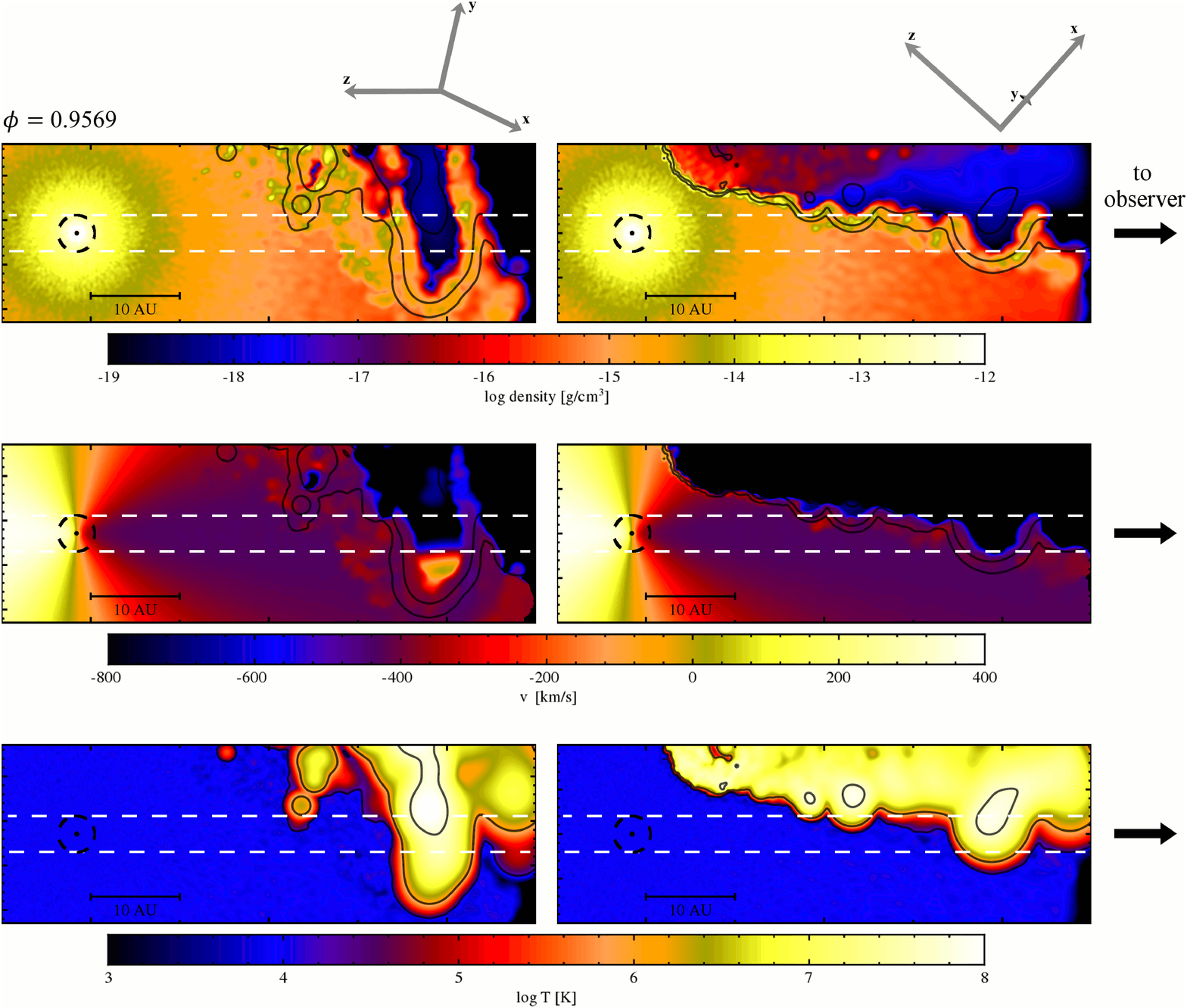}
\caption{\label{phase0p9569} Same as Fig.~\ref{phase0p9140}, but at $\phi = 0.9569$.}
\end{figure*}

\subsubsection{Absorptions at $\phi =$ 0.9665 and Feature \#4}

Interestingly, we find no evidence of any high-velocity ($\lesssim -600$~km~s$^{-1}$) material in the line-of-sight in Figs.~10--12 that could explain the relatively faint high-velocity absorption feature observed between HJD 2,456,770 ($\phi = 12.9484$) and HJD 2,456,810 ($\phi = 12.9682$, feature \#4 in Fig.~6). As far as we are aware, no significant high-velocity absorption is observed at other wavelengths (e.g. UV and near-IR) in this phase range either (Groh et al. 2010). The large OPD/LNA, \emph{HST}/STIS, and VLT/CRIRES observational datasets in Groh et al. (2010) confine the high-velocity absorption (more negative than $-$900~km~s$^{-1}$) at other wavelengths to roughly the phase range $0.976 \lesssim \phi \lesssim 1.023$, which is at least two weeks after the absorption feature we observe. Together, these support the idea that feature \#4 is stochastic in origin and not a periodic feature repeatable from event to event. Instead, given the very unstable and turbulent nature of the WWC region, we speculate that feature \#4 may be due to a `blob' of dense, post-shock primary wind from the WWC zone that separated from the WWC zone and was thereafter accelerated by the faster impinging wind of the secondary. Such an accelerated blob crossing our line-of-sight at the appropriate time could potentially explain this faint, high-velocity blue-shifted absorption feature. Observations of the optical He~I lines across \ec's next periastron passage in 2020 can be used to test this hypothesis, and we predict that if we are correct, feature \#4 will not occur during the next event. However, we note that if such a feature is indeed due to a blob of post-shock material breaking free from the WWC zone and crossing our line-of-sight after being accelerated by the secondary's faster impinging wind, a feature like \#4 could occur and be observed at any time. Detection of such features at non-periodic intervals would also support our hypothesis and provide more evidence for a binary orientation in which the WWC zone opens toward the observer for most of the 5.54-yr orbit, with the secondary star on the observer's side of the system at apastron.

\begin{figure*}
\includegraphics[width=16cm]{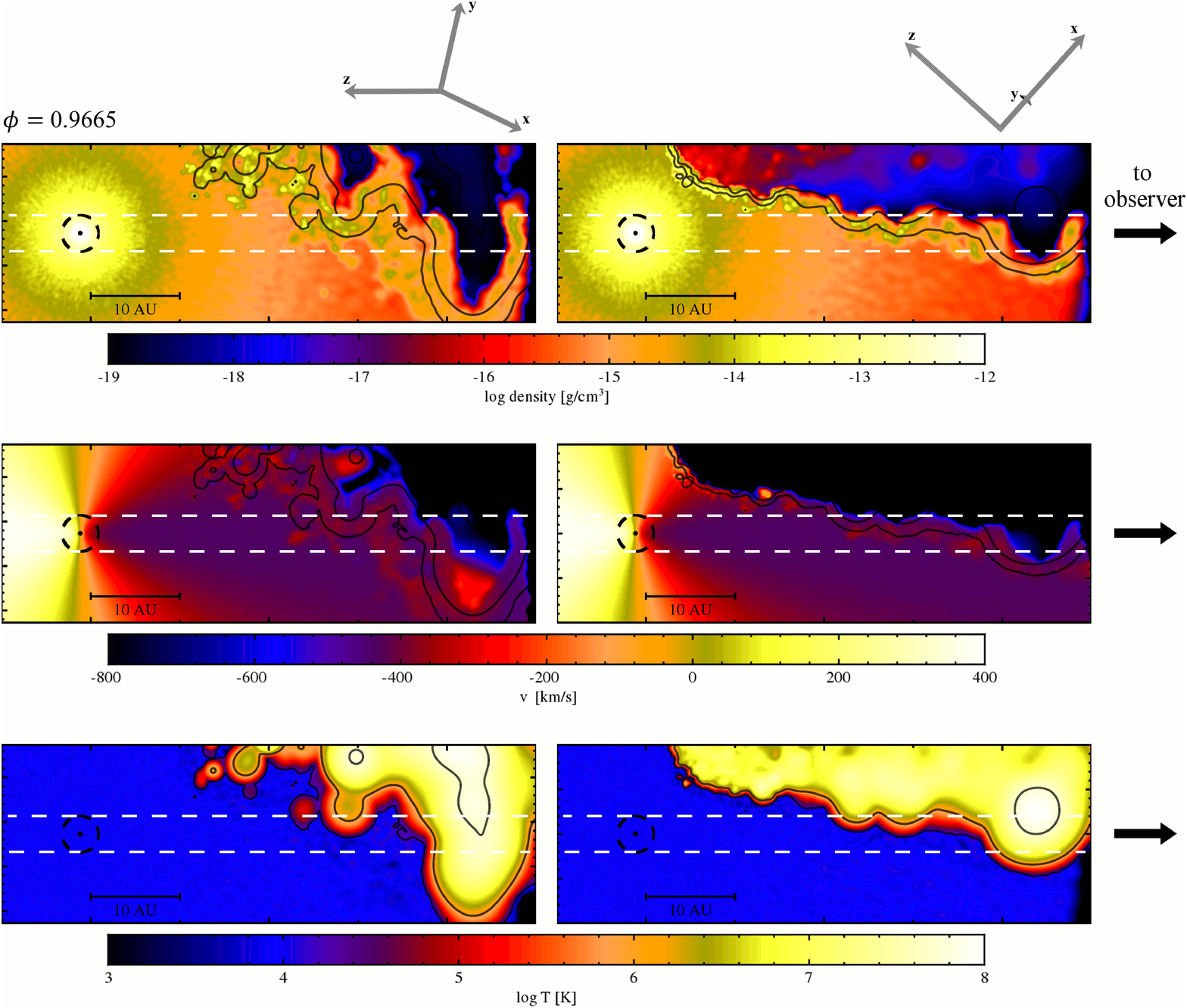}
\caption{\label{phase0p9665} Same as Fig.~\ref{phase0p9140}, but at $\phi = 0.9665$.}
\end{figure*}

\subsubsection{Absorptions at $\phi =$ 0.9800 and Feature \#3}

Given its centre velocity of $\sim -450$~km~s$^{-1}$ and its possible relation to feature \#2, it is difficult to determine exactly where absorption feature \#3 in Fig.~6 originates. As discussed in Section~5.2.2, the post-shock primary wind in the WWC zone is composed of material at a range of velocities, some of which are nearly identical to the pre-shock primary wind velocity (Fig.~12). The increased pre-shock wind velocity due to the increased orbital speeds of the stars at $\phi =$ 0.9800 is also very close to the observed centre velocity of feature \#3. Therefore, feature \#3 could arise in the pre-shock primary wind, the post-shock primary wind, or, most likely, a combination of both pre- and post-shock primary wind in the line-of-sight. Again, the slope of this feature, if connected to feature \#2, is likely indicative of the increasing orbital speeds as the stars rapidly enter periastron in their highly elliptical orbits.

Based on Fig.~12, one would also expect absorption to be observed at slower velocities ($\sim -300$ to $-350$~km~s$^{-1}$) arising from dense slower moving material in the post-shock primary wind region. This slower absorption is not visible in Fig.~6, but the line profiles of He~I $\lambda$4713 in Fig.~B1 do show that the absorption extends to such velocities at phases around 12.98.

\begin{figure*}
\includegraphics[width=16cm]{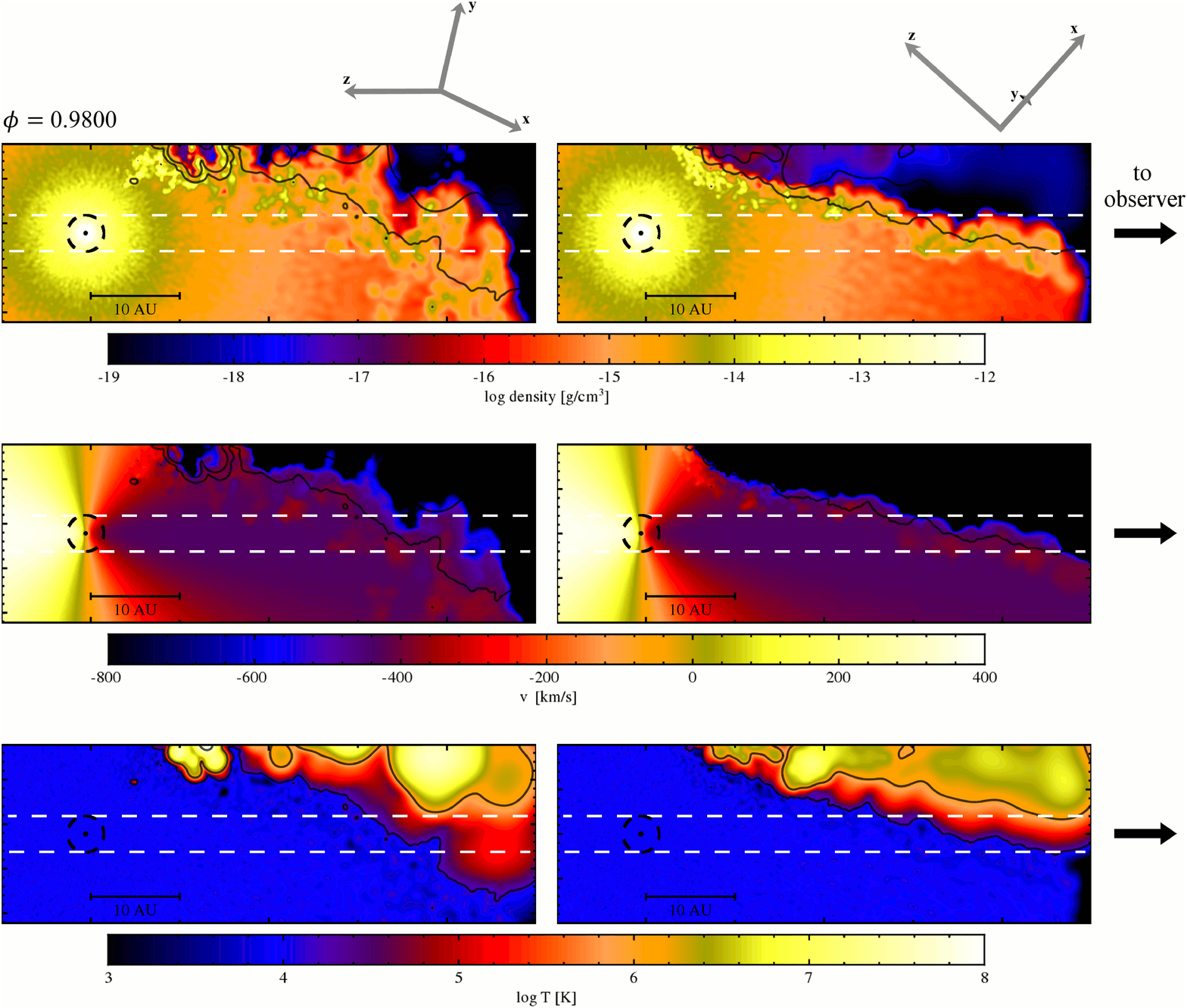}
\caption{\label{phase0p9800} Same as Fig.~\ref{phase0p9140}, but at $\phi = 0.9800$.}
\end{figure*}

\begin{figure*}
\includegraphics[width=16cm]{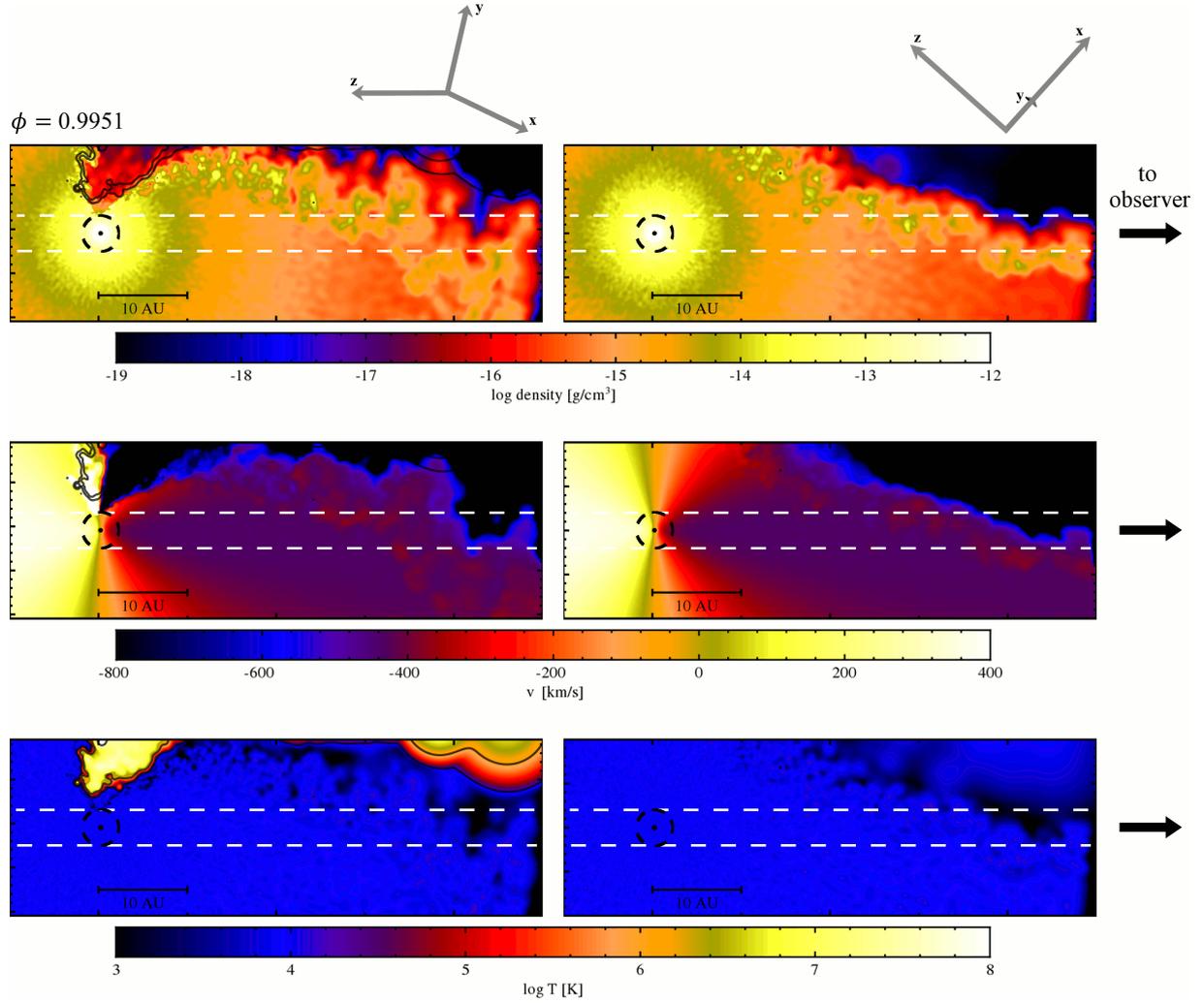}
\caption{\label{phase0p9951} Same as Fig.~\ref{phase0p9140}, but at $\phi = 0.9951$.}
\end{figure*}

\begin{figure*}
\includegraphics[width=16cm]{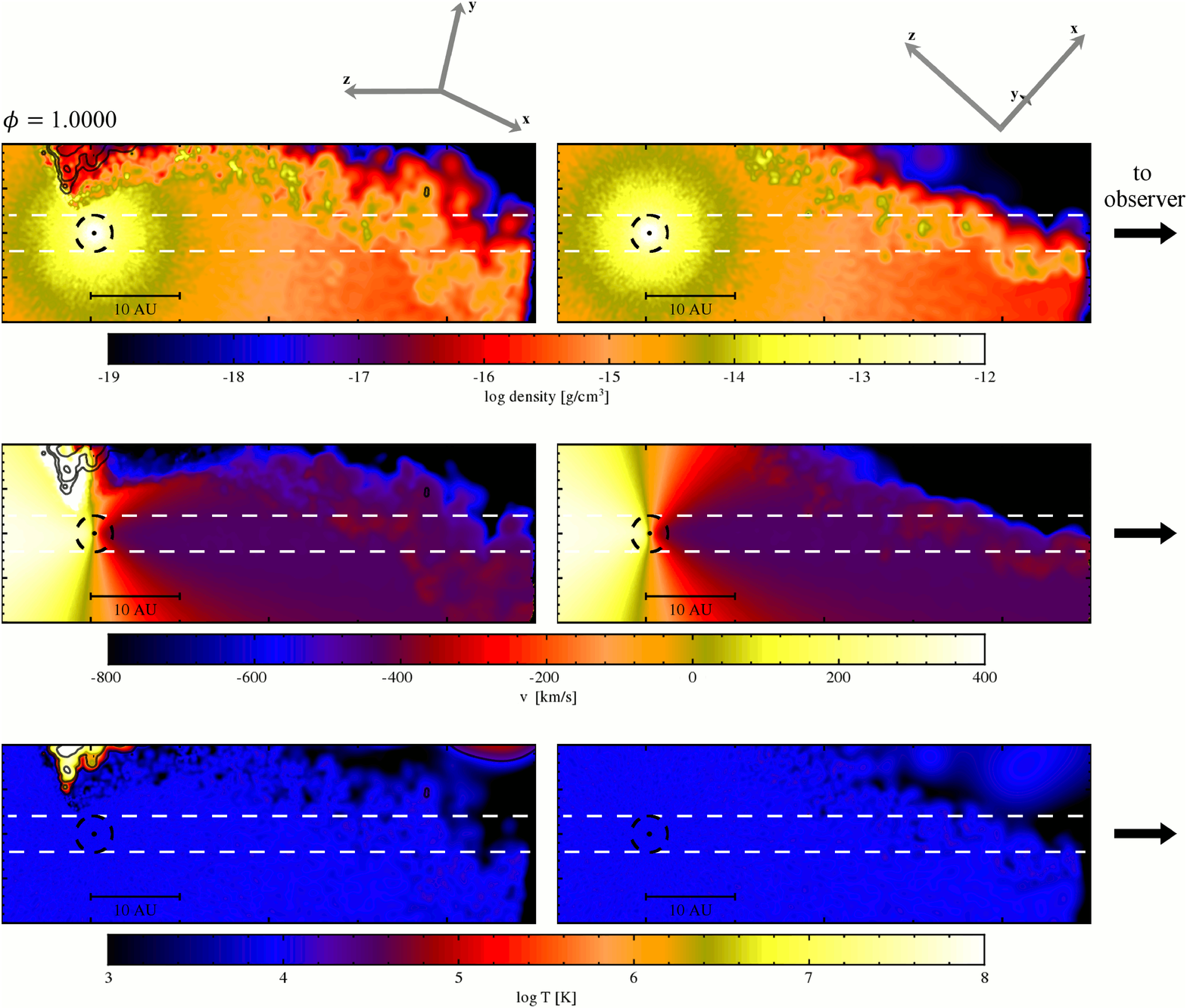}
\caption{\label{phase1p000} Same as Fig.~\ref{phase0p9140}, but at $\phi = 1.000$ (periastron).}
\end{figure*}

\subsubsection{Absorptions at $\phi =$ 0.9950 and periastron}

At $\phi \approx 0.995$ the situation starts to change drastically. Fig.~13 shows that as the system gets very close to periastron, the WWC region becomes highly distorted in the direction of orbital motion and the secondary star begins to become deeply embedded within the dense wind of the primary. As discussed in Madura et al.~(2013, 2015), around this phase and leading into periastron, the receding secondary wind is unable to collide with the receding primary wind, leading to the lack of an extended hot ($>$ $10^{6}$~K) trailing arm to the WWC zone. The models indicate that the post-shock secondary wind located near the apex of the WWC zone also has switched to the radiative regime at this time and is now much cooler ($T \lesssim 10^{4}$~K). The result is zero hot gas in the line-of-sight to the primary at $\phi = 0.995$ and periastron (Figs.~13 and 14). However, the dense, cool remnants of the WWC trailing arm and the receding unshocked secondary wind \emph{are} in the line-of-sight. 

The middle rows of Figs.~13 and 14 show that the remnants of the WWC trailing arm and the receding secondary wind have radial velocities in the line-of-sight ranging from $\sim$ $-350$~km~s$^{-1}$ to $<$$-800$~km~s$^{-1}$. The denser pre- and post-shock primary wind in the line-of-sight explains why the strongest absorption observed at these phases is centred at velocities around $-400$~km~s$^{-1}$ to $-500$~km~s$^{-1}$ (Fig.~1). Note also that due to the extreme curvature of the WWC region at this phase, the observer's line-of-sight samples a thicker column of compressed post-shock material, which also helps explain the increased strength of the absorption at such velocities at this phase.

Absorption by the lower-density column of fast secondary wind in the line-of-sight explains the fainter, but much higher velocity ($\sim -800$~km~s$^{-1}$ to $-900$~km~s$^{-1}$) absorption observed between $\phi = 0.995$ and periastron (Fig.~1 and Appendix~B). Moreover, because this process is governed by the dynamics of the orbit (e.g. the high eccentricity) and our viewing angle, this process should be repeatable from cycle to cycle (assuming that the stellar mass loss rates, wind velocities, and ionizing fluxes have not changed drastically from cycle to cycle), explaining why such high-velocity absorbing material has been consistently observed near $\phi \approx 0.995$ for several periastron events (those of 1992.5, 2009.1, and 2014.6, see e.g. Fig.~3). We suggest that this faint, repeatable high-velocity absorption is a direct signature of the high-velocity wind from the companion star located near the trailing arm of the WWC region, which crosses our line-of-sight just prior to periastron. We furthermore suggest that since this high-velocity He~I absorption feature has been so repeatable over the last several periastron events, there have been no significant changes in \ec's stellar mass loss rates, wind velocities, or ionizing fluxes over approximately the last twenty years, in agreement with the conclusions of Madura et al.~(2013).

\subsubsection{Absorptions after periastron}

Following periastron, until $\phi \gtrsim 1.02$, the secondary remains deeply embedded within the dense wind of the primary on the far side of the system opposite the observer. At these times, only unshocked primary wind flows toward the observer (see figs. in Madura et al.~2013). Moreover, because He in the outer portions of the primary wind is neutral (Hillier et al.~2001, 2006; Groh et al.~2012a), He~I absorption can only arise in the innermost primary wind where He is singly ionized. The photoionizing flux from the secondary is trapped within the dense wind of the primary on the far side of the system (Clementel et al.~2015b). Therefore, between periastron and $\phi \approx 1.02$, there should be little or no He~I absorption, and any weak absorption present should be at velocities comparable to the primary's wind terminal velocity. After $\phi \approx 1.02$, the WWC region has reestablished itself and returned to the observer's side of the system. Across the broad part of the orbit until near the next periastron, the He~I absorption should again arise in ionized regions of the pre-shock primary wind located just beyond the WWC zone, at velocities of $\sim -420$~km~s$^{-1}$ (Clementel et al.~2015a), and portions of the dense post-shock primary wind in the WWC region at velocities near $-350$~km~s$^{-1}$. This is exactly what is observed in Fig.~1 and during the 2009.0 event (Richardson et al.~2015).

\section{Conclusions and Future Work}

We monitored \ec with the CTIO 1.5 m and the CHIRON spectrograph through the 2014.6 spectroscopic event. Prior to the 2014.6 event, no time-series of this intensity or spectral resolution has been obtained on \ec through a periastron passage. Here we presented an analysis of the He~I P Cygni absorption profiles. We concentrated on the He~I $\lambda$4713, 5876, 7065 transitions due to line blending and the spectroscopic coverage. In summary, we find:

\begin{itemize}
\item The absorption troughs vary in depth from roughly absent to nearly 40\% of the continuum. Further, during the times prior to the event, there was a secondary absorption feature that appeared in all absorption profiles and accelerated to higher velocities in our line-of-sight. Through a comparison to simultaneous {\it HST}/STIS spectroscopy, we know this feature must originate in the central stellar winds or wind-wind collision zone.
\item Comparing the absorption profiles of He~I $\lambda$5876 across the last five periastron passages, the absorption strength seems relatively repeatable from cycle-to-cycle. However, there are small deviations, including the secondary absorption component we observed between $\phi \approx 0.95$ and 0.97.
\item Between $\phi = 0.99$ and 1.00, the He~I $\lambda$5876 absorption profile has been indistinguishable across the last several events. The equivalent width and feature kinematics are well-reproduced between the well-observed 2009.0 event (Richardson et al.~2015) and the 2014.6 event.
\item We qualitatively compared the observations to 3D SPH simulations of the system (e.g., Madura et al.~2013, 2015) and find that the secondary absorption variability can be well explained if we look across the edge of the shock cone near periastron. This geometry is dependent on the inclination of the system and the ratio of the mass-loss rates between $\eta$~Car$_{\rm A}$ and $\eta$~Car$_{\rm B}$, and there is a degeneracy between these parameters. As the secondary absorption feature was not observed in previous cycles, this may show a small change in the mass-loss rate of one or both of the stars, as speculated by Mehner et al.~(2010b) and Martin et al.~(2010). However, it is perhaps more likely that the mass-loss rate(s) have not changed significantly and we just happened this cycle to observe a small blob of dense material break free from the WWC region and cross our line-of-sight as it was accelerated by the much faster impinging wind from the secondary star. This is supported by the remarkable repeatability of the high-velocity ($\sim$$-800$~km~s$^{-1}$) absorption feature observed near $\phi = 0.995$.
\item We speculate that the high-velocity He~I absorption observed regularly near $\phi = 0.995$ is a direct signature of the high-velocity wind from the companion star located near the trailing arm of the WWC region, which crosses our line-of-sight just prior to periastron.
\item As no time-series of this temporal resolution was ever obtained during past events, we encourage observers to obtain frequent observations from 2019--2021 in order to best observe the 2020 periastron passage and search for a repeatability in the variability we found in these absorption profiles.
\end{itemize}

In the future, more detailed 3D hydrodynamical and radiative transfer simulations similar to those in Clementel~et al.~(2015a,b) will be extremely useful in determining the helium ionization balance at the times of the observed excess He~I P~Cygni absorption variability. Such simulations will help us better physically constrain the formation regions of the absorption and aid our understanding of colliding winds in this nearest very massive binary system. Radiative transfer of these simulations will allow for a direct comparison between the observational record and the theoretical predictions. Future studies will need similar observations for the time period around periastron to determine the repeatability and differences in the periastron passages of the system. With efforts that combine radiative transfer and multiple observable diagnostics including the He~I profiles and the He~II $\lambda 4686$ emission, we should be able to determine empirically how the wind momentum ratio varies with subtle changes in the system with time. Such observations and modeling will help reveal the masses and fundamental parameters of this most unique massive binary in the Galaxy.

\section*{Acknowledgements}

We are thankful to our referee, Ian Howarth, for a thoughtful and critical review that greatly improved this paper. This work was based on observations at Cerro Tololo Inter-American Observatory, National Optical Astronomy Observatory (NOAO Prop. IDs: 2012A-0216, 2012B-0194, 2013B-0328, and 2015A-0109; PI: N. Richardson), which is operated by the Association of Universities for Research in Astronomy (AURA) and the Small and Moderate Aperture Research Telescope System (SMARTS) under a cooperative agreement with the National Science Foundation. This work was largely helped through previous discussions with Douglas Gies (Georgia State University) which led to successful observing proposals.
Some SMARTS observations with CHIRON were prompted by conversations with Roberta Humphreys and Kris Davidson (University of Minnesota), whom we thank for their contributions. Some of the spectra were obtained under the aegis of Stony Brook University, whose participation has been supported by the office of the Provost. These observations would not have been made possible without the careful scheduling by the SMARTS staff at Yale University, most notably Emily MacPherson and Imran Hasan and the queue observing expertly done by Carlos Corco, Alberto Miranda, Rodrigo Hernandez, and Manuel Hernandez. Resources supporting this work were provided by the NASA High-End Computing (HEC) Program through the NASA Advanced Supercomputing (NAS) Division at Ames Research Center. This research has made use of the data archive for the HST Treasury Program on Eta Carinae (GO 9973) which is available online at http://etacar.umn.edu. The archive is supported by the University of Minnesota and the Space Telescope Science Institute under contract with NASA.

NDR is grateful for his former CRAQ (Centre de Recherche en Astrophysique du Qu\'ebec) postdoctoral fellowship and for postdoctoral support by the University of Toledo and by the Helen Luedtke Brooks Endowed Professorship.
AFJM is grateful for financial support from NSERC (Canada) and FRQNT (Qu\'{e}bec). CMPR is and TIM was supported by an appointment to the NASA Postdoctoral Program at the Goddard Space Flight Center, administered by Oak Ridge Associated Universities through a contract with NASA.  AD thanks FAPESP for financial support through grant 2011/51680-6. DJH acknowledges support from HST-GO 12508.02-A.

\appendix

\section{Measurements}

\begin{table*}
\centering
\begin{minipage}{160mm}
\caption{He~I $\lambda$5876 Measurements}
\begin{tabular}{lrcccc}
\hline
      HJD - 2,450,000	&	    $W_\lambda$ 	&	      Abs. Depth 	&	     Abs. Depth & $V_{\rm min}$ & $V_{\rm min}$	\\

      		&			&		(observed) & (gaussian) & (observed) & (gaussian) \\
   (d)			&		(\AA)	 & (norm.) & (norm.) & (km s$^{-1}$) & (km s$^{-1}$) \\ \hline

5125.8682 & -13.16 & 0.251 & \ldots & -427.8 & \ldots \\
5139.8436 & -12.90 & 0.232 & \ldots & -431.4 & \ldots \\
5140.8246 & -12.65 & 0.248 & \ldots & -431.4 & \ldots \\
5164.8118 & -11.36 & 0.314 & \ldots & -434.9 & \ldots \\
5213.7590 & -10.13 & 0.311 & \ldots & -434.9 & \ldots \\
5226.7402 & -11.96 & 0.309 & \ldots & -427.8 & \ldots \\
5230.6255 & -12.03 & 0.325 & \ldots & -459.7 & \ldots \\
5240.6474 & -11.57 & 0.317 & \ldots & -420.7 & \ldots \\
5250.6127 & -12.05 & 0.315 & \ldots & -417.1 & \ldots \\
5254.5753 & -13.12 & 0.291 & \ldots & -427.8 & \ldots \\
5259.6044 & -12.40 & 0.300 & \ldots & -431.4 & \ldots \\
5270.5564 & -11.87 & 0.328 & \ldots & -438.5 & \ldots \\
5278.4984 & -13.11 & 0.324 & \ldots & -431.4 & \ldots \\
5284.5346 & -11.74 & 0.361 & \ldots & -434.9 & \ldots \\
5293.5099 & -11.83 & 0.377 & \ldots & -431.4 & \ldots \\
5306.5008 & -12.80 & 0.348 & \ldots & -431.4 & \ldots \\
5311.5186 & -12.02 & 0.334 & \ldots & -417.1 & \ldots \\
5314.5970 & -11.90 & 0.354 & \ldots & -424.2 & \ldots \\
5317.4860 & -2.298 & 0.350 & \ldots & -417.1 & \ldots \\
5327.4623 & -11.82 & 0.333 & \ldots & -417.1 & \ldots \\
5335.4752 & -12.44 & 0.341 & \ldots & -413.7 & \ldots \\
5360.3720 & -11.31 & 0.367 & \ldots & -434.9 & \ldots \\
5989.6915 & -12.61 & 0.246 & \ldots & -423.2 & \ldots \\
5993.7546 & -13.07 & 0.245 & 0.257 & -424.7 & -420.8 \\
6001.7288 & -13.36 & 0.224 & 0.221 & -418.8 & \ldots \\
6013.6892 & -14.33 & 0.199 & 0.150 & -409.7 & -408.8 \\
6221.8876 & -13.75 & 0.340 & 0.317 & -432.2 & -431.9 \\
6236.8436 & -12.94 & 0.351 & \ldots & -429.1 & \ldots \\
6238.8674 & -13.08 & 0.350 & 0.341 & -436.7 & -435.4 \\
6248.8123 & -13.39 & 0.340 & 0.338 & -430.6 & \ldots \\
6254.8795 & -13.27 & 0.313 & \ldots & -438.2 & \ldots \\
6260.7808 & -14.44 & 0.302 & \ldots & -432.2 & \ldots \\
6275.8175 & -11.71 & 0.378 & 0.352 & -439.6 & -439.4 \\
6289.7971 & -12.84 & 0.373 & \ldots & -438.2 & \ldots \\
6361.6472 & -13.44 & 0.414 & 0.397 & -448.7 & -448.3 \\
6401.5905 & -14.36 & 0.393 & 0.386 & -433.7 & \ldots \\
6417.5924 & -15.14 & 0.371 & 0.343 & -433.7 & \ldots \\
6428.6435 & -14.37 & 0.292 & \ldots & -439.6 & \ldots \\
6607.8431 & -14.64 & 0.362 & 0.314 & -511.4 & -519.0 \\
6612.8653 & -15.45 & 0.369 & 0.390 & -511.4 & \ldots \\
6656.7032 & -16.64 & 0.239 & 0.219 & -526.4 & -526.9 \\
6659.7093 & -17.02 & 0.199 & 0.193 & -536.9 & -526.9 \\
6664.6745 & -16.28 & 0.191 & 0.162 & -526.4 & -526.5 \\
6670.7826 & -15.86 & 0.147 & 0.171 & -532.5 & \ldots \\
6672.8378 & -15.89 & 0.145 & 0.128 & -510.0 & -510.2 \\
6677.7595 & -15.46 & 0.155 & 0.139 & -508.5 & -494.0 \\
6687.7085 & -16.75 & 0.148 & 0.101 & -454.6 & -454.5 \\
6690.6917 & -16.40 & 0.149 & 0.103 & -448.7 & -448.9 \\
6697.7271 & -17.40 & 0.157 & 0.145 & -496.5 & -496.5 \\
6698.7748 & -16.32 & 0.139 & 0.131 & -496.5 & -496.0 \\
6710.6697 & -17.36 & 0.171 & 0.162 & -524.9 & -525.2 \\
6712.6945 & -16.69 & 0.176 & 0.138 & -512.9 & -512.5 \\
6718.7120 & -17.06 & 0.195 & 0.211 & -510.0 & \ldots \\
6724.5925 & -16.86 & 0.149 & 0.145 & -472.7 & -473.2 \\
6725.5713 & -17.27 & 0.149 & 0.153 & -465.1 & -482.4 \\
6729.5623 & -16.76 & 0.149 & 0.156 & -454.6 & -465.4 \\
6732.6062 & -16.81 & 0.155 & 0.125 & -456.1 & -455.5 \\
6739.5490 & -17.31 & 0.138 & 0.104 & -450.1 & -450.3 \\
6746.5203 & -18.55 & 0.108 & 0.084 & -448.7 & -449.3 \\
6750.5418 & -17.46 & 0.154 & 0.133 & -478.6 & -477.9 \\
6750.5454 & -17.58 & 0.125 & 0.107 & -477.1 & \ldots \\
6754.5540 & -17.90 & 0.183 & 0.182 & -484.5 & \ldots \\
\hline
\end{tabular}
\end{minipage}
\end{table*}

\begin{table*}
\begin{minipage}{160mm}
\contcaption{He~I $\lambda$5876 Measurements}
\begin{tabular}{lrcccc}
\hline
      HJD - 2,450,000	&	    $W_\lambda$ 	&	      Abs. Depth 	&	     Abs. Depth & $V_{\rm min}$ & $V_{\rm min}$	\\

      		&			&		(observed) & (gaussian) & (observed) & (gaussian) \\
   (d)			&		(\AA)	 & (norm.) & (norm.) & (km s$^{-1}$) & (km s$^{-1}$) \\  \hline

6765.5903 & -17.27 & 0.324 & 0.331 & -523.4 & -527.3 \\
6766.5463 & -17.10 & 0.324 & 0.314 & -524.9 & -525.9 \\
6767.5116 & -17.37 & 0.337 & 0.335 & -514.4 & \ldots \\
6774.5470 & -17.54 & 0.333 & 0.329 & -523.4 & \ldots \\
6775.4705 & -16.89 & 0.333 & 0.299 & -529.5 & -531.0 \\
6777.5390 & -16.72 & 0.341 & 0.336 & -532.5 & -533.2 \\
6781.4742 & -16.23 & 0.371 & 0.345 & -536.9 & -536.4 \\
6786.5036 & -16.35 & 0.345 & 0.348 & -538.4 & -546.3 \\
6787.5061 & -16.02 & 0.339 & 0.337 & -545.9 & \ldots \\
6791.5527 & -15.91 & 0.307 & 0.295 & -544.5 & -526.1 \\
6795.4970 & -15.16 & 0.280 & 0.281 & -554.8 & -565.4 \\
6795.5095 & -15.29 & 0.283 & 0.275 & -554.8 & -554.6 \\
6800.4644 & -15.12 & 0.273 & \ldots & -559.4 & -562.0 \\
6801.5000 & -15.09 & 0.276 & 0.255 & -565.3 & \ldots \\
6802.5067 & -14.79 & 0.282 & \ldots & -560.9 & \ldots \\
6803.5294 & -14.55 & 0.299 & 0.263 & -560.9 & -560.8 \\
6806.5318 & -14.19 & 0.290 & 0.315 & -574.4 & -557.5 \\
6807.6147 & -14.21 & 0.286 & 0.252 & -562.4 & -561.7 \\
6809.4604 & -13.90 & 0.272 & 0.269 & -553.3 & \ldots \\
6810.4839 & -14.10 & 0.274 & 0.243 & -565.3 & -565.0 \\
6811.4890 & -13.98 & 0.255 & 0.247 & -559.4 & -559.7 \\
6818.5059 & -15.59 & 0.198 & 0.153 & -535.4 & \ldots \\
6818.5097 & -15.54 & 0.198 & \ldots & -559.4 & \ldots \\
6819.5294 & -15.03 & 0.210 & 0.179 & -554.8 & -552.6 \\
6823.5395 & -15.19 & 0.188 & 0.169 & -554.8 & -546.7 \\
6824.5183 & -14.62 & 0.198 & 0.190 & -557.9 & -557.0 \\
6825.5069 & -14.41 & 0.195 & \ldots & -556.4 & \ldots \\
6829.4942 & -13.29 & 0.204 & 0.195 & -519.0 & -520.0 \\
6832.5015 & -13.15 & 0.232 & 0.220 & -536.9 & -536.6 \\
6833.5015 & -13.26 & 0.215 & 0.206 & -553.3 & -553.1 \\
6835.5347 & -15.46 & 0.217 & 0.177 & -538.4 & -538.0 \\
6836.4977 & -13.24 & 0.224 & 0.231 & -539.8 & \ldots \\
6837.4984 & -13.41 & 0.213 & 0.215 & -553.3 & -560.8 \\
6838.5142 & -13.09 & 0.222 & 0.202 & -553.3 & \ldots \\
6839.5007 & -12.98 & 0.227 & 0.218 & -551.9 & -551.5 \\
6840.4992 & -12.46 & 0.231 & 0.224 & -548.9 & -549.1 \\
6845.4618 & -12.38 & 0.237 & 0.242 & -563.8 & -571.3 \\
6847.4567 & -11.92 & 0.223 & 0.215 & -571.4 & -570.2 \\
6850.5099 & -11.59 & 0.221 & 0.193 & -565.3 & -566.2 \\
6855.4628 & -8.489 & 0.231 & 0.230 & -586.3 & \ldots \\
6859.4699 & -6.609 & 0.265 & 0.269 & -557.9 & \ldots \\
6863.4696 & -4.399 & 0.278 & 0.271 & -547.4 & -548.1 \\
6864.4670 & -3.938 & 0.274 & 0.262 & -533.9 & -534.7 \\
6864.4717 & -3.956 & 0.262 & 0.263 & -550.4 & -558.9 \\
6864.4835 & -3.903 & 0.262 & 0.263 & -550.4 & -559.5 \\
6866.5170 & -2.522 & 0.259 & 0.254 & -551.9 & -552.3 \\
6867.4637 & -2.257 & 0.274 & \ldots & -501.1 & \ldots \\
6870.4592 & -3.097 & 0.185 & 0.178 & -501.1 & -501.3 \\
6871.4663 & -3.580 & 0.156 & 0.143 & -512.9 & -512.6 \\
6872.4634 & -3.336 & 0.162 & \ldots & -496.5 & \ldots \\
6873.4627 & -3.797 & 0.135 & 0.124 & -514.4 & \ldots \\
6874.4542 & -3.970 & 0.122 & \ldots & -489.1 & -485.9 \\
6876.4581 & -4.250 & 0.119 & \ldots & -510.0 & \ldots \\
6878.4816 & -4.230 & 0.092 & 0.050 & -445.7 & \ldots \\
6879.4665 & -3.837 & 0.091 & 0.086 & -426.1 & \ldots \\
6880.4665 & -3.745 & 0.094 & 0.076 & -442.6 & -441.9 \\
6881.4646 & -3.659 & 0.082 & 0.072 & -436.7 & -429.8 \\
6882.4610 & -3.692 & 0.073 & 0.042 & -423.2 & -411.8 \\
6883.4684 & -3.593 & 0.055 & 0.043 & -424.7 & -423.7 \\
6885.4659 & -3.827 & 0.018 & 0.015 & -408.2 & -406.4 \\
6886.4726 & -3.923 & 0.010 & 0.006 & -412.7 & -412.1 \\
6887.4599 & -4.362 & \ldots & 0.001 & -402.2 & -407.5 \\
\hline
\end{tabular}
\end{minipage}
\end{table*}

\begin{table*}
\begin{minipage}{160mm}
\contcaption{He~I $\lambda$5876 Measurements}
\begin{tabular}{lrcccc}
\hline
      HJD - 2,450,000	&	    $W_\lambda$ 	&	      Abs. Depth 	&	     Abs. Depth & $V_{\rm min}$ & $V_{\rm min}$	\\

      		&			&		(observed) & (gaussian) & (observed) & (gaussian) \\
   (d)			&		(\AA)	 & (norm.) & (norm.) & (km s$^{-1}$) & (km s$^{-1}$) \\  \hline

6944.8969 & -8.481 & 0.212 & \ldots & -375.4 & \ldots \\
6950.8754 & -10.96 & 0.169 & 0.148 & -412.7 & -412.4 \\
6951.8733 & -10.97 & 0.174 & 0.141 & -408.2 & -409.2 \\
6952.8645 & -10.91 & 0.176 & 0.155 & -408.2 & -407.9 \\
6953.8796 & -10.73 & 0.190 & 0.163 & -409.7 & -410.5 \\
6954.8312 & -10.48 & 0.199 & 0.206 & -415.6 & -421.1 \\
6955.8613 & -10.59 & 0.183 & 0.168 & -411.2 & -410.4 \\
6956.8738 & -10.48 & 0.194 & 0.167 & -408.2 & -408.5 \\
6957.8902 & -10.59 & 0.188 & 0.168 & -412.7 & -413.2 \\
6958.8639 & -10.56 & 0.192 & 0.184 & -409.7 & -410.1 \\
6959.8471 & -10.78 & 0.192 & 0.199 & -417.3 & -421.7 \\
6961.8481 & -10.63 & 0.191 & 0.176 & -418.8 & -419.5 \\
6964.8650 & -10.50 & 0.196 & 0.181 & -423.2 & -423.6 \\
6965.8623 & -10.46 & 0.198 & 0.240 & -433.7 & -415.6 \\
6968.8670 & -10.56 & 0.192 & 0.165 & -420.2 & -420.6 \\
6969.8629 & -10.45 & 0.188 & 0.192 & -426.1 & \ldots \\
6972.8807 & -10.07 & 0.182 & 0.169 & -423.2 & -428.6 \\
6974.8472 & -10.32 & 0.184 & 0.175 & -432.2 & -432.7 \\
6977.8456 & -10.32 & 0.171 & 0.161 & -433.7 & -433.6 \\
6979.8393 & -10.67 & 0.171 & 0.165 & -438.2 & -436.1 \\
6980.8055 & -10.65 & 0.175 & 0.145 & -427.6 & \ldots \\
6982.8125 & -10.85 & 0.170 & 0.167 & -427.6 & \ldots \\
6984.8198 & -10.93 & 0.168 & 0.163 & -444.1 & -434.4 \\
6985.8277 & -10.82 & 0.170 & 0.170 & -442.6 & \ldots \\
6986.8716 & -10.75 & 0.182 & 0.162 & -441.1 & -440.8 \\
6988.7943 & -10.77 & 0.170 & 0.164 & -442.6 & -439.6 \\
6990.7638 & -10.60 & 0.172 & 0.167 & -438.2 & -450.6 \\
6992.7814 & -10.78 & 0.166 & 0.171 & -450.1 & -444.8 \\
6994.8169 & -10.70 & 0.167 & 0.148 & -444.1 & -443.4 \\
6996.8516 & -10.77 & 0.149 & 0.122 & -441.1 & -441.4 \\
6998.8053 & -10.96 & 0.144 & 0.119 & -439.6 & -439.2 \\
7000.8371 & -10.85 & 0.133 & 0.134 & -424.7 & \ldots \\
7007.8533 & -10.84 & 0.129 & 0.121 & -420.2 & \ldots \\
7008.7387 & -10.79 & 0.131 & 0.129 & -417.3 & \ldots \\
7009.7487 & -10.80 & 0.129 & 0.120 & -424.7 & -427.5 \\
7012.8169 & -10.82 & 0.139 & 0.118 & -423.2 & -423.1 \\
7013.8113 & -10.90 & 0.134 & 0.104 & -420.2 & -420.1 \\
7014.7384 & -10.87 & 0.135 & 0.113 & -421.7 & -420.4 \\
7015.7661 & -10.72 & 0.133 & \ldots & -417.3 & \ldots \\
7016.7963 & -10.83 & 0.141 & 0.106 & -415.6 & -416.7 \\
7018.7787 & -10.77 & 0.145 & 0.105 & -417.3 & -417.6 \\
7021.8358 & -10.89 & 0.122 & 0.079 & -418.8 & -418.8 \\
7022.7519 & -10.91 & 0.120 & 0.128 & -418.8 & -424.5 \\
7025.8280 & -10.94 & 0.112 & 0.088 & -418.8 & -418.6 \\
7026.7004 & -10.96 & 0.112 & 0.072 & -417.3 & -416.5 \\
7030.6911 & -11.19 & 0.095 & 0.092 & -414.2 & -427.0 \\
7032.8183 & -11.44 & 0.080 & 0.062 & -415.6 & -415.5 \\
7035.7243 & -11.54 & 0.090 & 0.097 & -415.6 & -406.9 \\
7037.8164 & -11.43 & 0.089 & 0.084 & -417.3 & \ldots \\
7038.7462 & -10.88 & 0.119 & 0.084 & -414.2 & -414.5 \\
7038.8521 & -11.32 & 0.085 & 0.096 & -411.2 & -406.9 \\
7039.8336 & -11.22 & 0.091 & \ldots & -417.3 & \ldots \\
7040.7121 & -11.20 & 0.091 & 0.060 & -417.3 & -416.8 \\
7043.8463 & -11.07 & 0.102 & 0.070 & -418.8 & -417.9 \\
7046.6847 & -10.89 & 0.105 & 0.070 & -415.6 & -415.6 \\
7047.6855 & -10.65 & 0.107 & 0.087 & -415.6 & -416.3 \\
7048.8103 & -10.64 & 0.108 & 0.076 & -417.3 & -417.2 \\
7050.7318 & -10.78 & 0.102 & 0.089 & -417.3 & -418.2 \\
7051.8465 & -10.83 & 0.111 & 0.073 & -414.2 & -416.5 \\
7059.7902 & -10.07 & 0.142 & 0.141 & -421.7 & \ldots \\
7060.6626 & -9.849 & 0.146 & 0.131 & -421.7 & -422.0 \\
7061.8510 & -9.693 & 0.159 & \ldots & -418.8 & -438.4 \\
\hline
\end{tabular}
\end{minipage}
\end{table*}

\begin{table*}
\begin{minipage}{160mm}
\contcaption{He~I $\lambda$5876 Measurements}
\begin{tabular}{lrcccc}
\hline
      HJD - 2,450,000	&	    $W_\lambda$ 	&	      Abs. Depth 	&	     Abs. Depth & $V_{\rm min}$ & $V_{\rm min}$	\\

      		&			&		(observed) & (gaussian) & (observed) & (gaussian) \\
   (d)			&		(\AA)	 & (norm.) & (norm.) & (km s$^{-1}$) & (km s$^{-1}$) \\  \hline
7062.6967 & -9.603 & 0.160 & 0.151 & -421.7 & \ldots \\
7063.7354 & -9.664 & 0.152 & \ldots & -414.2 & \ldots \\
7064.7755 & -9.475 & 0.173 & 0.142 & -418.8 & -419.4 \\
7068.7710 & -9.706 & 0.173 & 0.162 & -423.2 & \ldots \\
7070.6468 & -9.835 & 0.168 & 0.172 & -421.7 & \ldots \\
7074.8003 & -10.31 & 0.176 & 0.177 & -417.3 & -432.7 \\
7075.6999 & -10.10 & 0.183 & 0.169 & -429.1 & -429.3 \\
7078.8993 & -10.24 & 0.190 & 0.180 & -436.7 & -436.4 \\
7083.6618 & -9.218 & 0.235 & 0.213 & -427.6 & -427.4 \\
7088.6417 & -9.299 & 0.258 & \ldots & -424.7 & \ldots \\
7091.5928 & -9.330 & 0.261 & 0.254 & -426.1 & \ldots \\
7092.7233 & -9.320 & 0.265 & 0.251 & -427.6 & \ldots \\
7093.7108 & -9.369 & 0.271 & 0.251 & -426.1 & -426.0 \\
7095.8005 & -9.777 & 0.266 & 0.228 & -427.6 & -427.7 \\
7097.6962 & -10.28 & 0.267 & 0.231 & -427.6 & -427.4 \\
7104.5614 & -10.13 & 0.293 & 0.262 & -423.2 & -423.9 \\
7115.7489 & -10.10 & 0.296 & 0.263 & -423.2 & -423.7 \\
7118.5573 & -10.40 & 0.299 & 0.270 & -423.2 & -423.6 \\
7120.5391 & -10.27 & 0.306 & 0.296 & -421.7 & \ldots \\
7124.4992 & -10.19 & 0.323 & 0.307 & -424.7 & -425.3 \\
7125.5235 & -10.43 & 0.319 & 0.297 & -420.2 & -420.8 \\
7127.6192 & -9.920 & 0.332 & 0.337 & -424.7 & -416.8 \\
7130.5443 & -10.15 & 0.329 & 0.332 & -424.7 & \ldots \\
7132.5158 & -9.916 & 0.334 & 0.323 & -426.1 & -426.4 \\
7134.6264 & -9.934 & 0.341 & 0.327 & -423.2 & -422.7 \\
7144.6931 & -9.405 & 0.358 & \ldots & -411.2 & \ldots \\
7145.6268 & -9.668 & 0.354 & 0.364 & -427.6 & -435.5 \\
7147.6292 & -9.297 & 0.314 & \ldots & -436.7 & -430.4 \\
7150.6168 & -9.722 & 0.349 & 0.326 & -426.1 & \ldots \\
7152.5910 & -9.674 & 0.383 & \ldots & -445.7 & -445.8 \\
7155.5746 & -10.31 & 0.344 & 0.333 & -432.2 & -432.1 \\
7161.5532 & -10.30 & 0.344 & 0.329 & -439.6 & -438.8 \\
7164.5506 & -10.24 & 0.356 & 0.363 & -436.7 & -441.9 \\
7166.5332 & -10.55 & 0.352 & 0.346 & -441.1 & \ldots \\
7172.5215 & -10.27 & 0.343 & 0.342 & -445.7 & \ldots \\
7174.5941 & -10.30 & 0.322 & 0.324 & -438.2 & \ldots \\
7177.4899 & -10.40 & 0.334 & 0.340 & -441.1 & -451.1 \\
7181.4486 & -10.66 & 0.312 & 0.309 & -450.1 & \ldots \\
7185.5031 & -11.27 & 0.305 & \ldots & -453.1 & \ldots \\
7189.5032 & -11.60 & 0.300 & \ldots & -454.6 & \ldots \\
7191.4615 & -11.55 & 0.295 & 0.276 & -453.1 & \ldots \\
7194.5003 & -11.33 & 0.298 & \ldots & -451.6 & \ldots \\
7212.4449 & -9.997 & 0.348 & 0.339 & -445.7 & \ldots \\
\hline

\end{tabular}
\end{minipage}
\end{table*}


\section{Observed Line Profiles}

\begin{figure*}
 \includegraphics[width=50mm, angle=0]{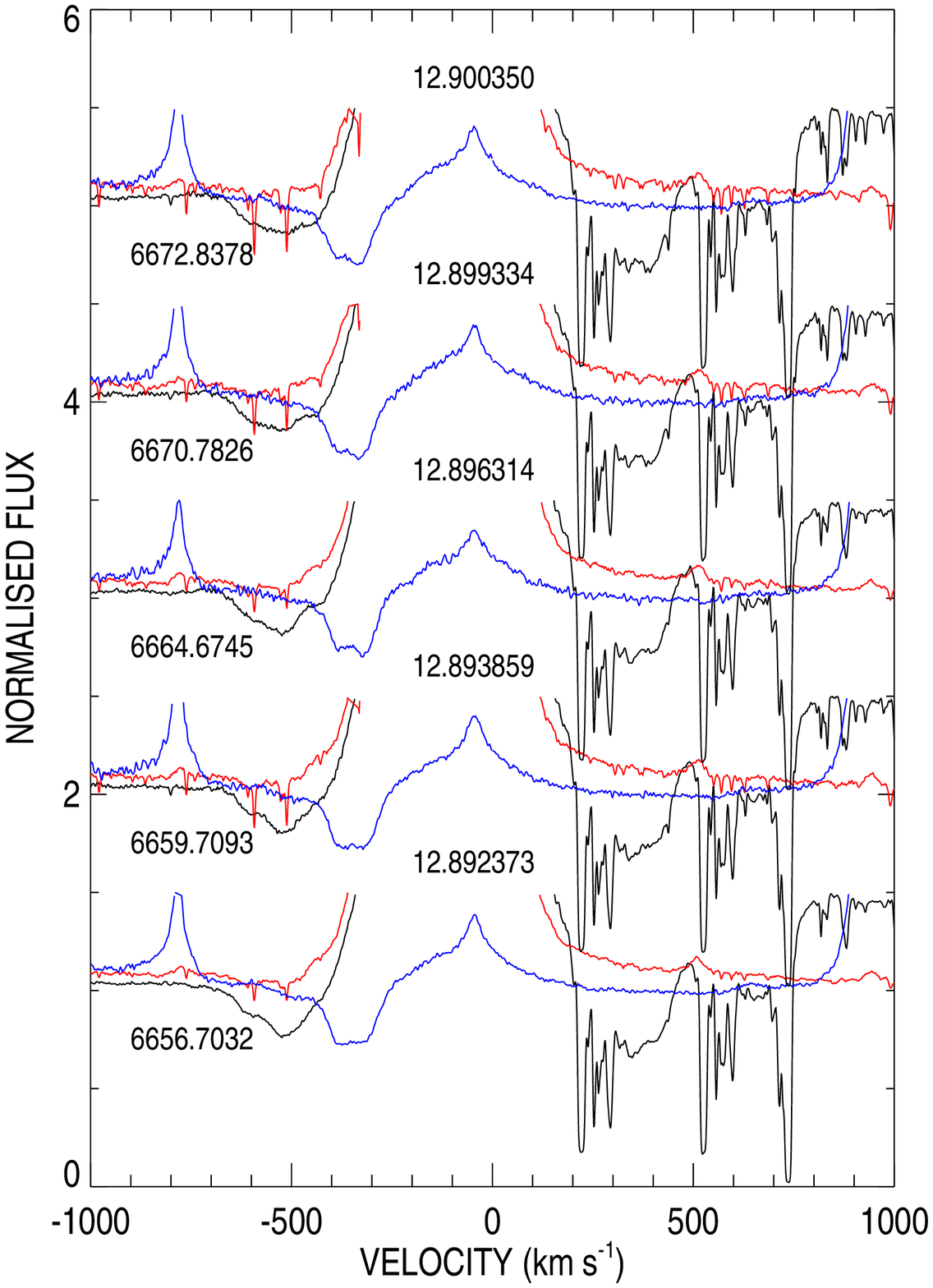}
 \includegraphics[width=50mm, angle=0]{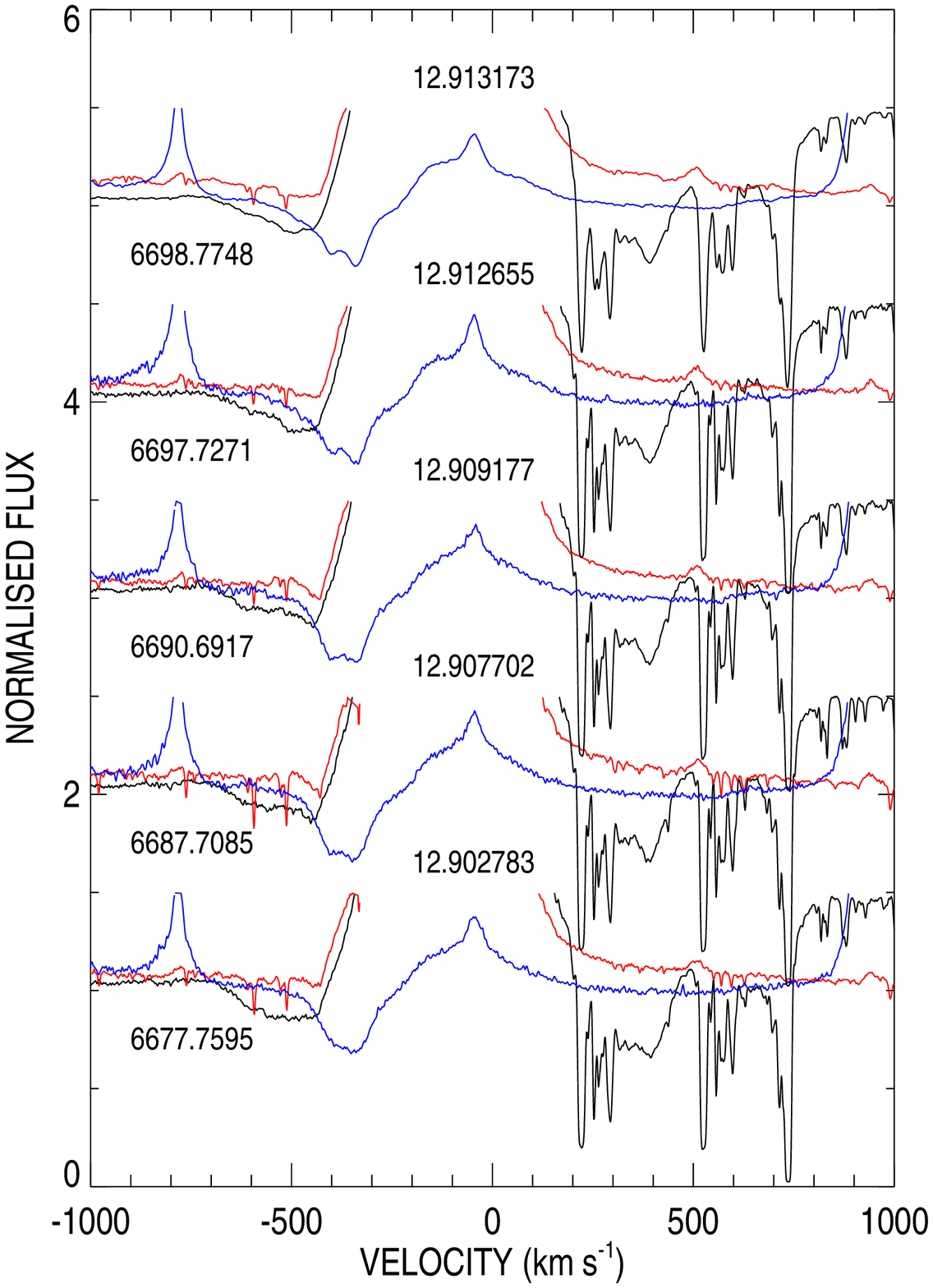}
 \includegraphics[width=50mm, angle=0]{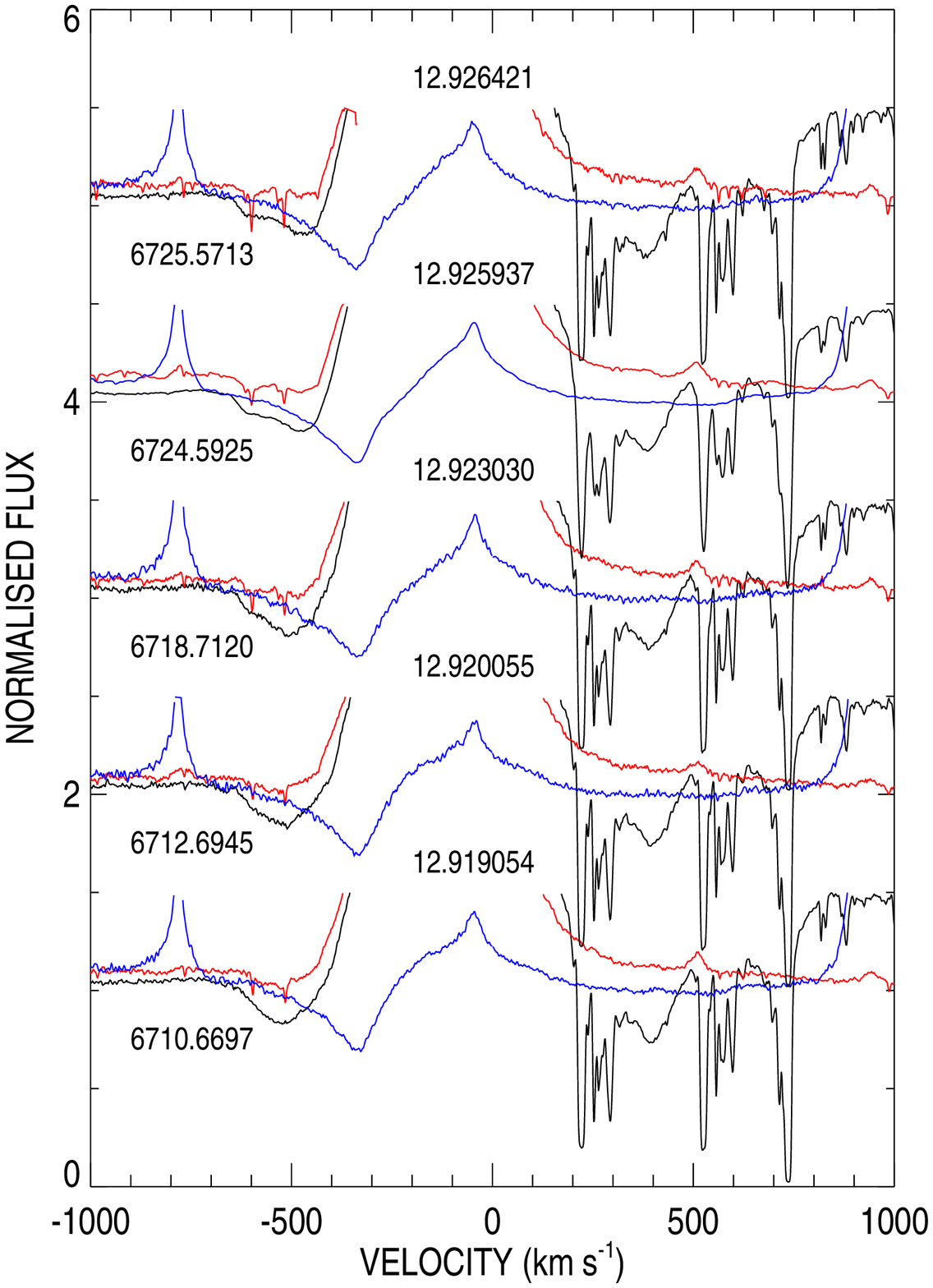}
 \includegraphics[width=50mm, angle=0]{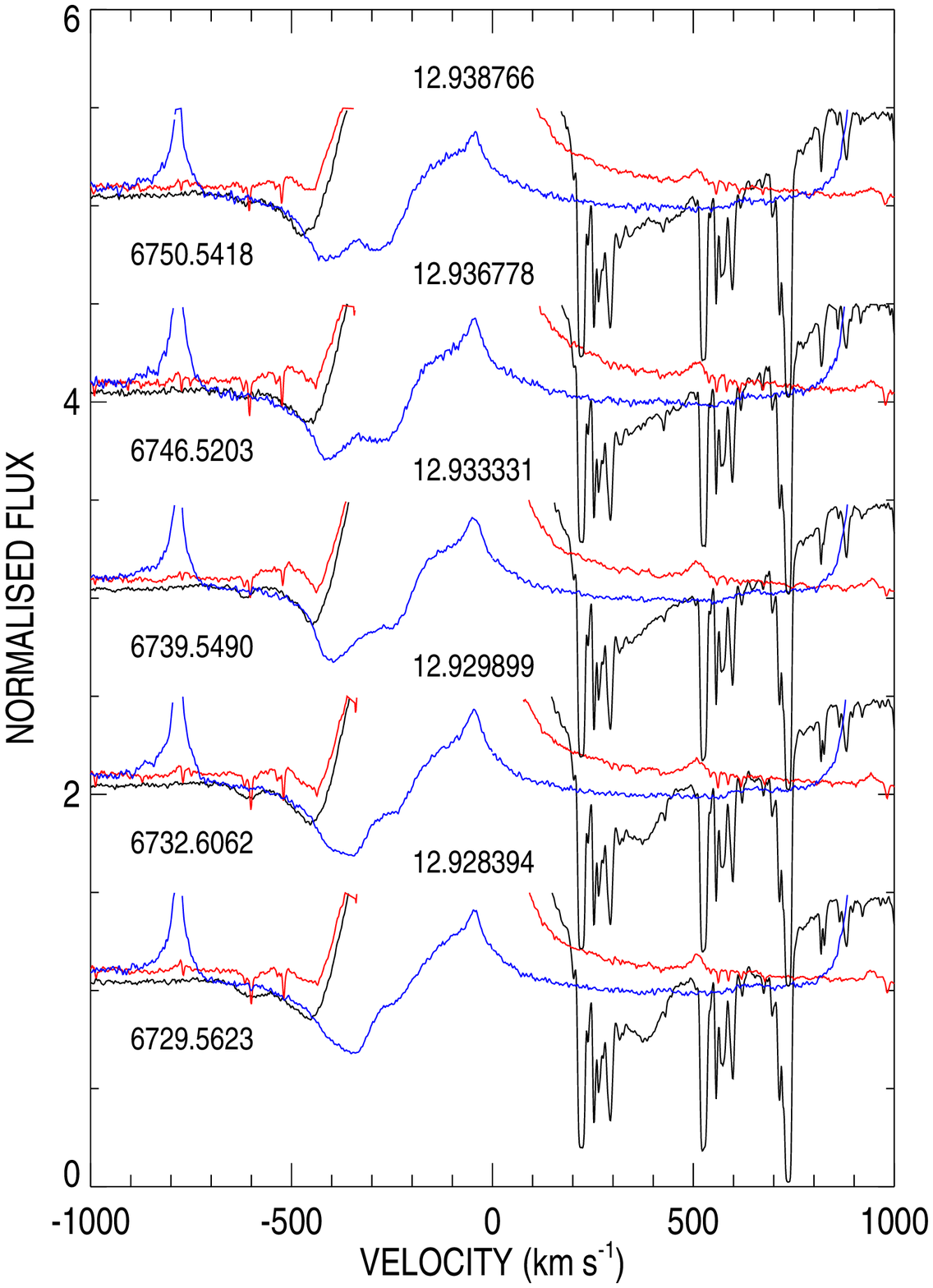}
 \includegraphics[width=50mm, angle=0]{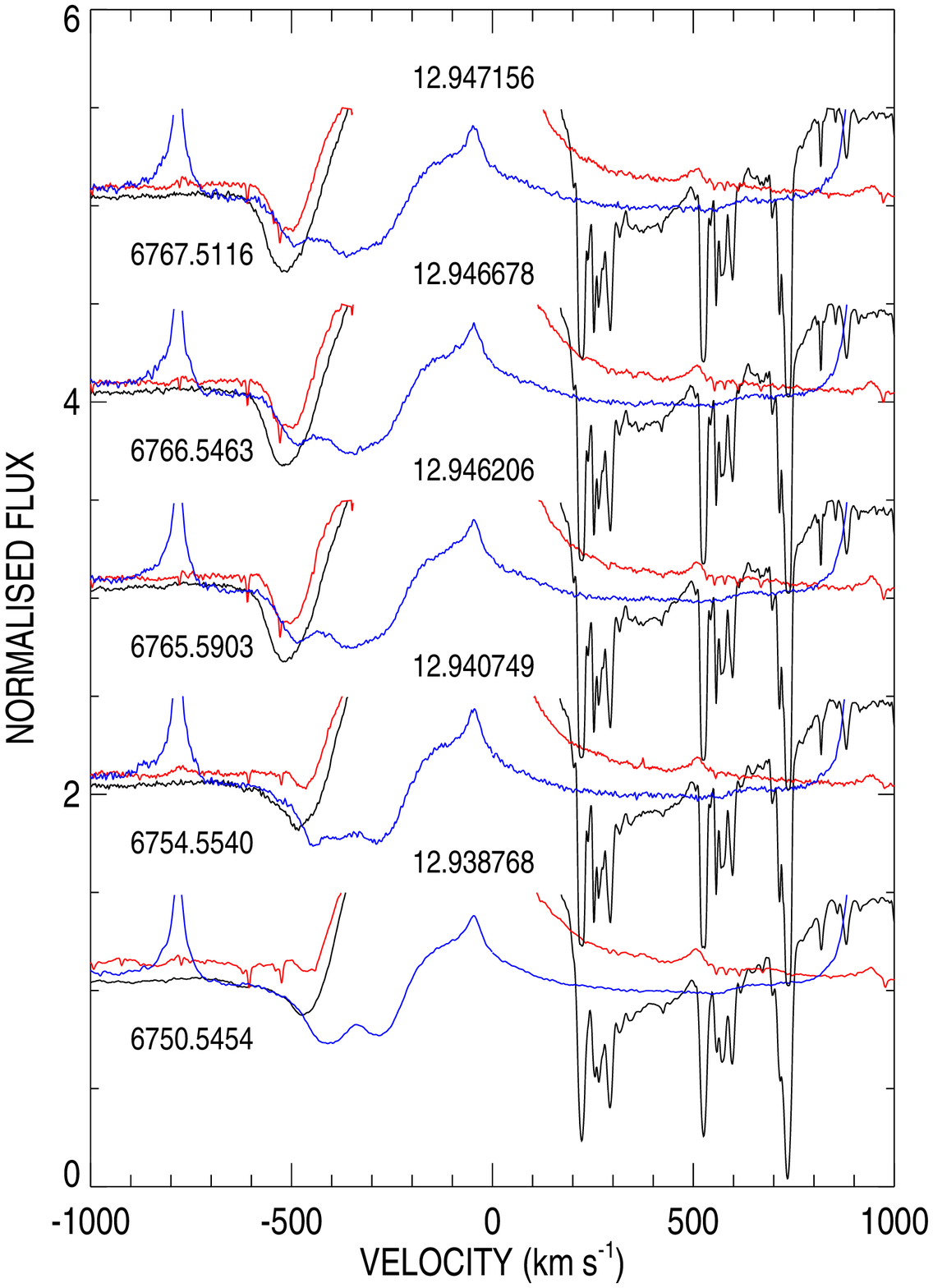}
 \includegraphics[width=50mm, angle=0]{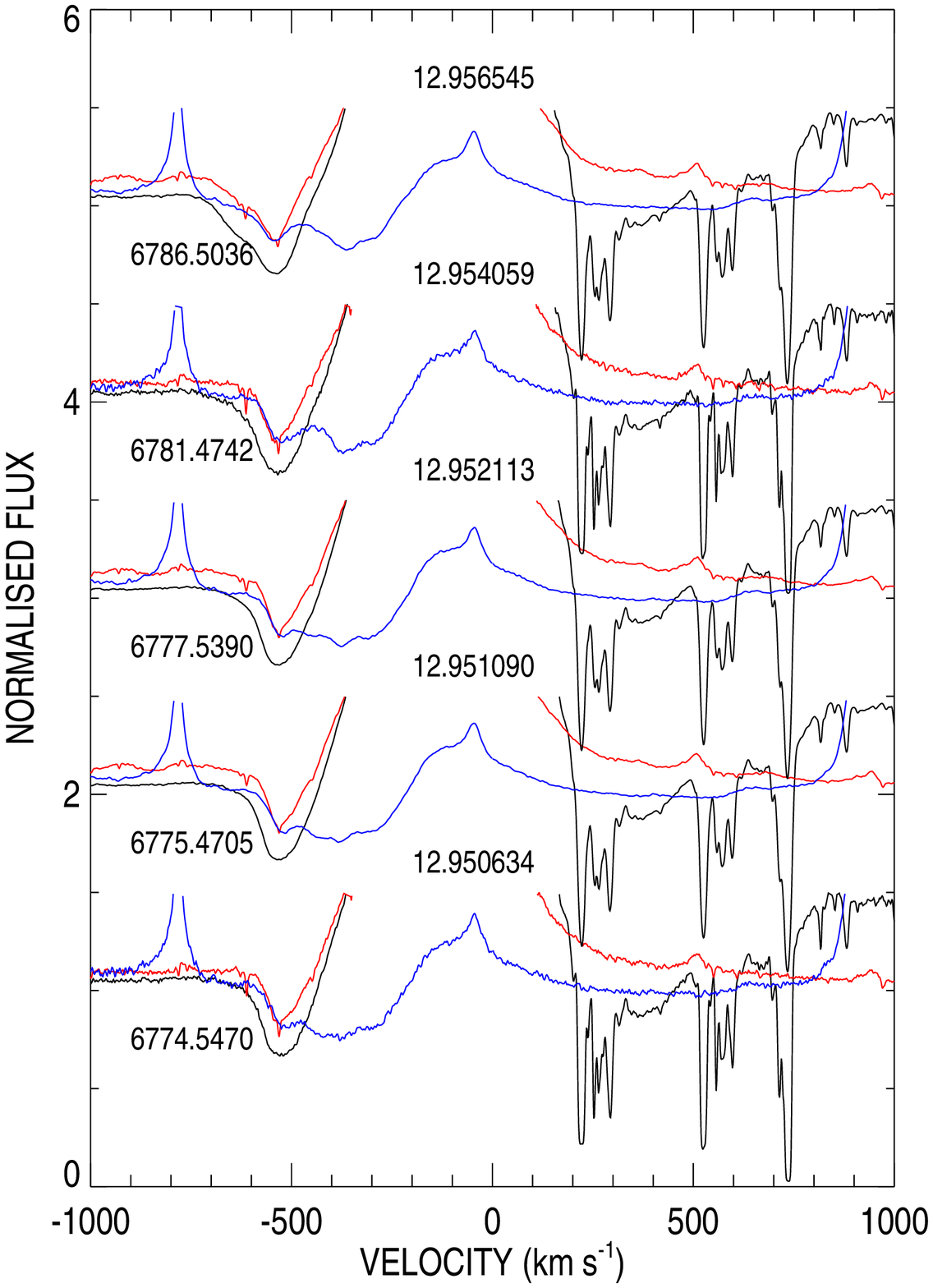}
 \includegraphics[width=50mm, angle=0]{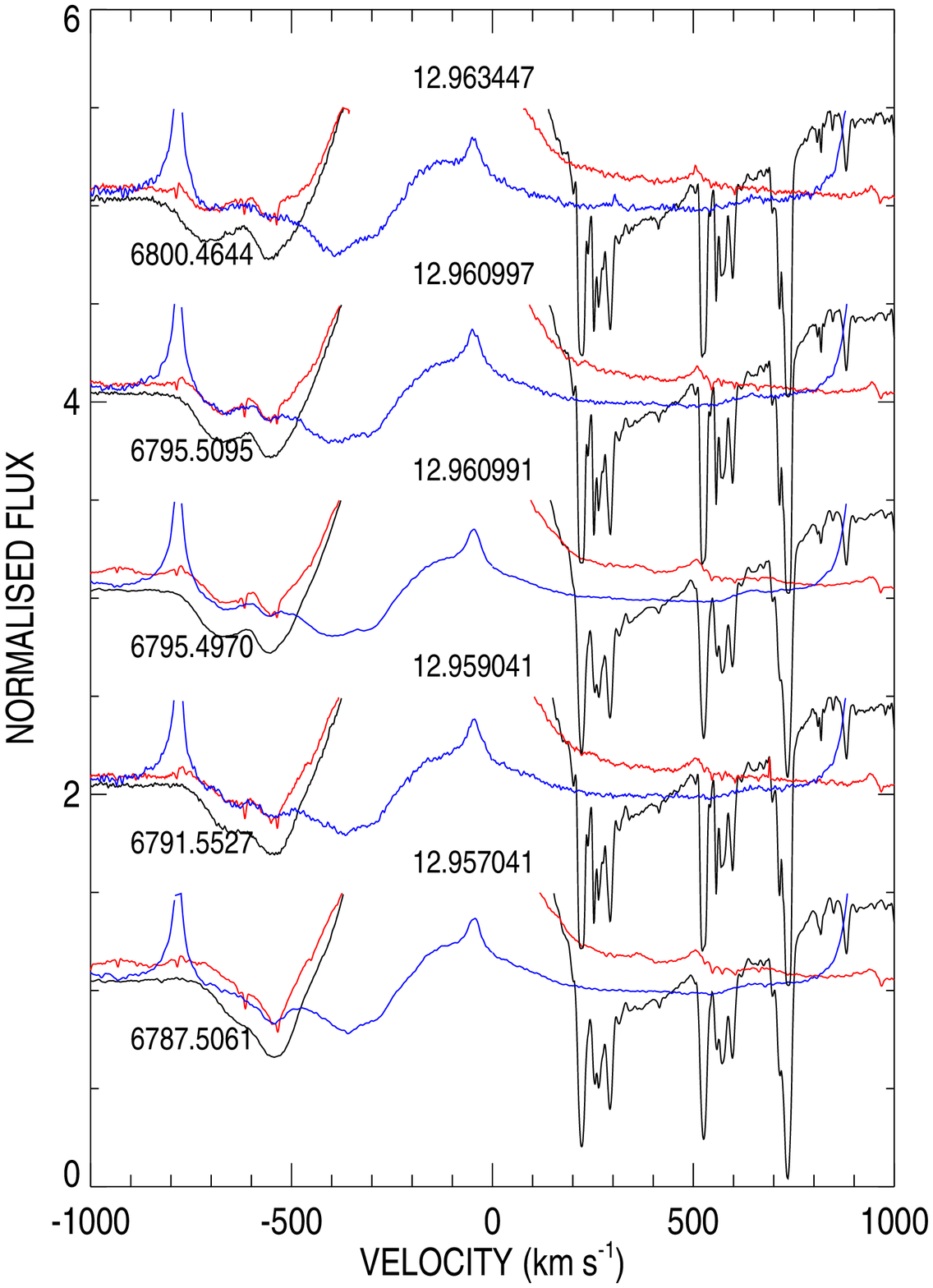}
 \includegraphics[width=50mm, angle=0]{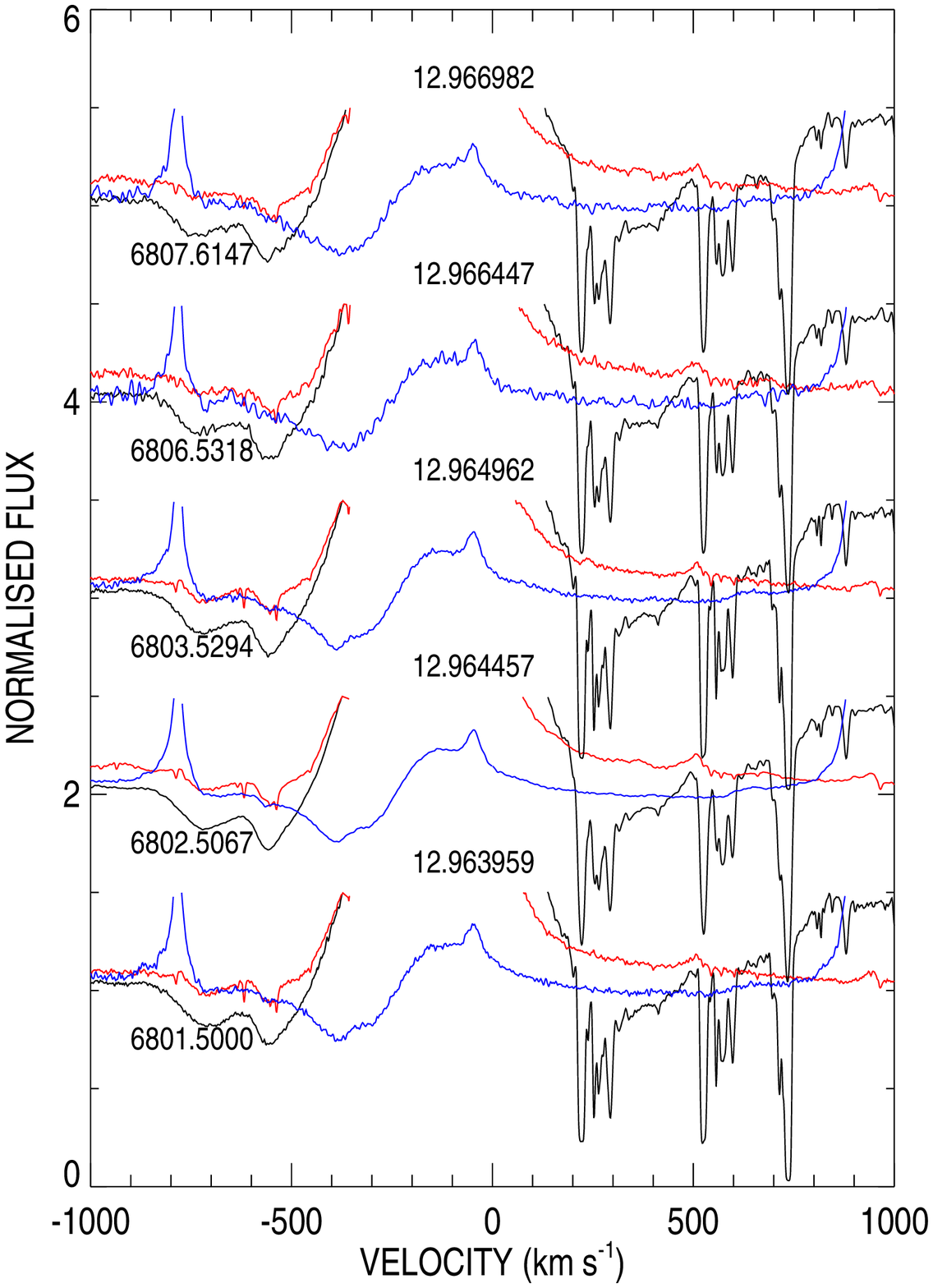}
 \includegraphics[width=50mm, angle=0]{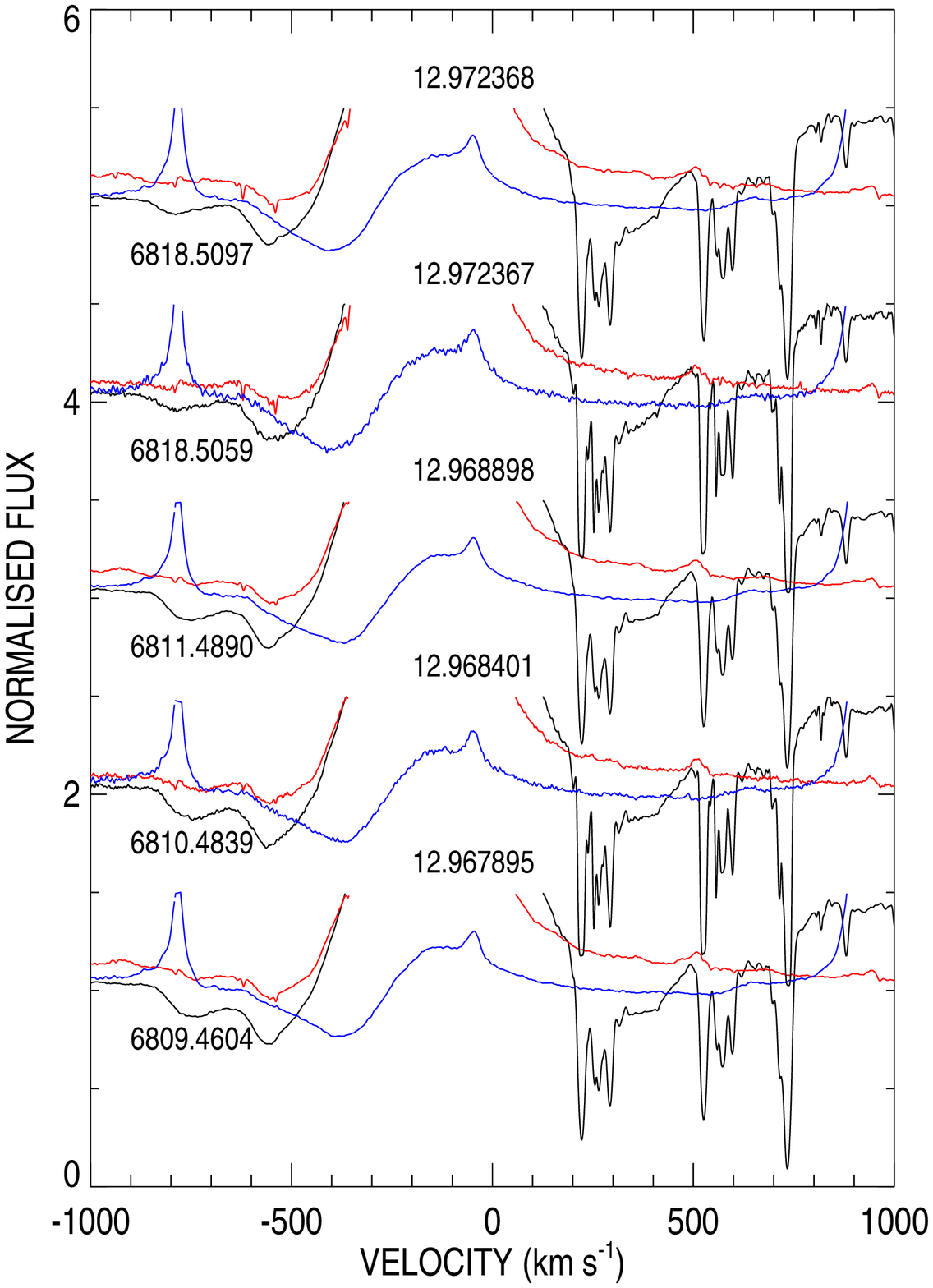}
\caption{\label{figprofiles_app} All observed (absorption) line profiles of the He~I 4713 (blue), He~I 5876 (black) and He~I 7065 (red) transitions from just before the onset of the high-velocity variability to just following the large variations. All profiles are set to a heliocentric frame and large emission portions are cut off for clarity of the absorption variations. The dates (HJD$-$2,450,000) are indicated on the left, while the phases computed using the ephemeris from Teodoro et al.~(2015, in prep.) are shown above each set of observations. In each frame, the different dates are offset for clarity.}
\end{figure*}

\begin{figure*}
 \includegraphics[width=50mm, angle=0]{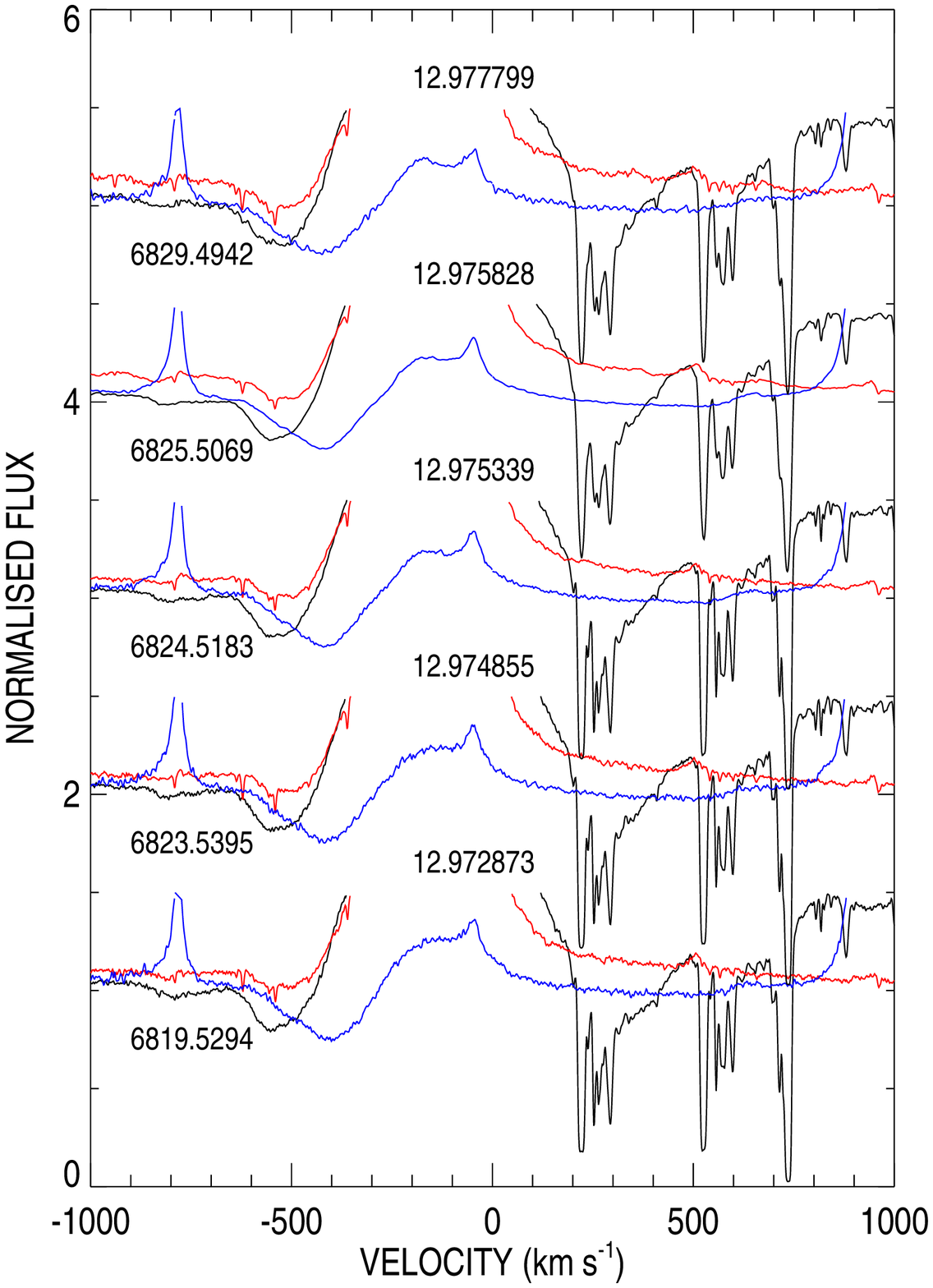}
 \includegraphics[width=50mm, angle=0]{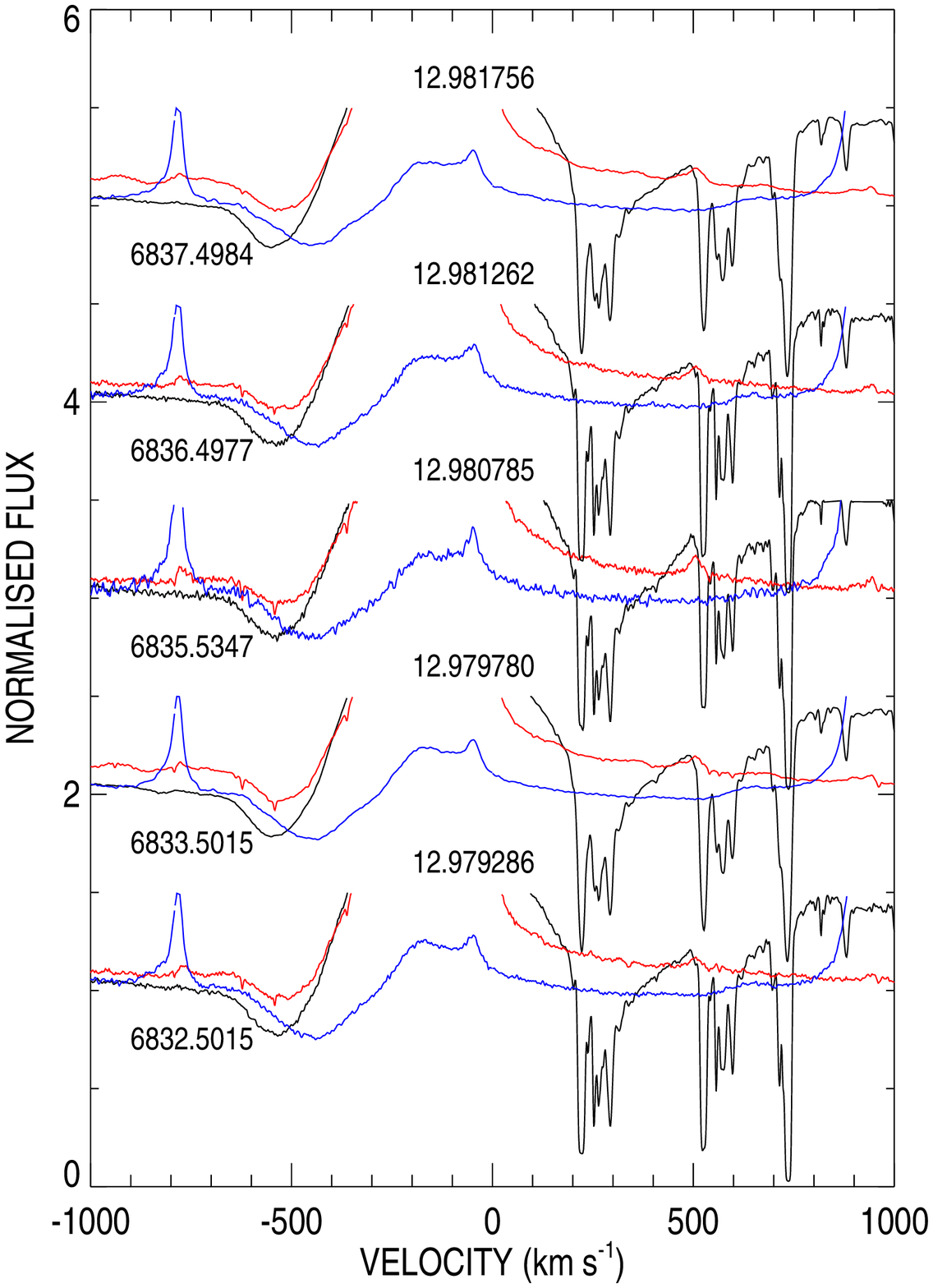}
 \includegraphics[width=50mm, angle=0]{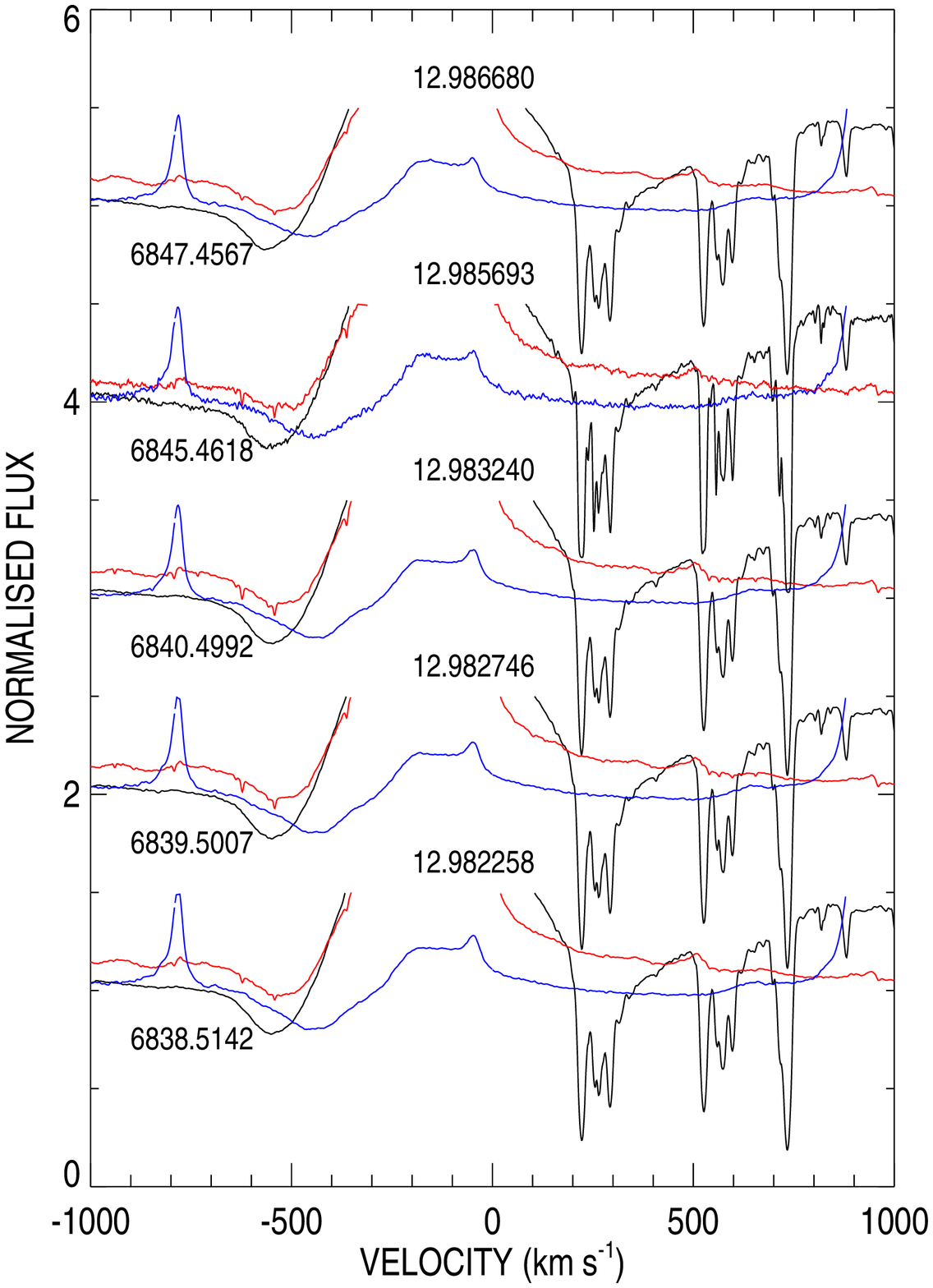}
 \includegraphics[width=50mm, angle=0]{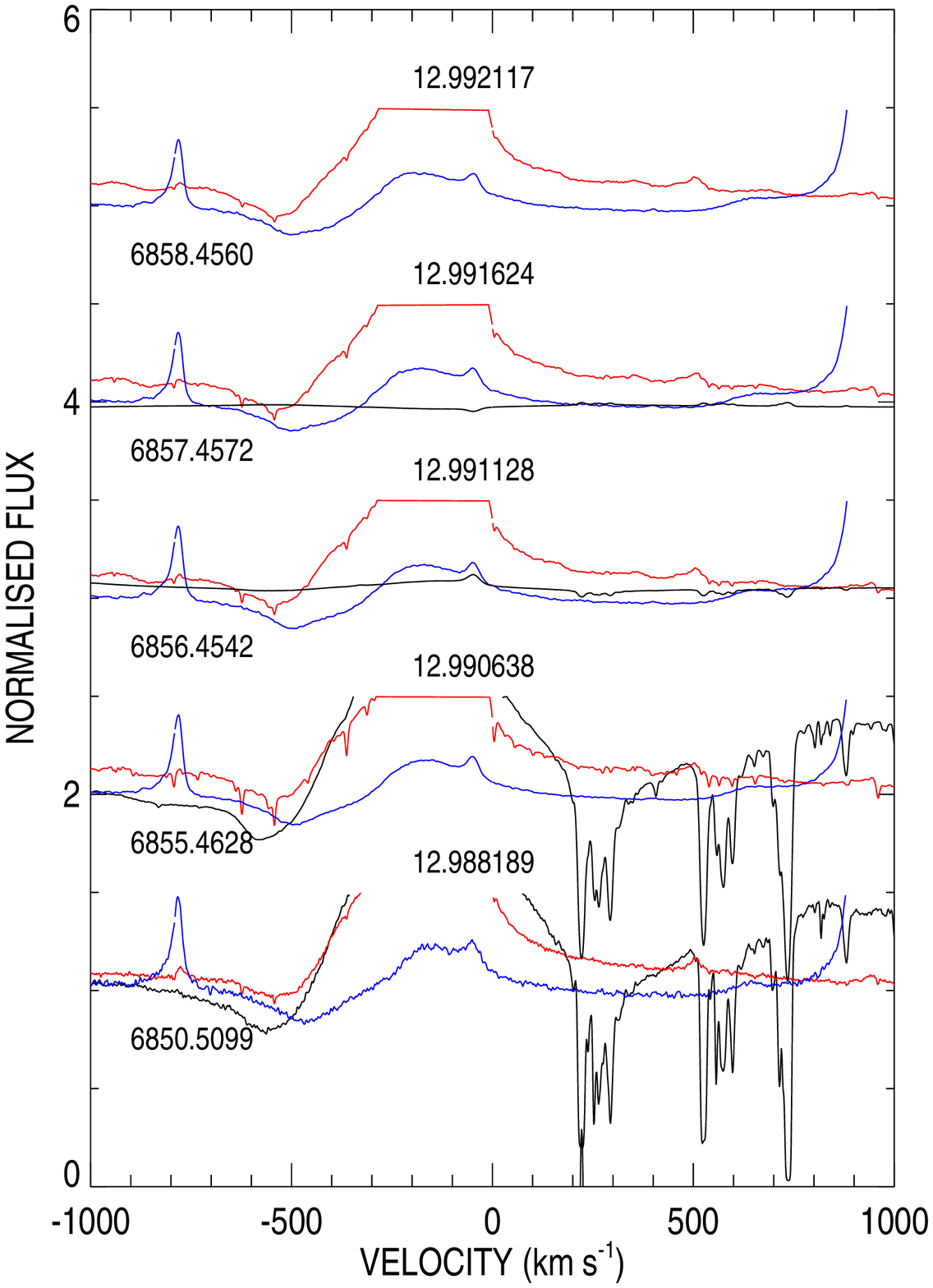}
 \includegraphics[width=50mm, angle=0]{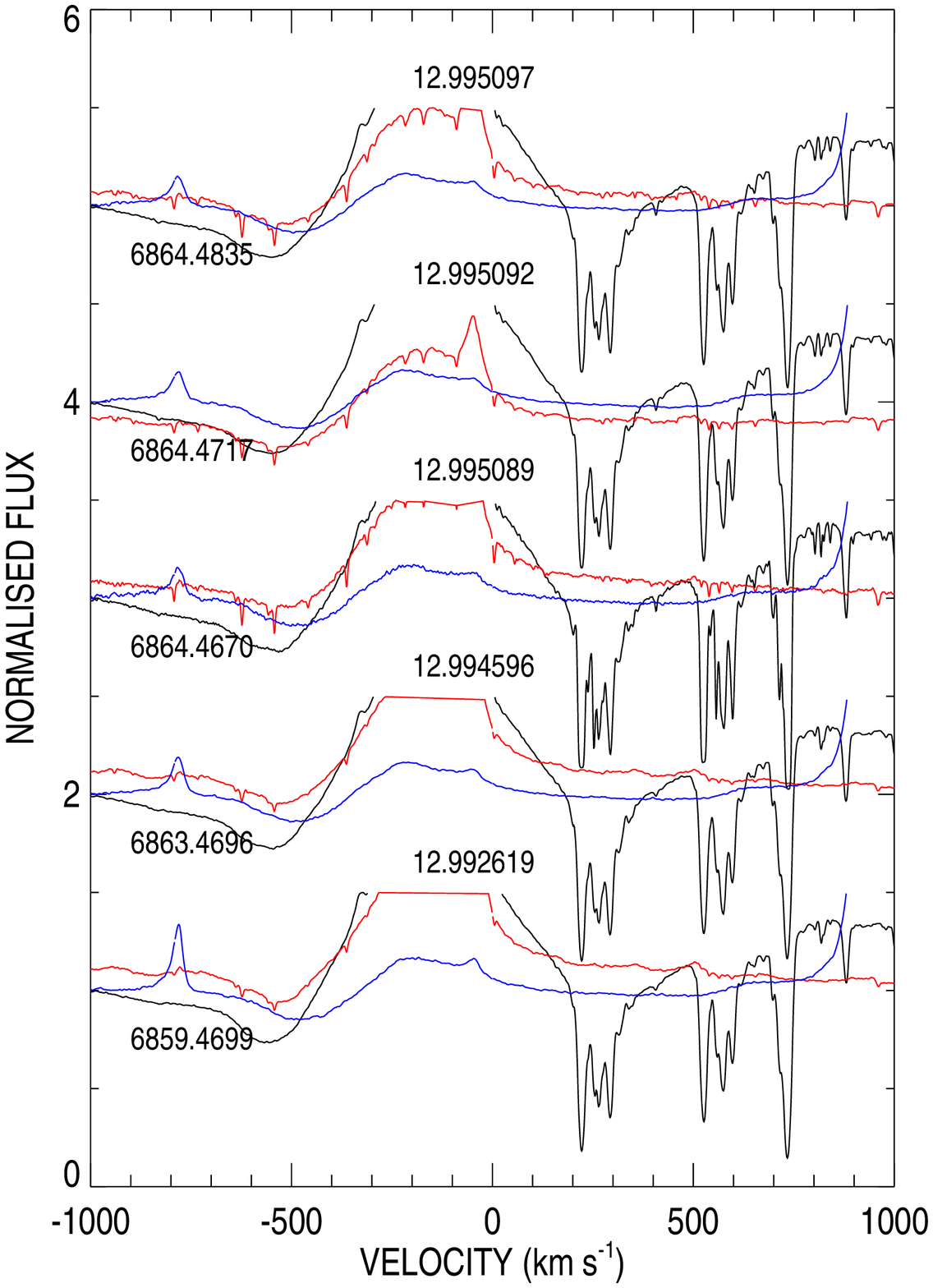}
 \includegraphics[width=50mm, angle=0]{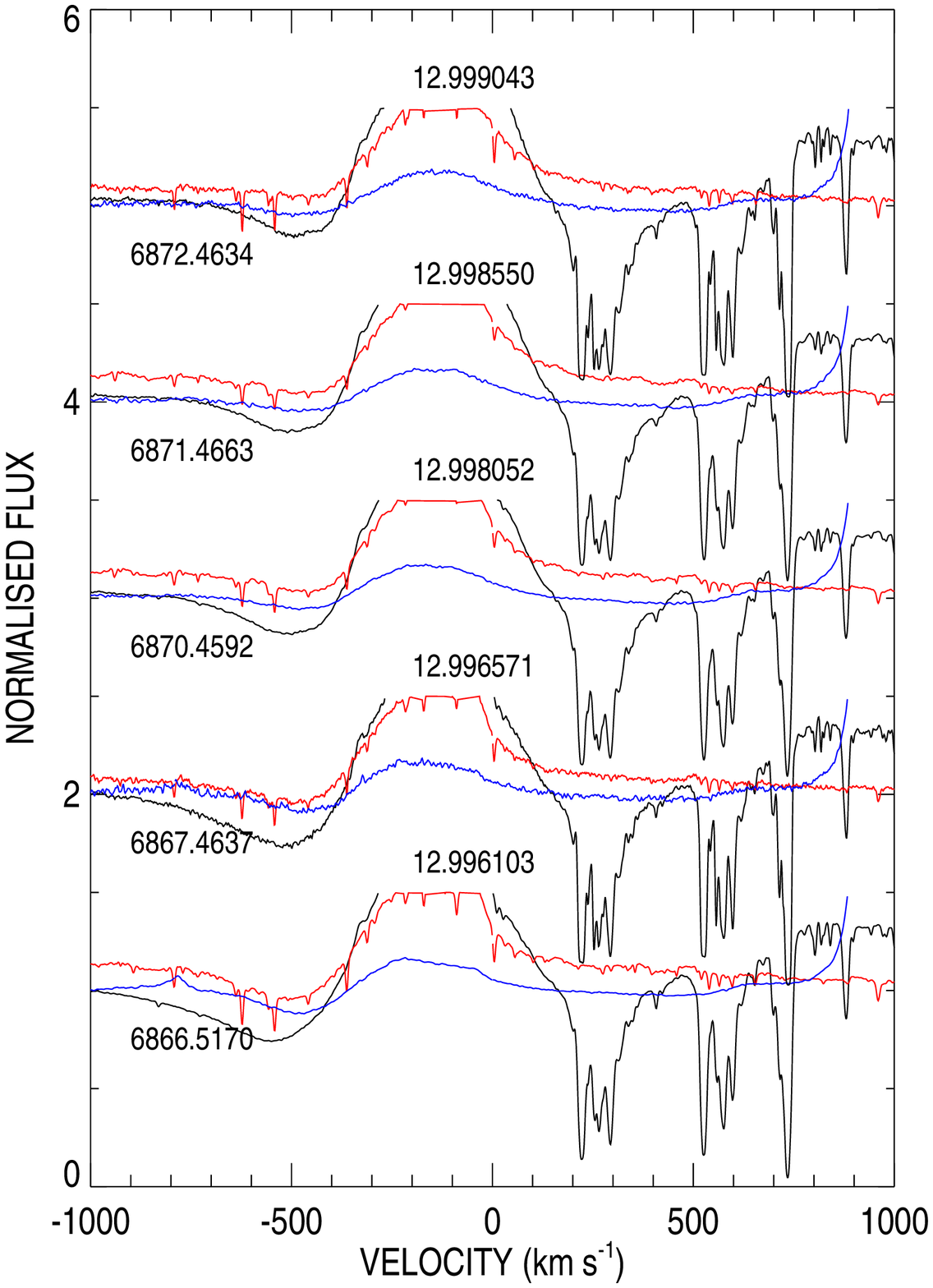}
 \includegraphics[width=50mm, angle=0]{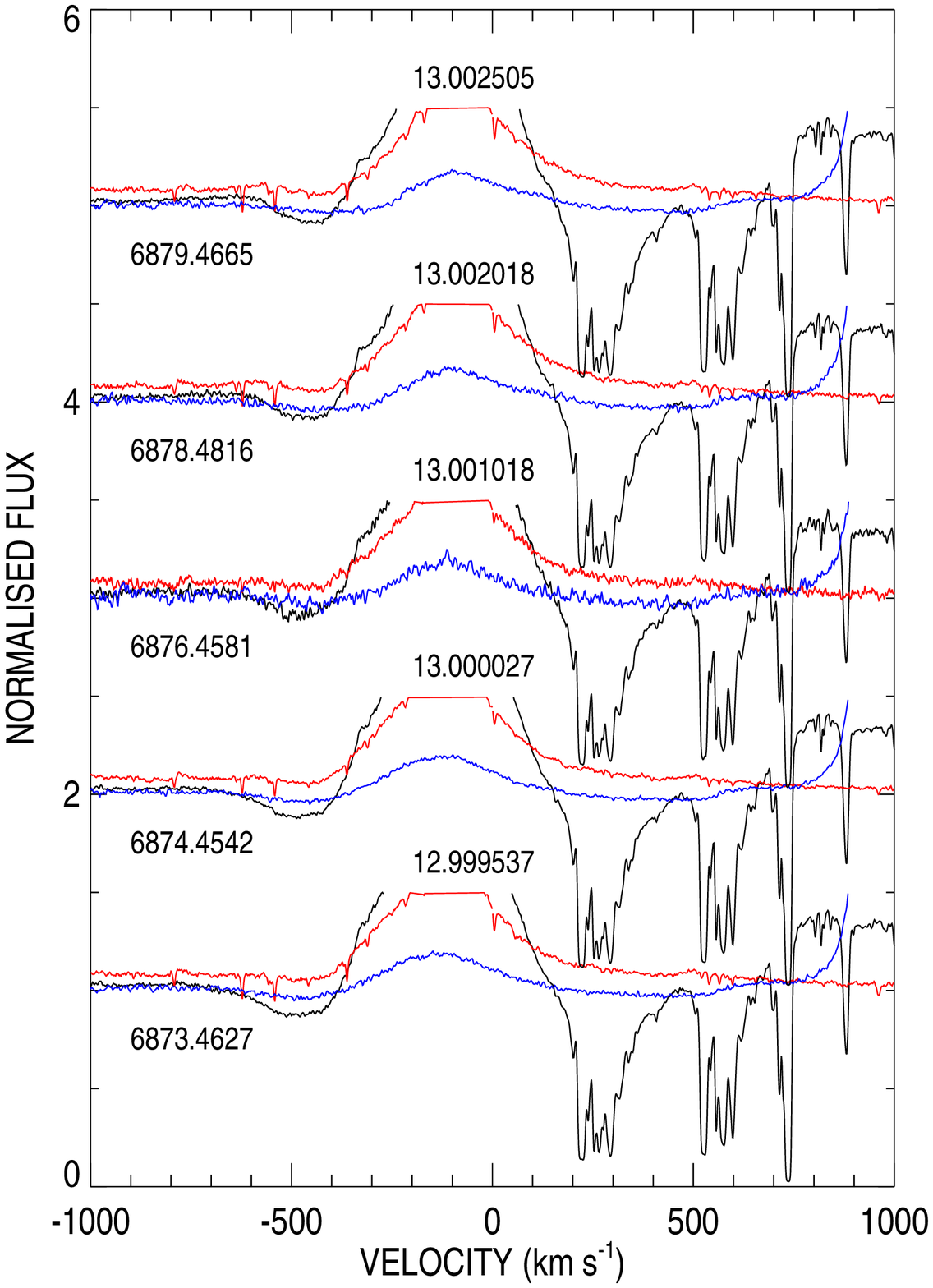}
 \includegraphics[width=50mm, angle=0]{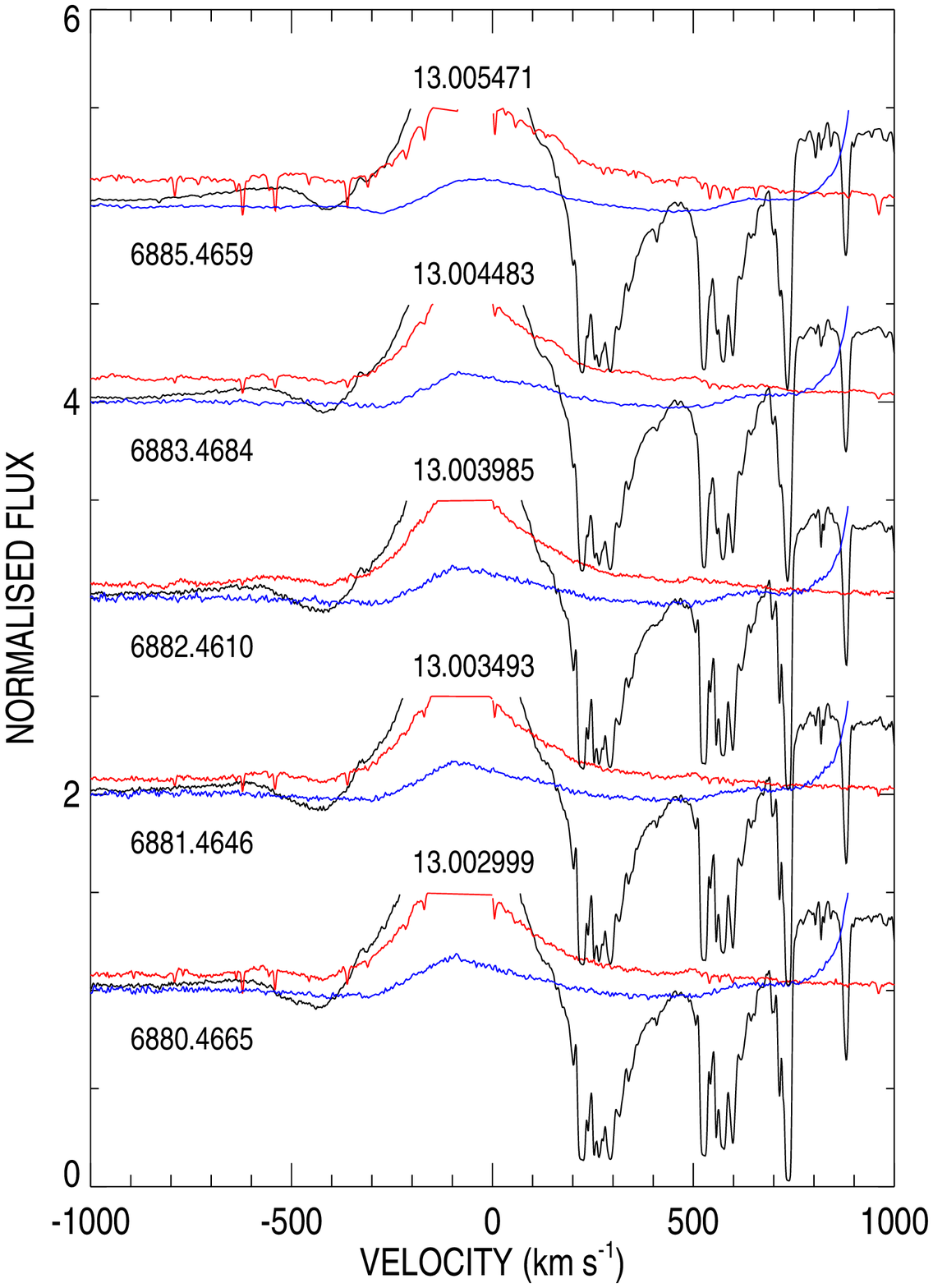}
 \includegraphics[width=50mm, angle=0]{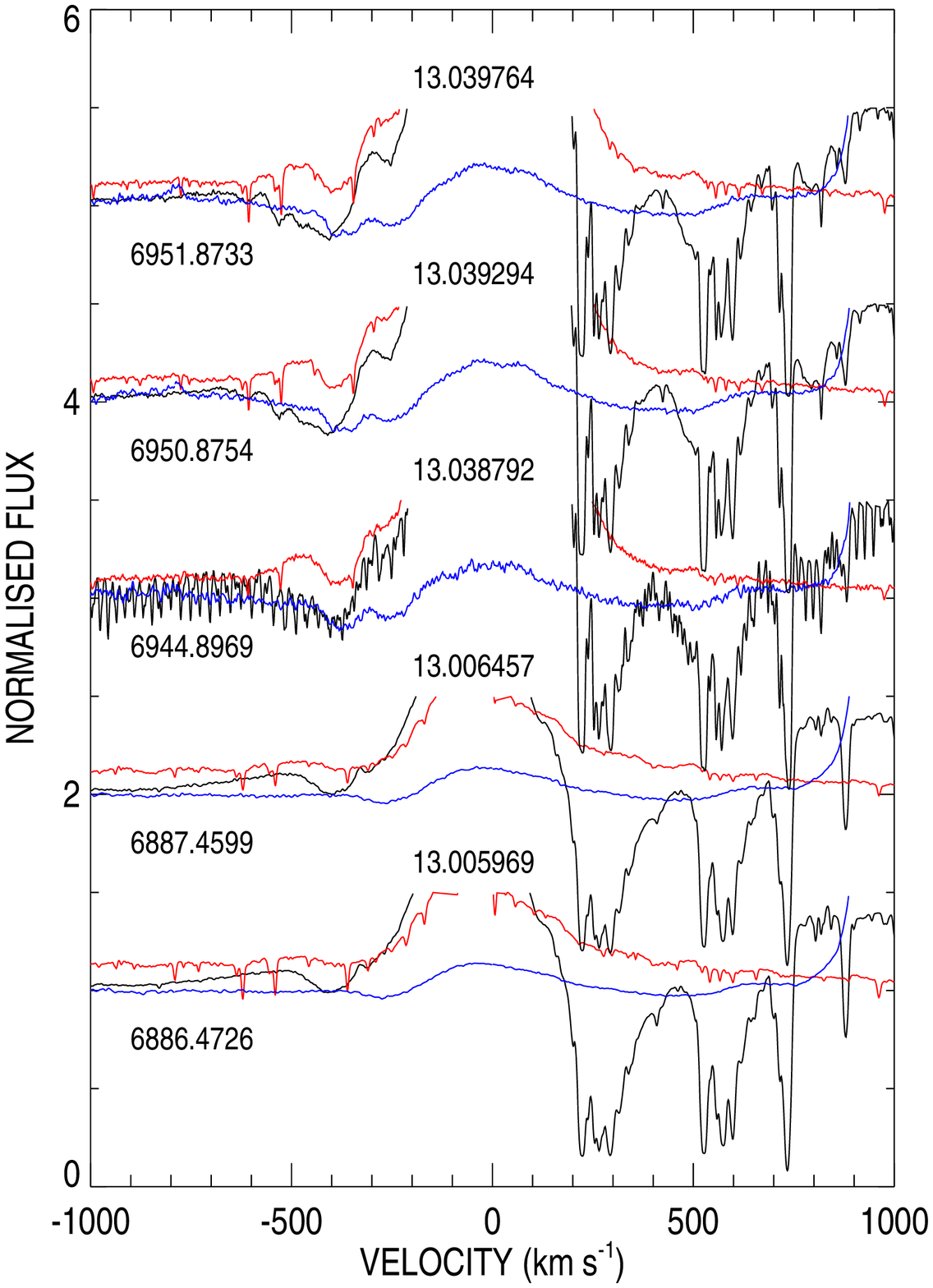}
\contcaption{\label{figprofiles_app}} 
\end{figure*}

\begin{figure*}
 \includegraphics[width=50mm, angle=0]{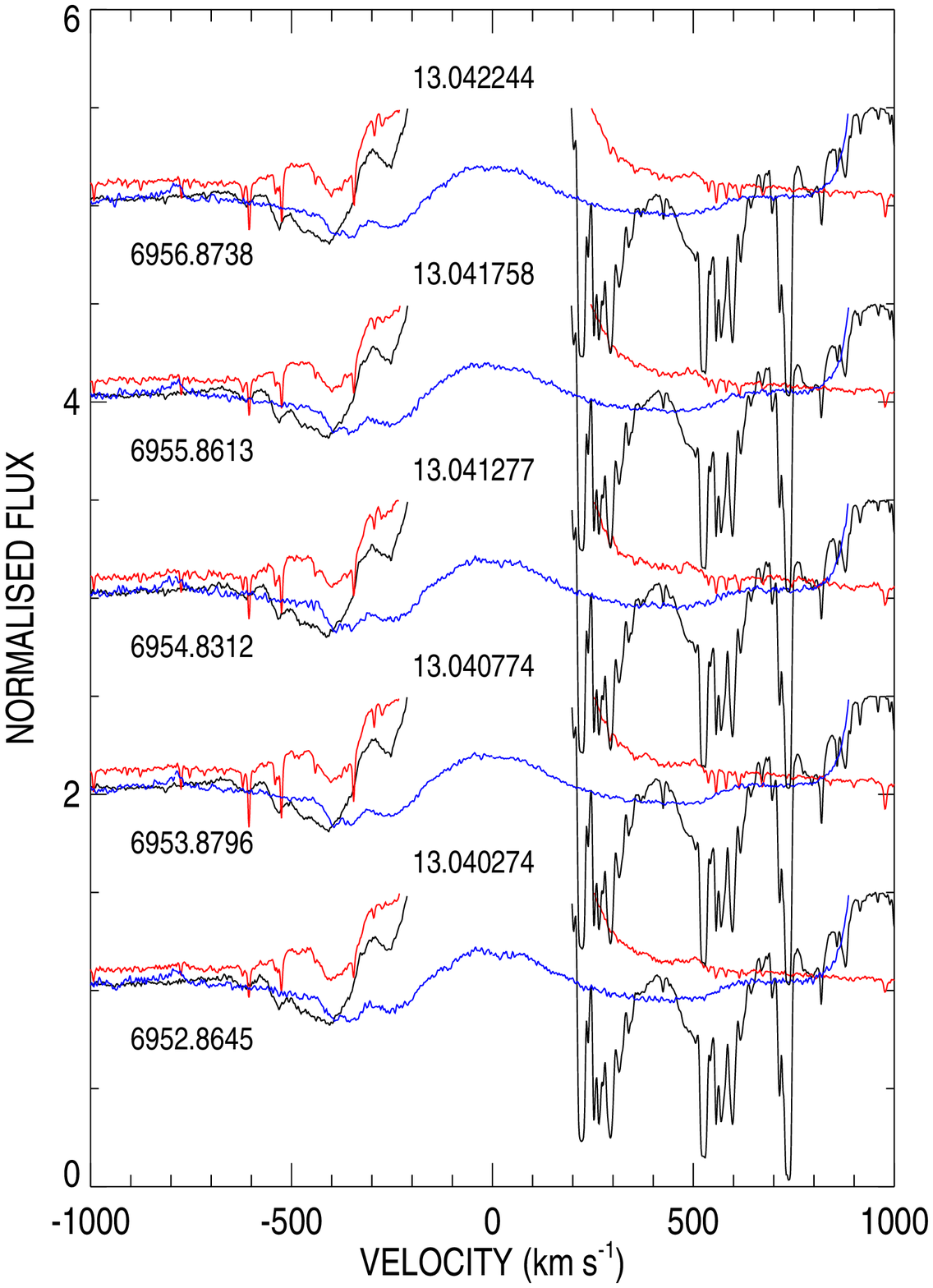}
 \includegraphics[width=50mm, angle=0]{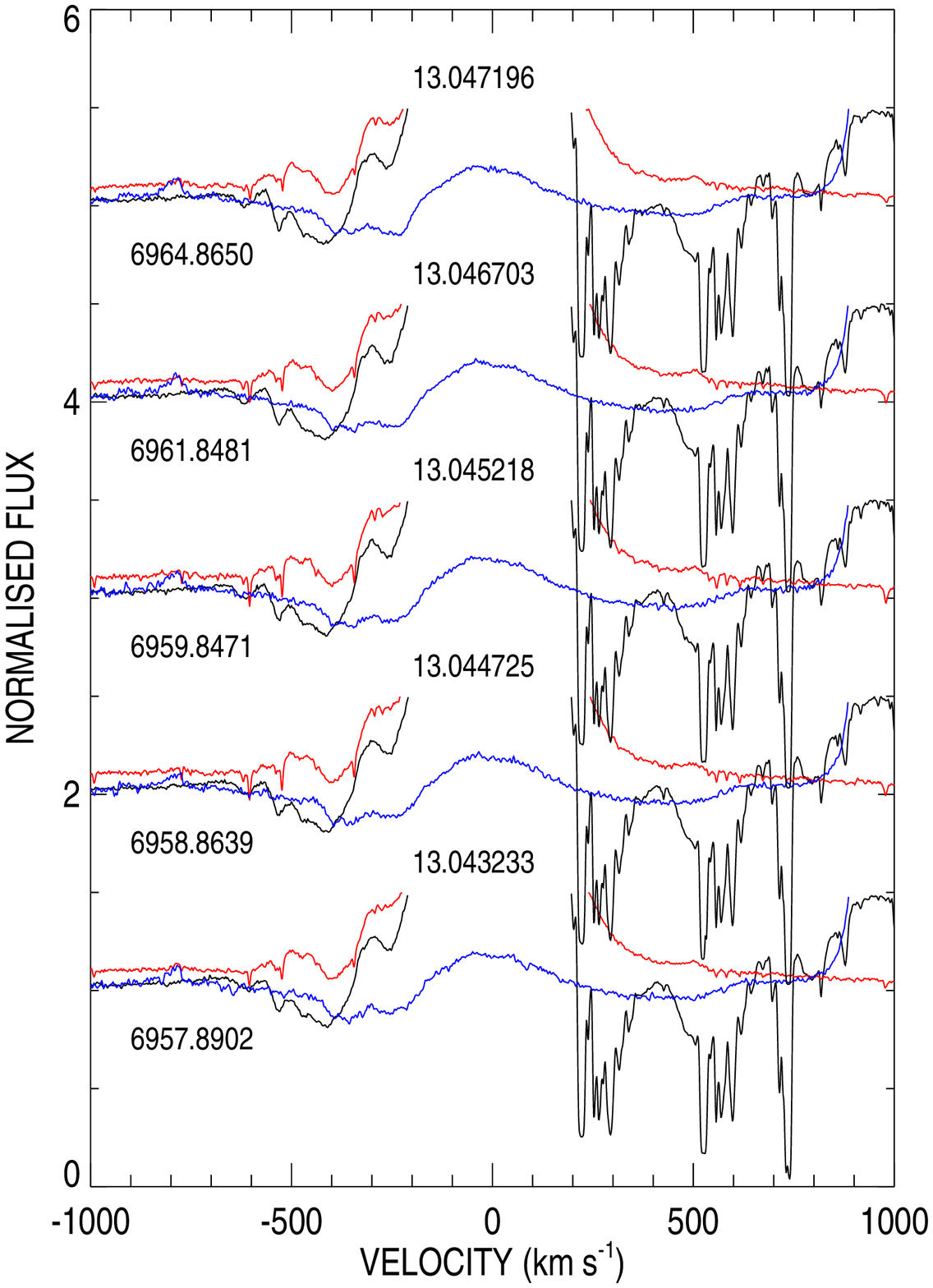}
 \includegraphics[width=50mm, angle=0]{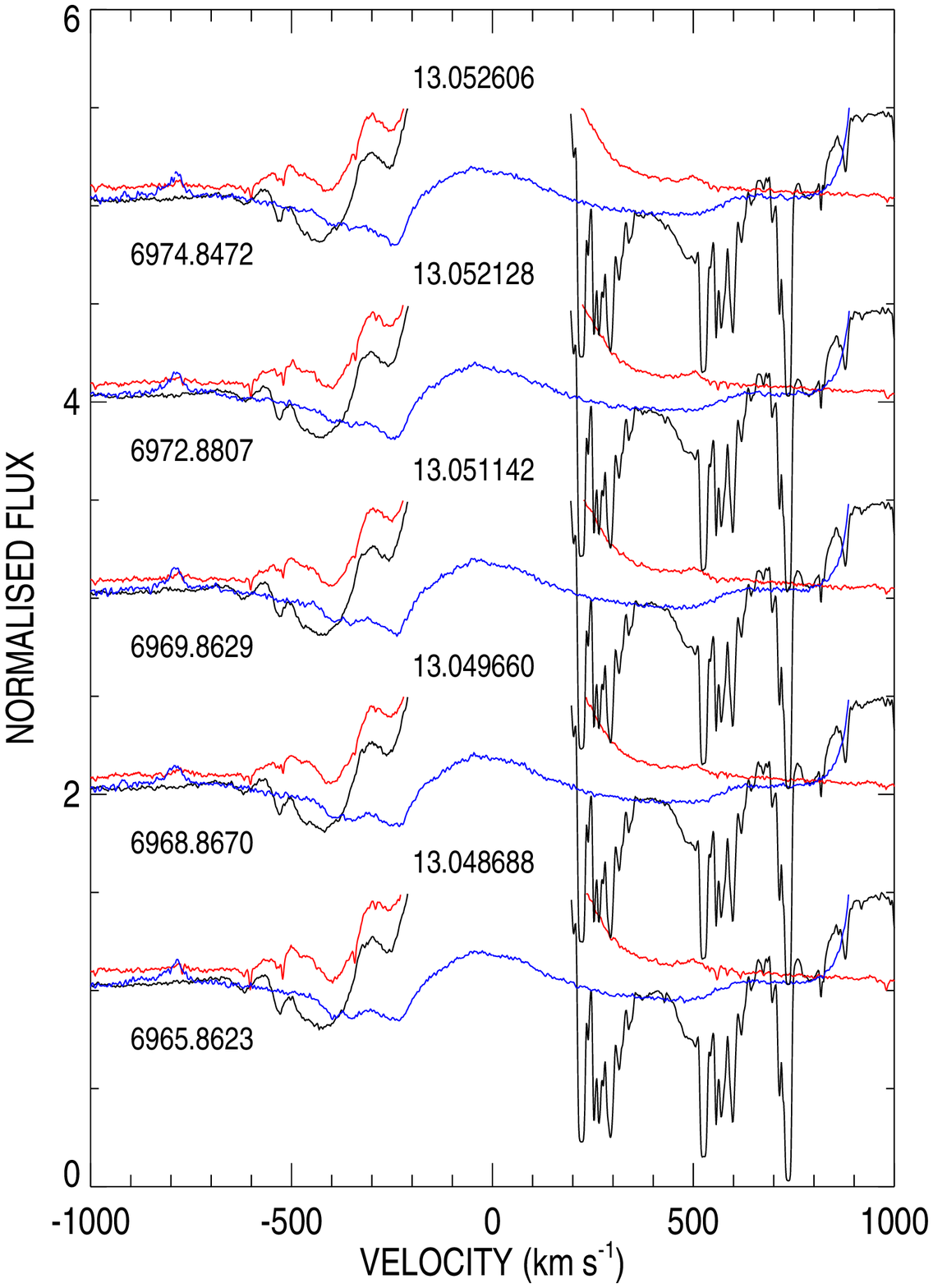}
 \includegraphics[width=50mm, angle=0]{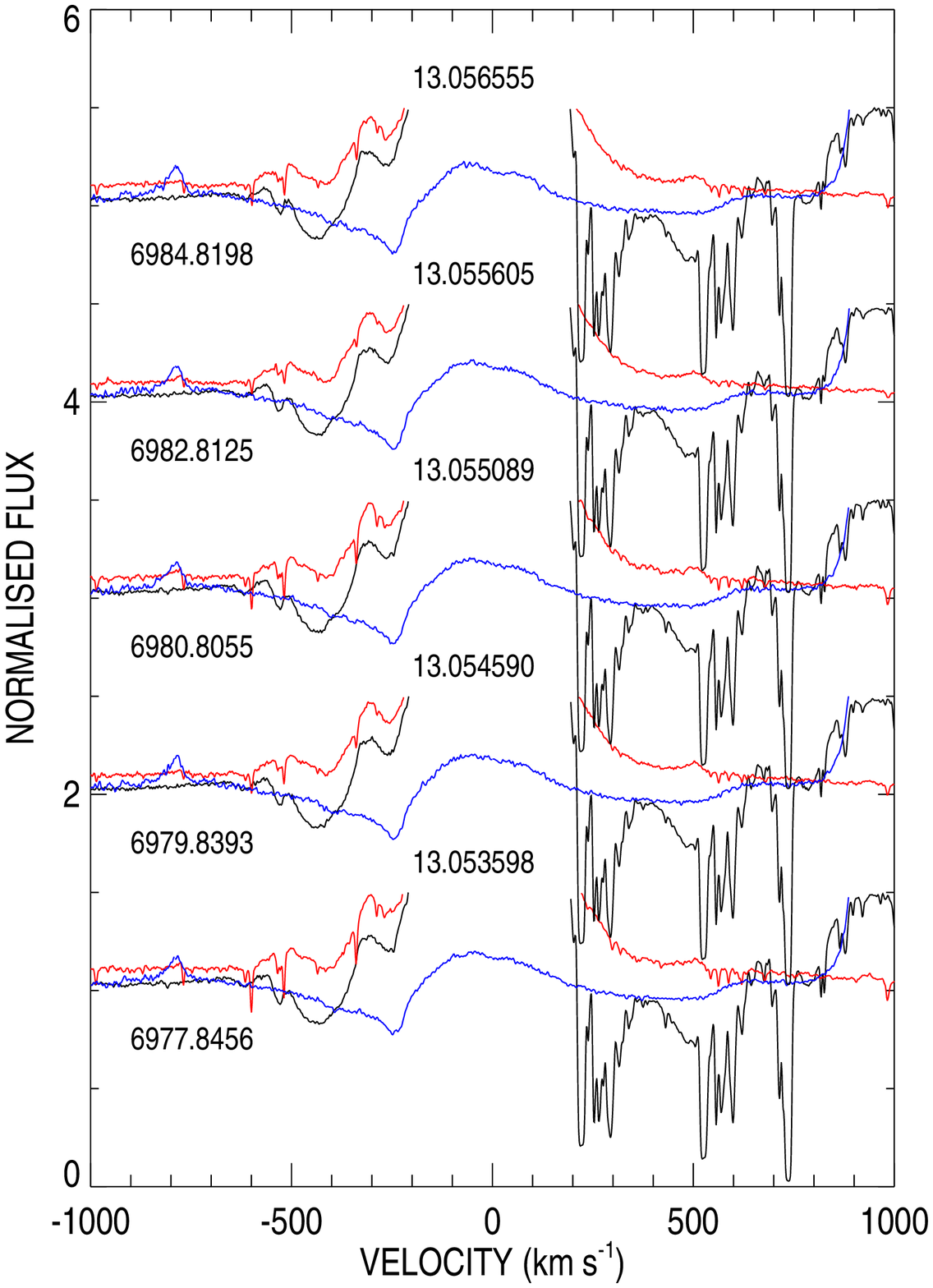}
 \includegraphics[width=50mm, angle=0]{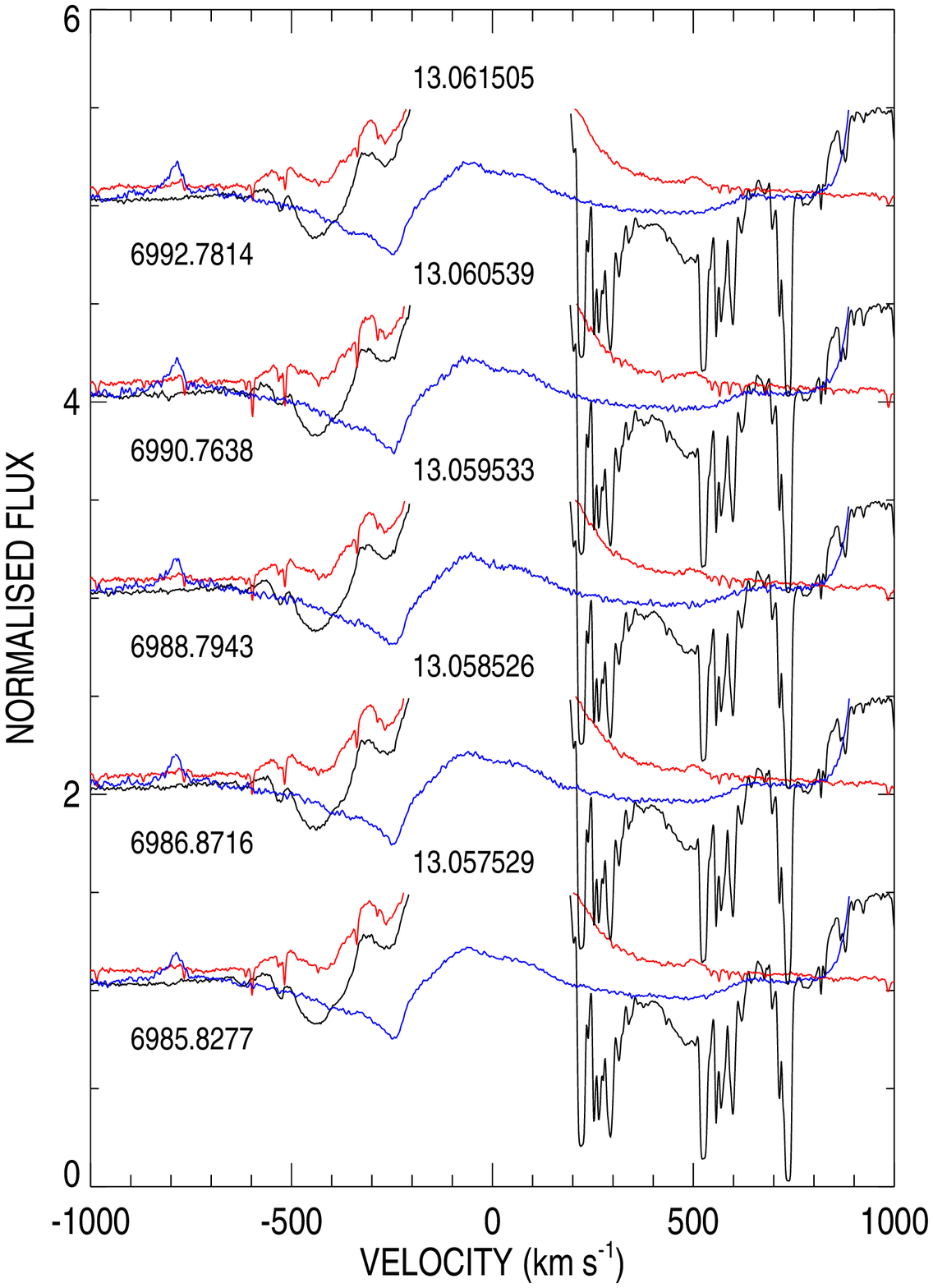}
 \includegraphics[width=50mm, angle=0]{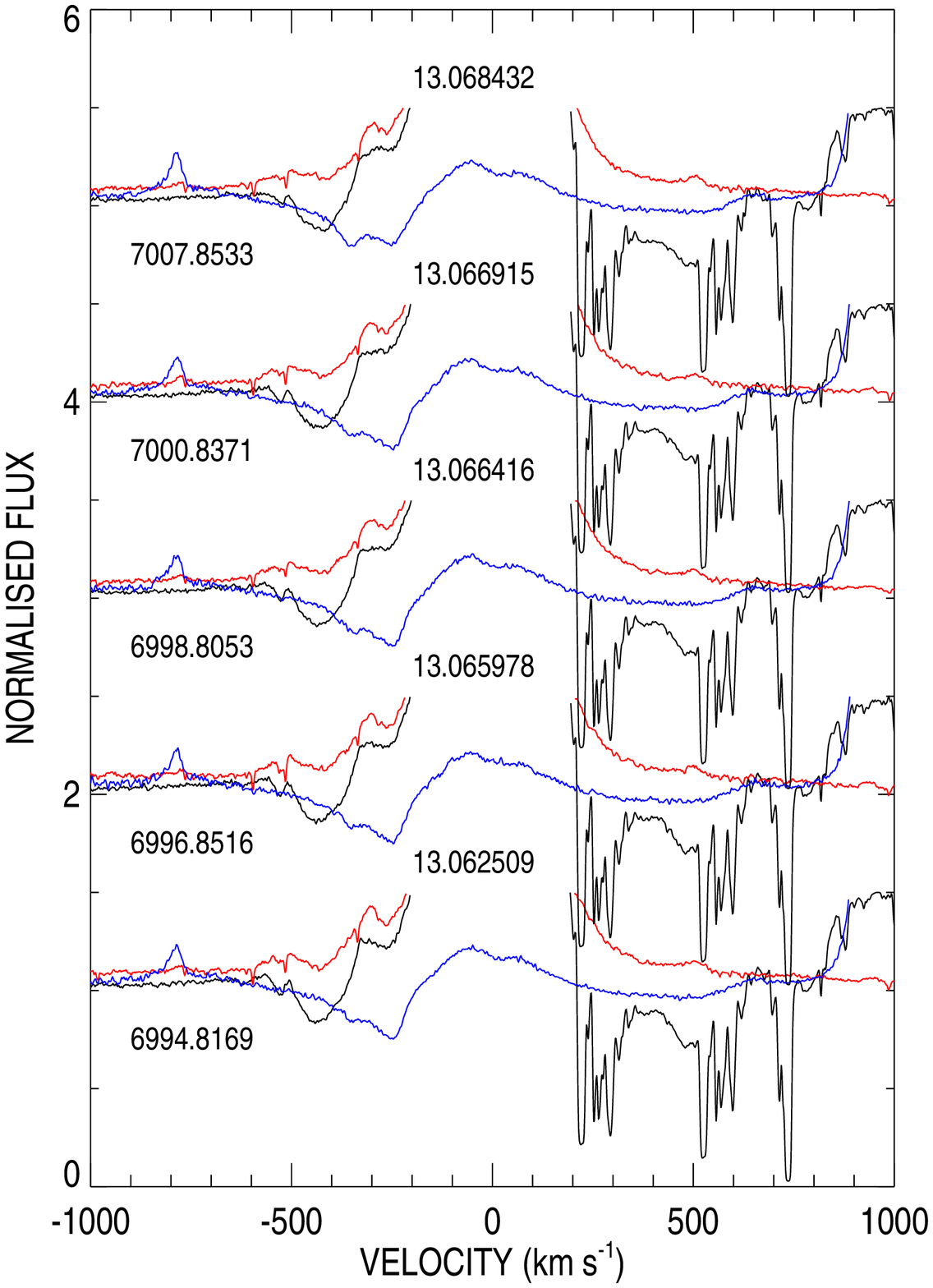}
 \includegraphics[width=50mm, angle=0]{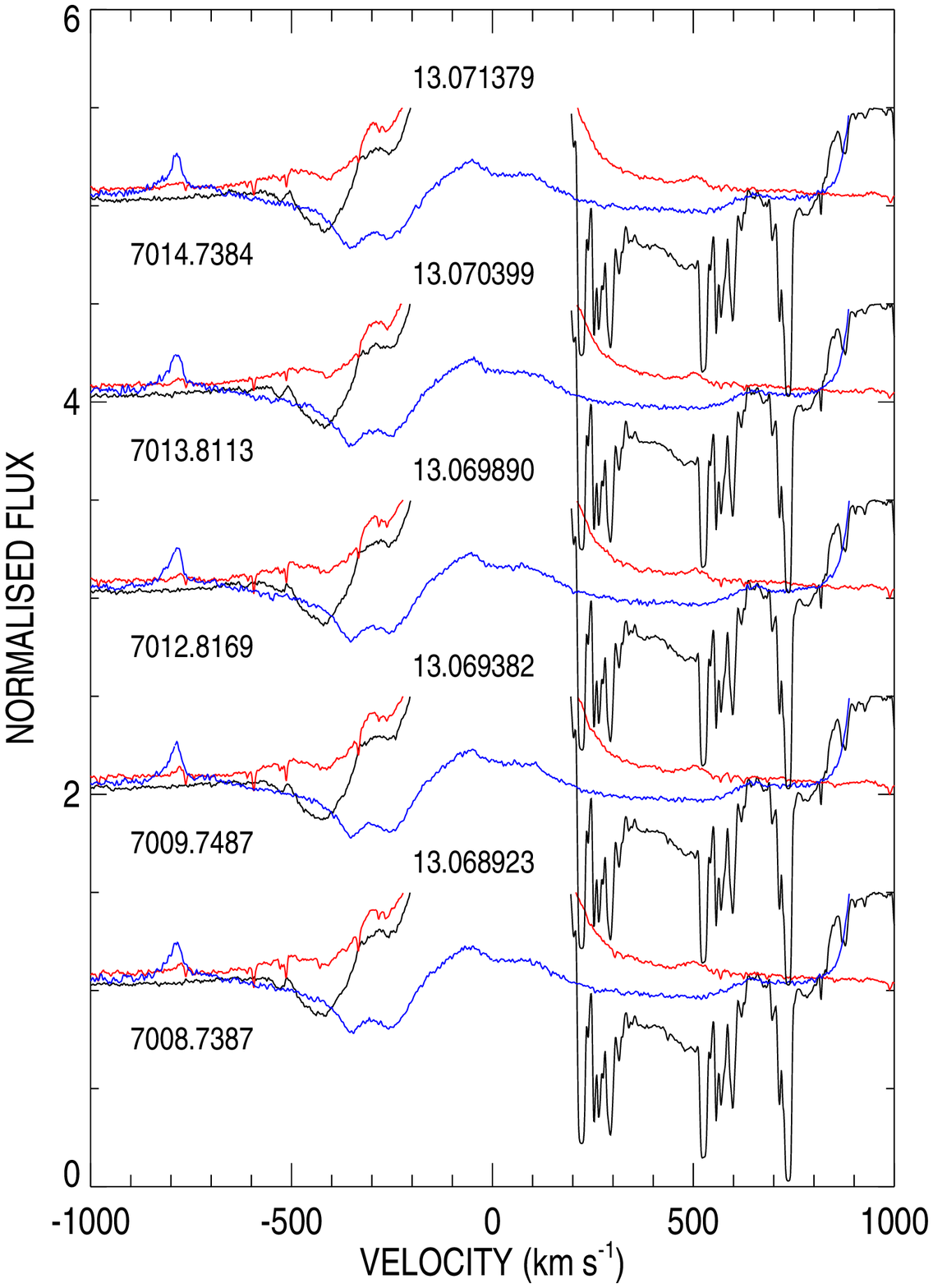}
 \includegraphics[width=50mm, angle=0]{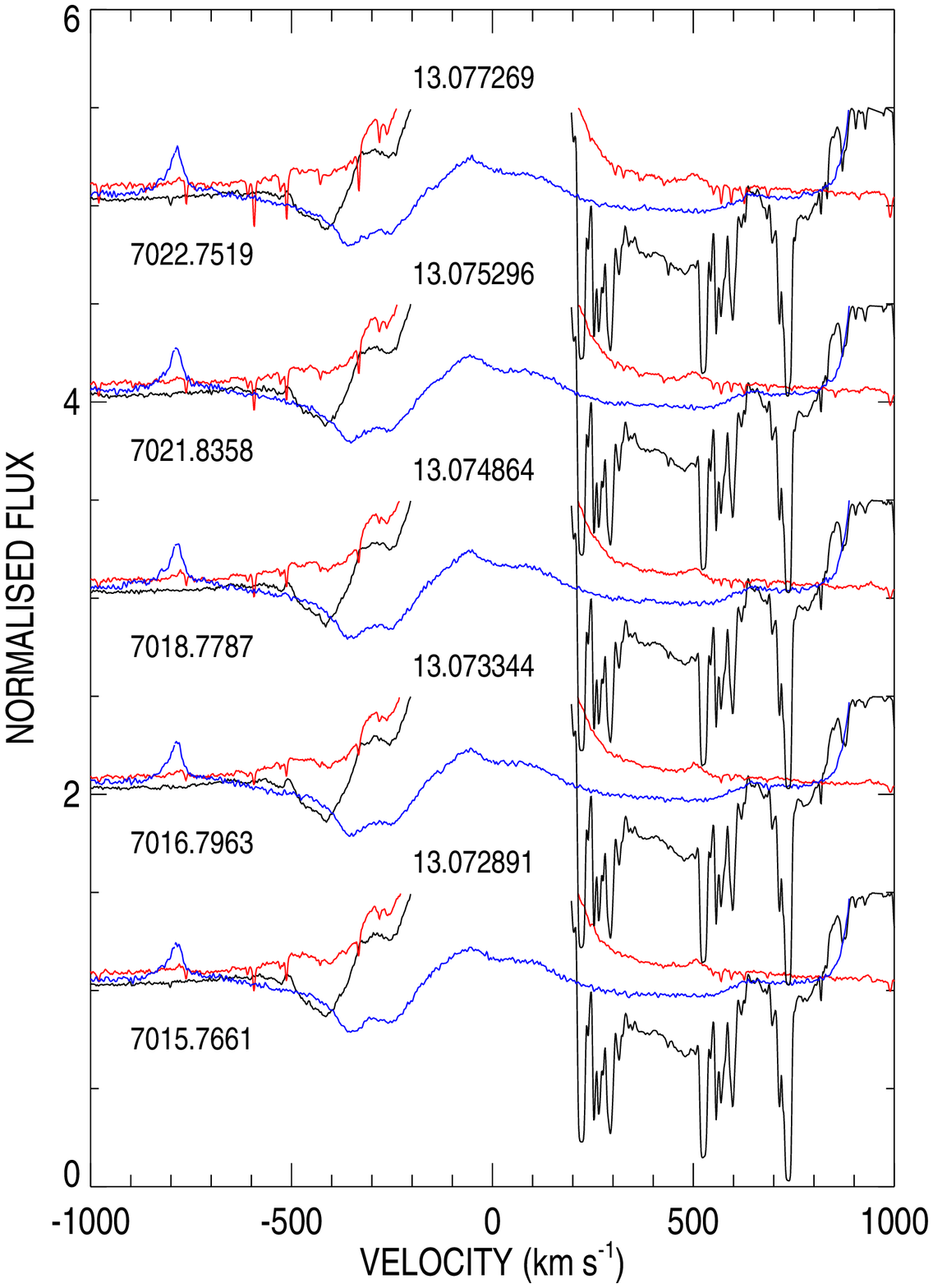}
 \includegraphics[width=50mm, angle=0]{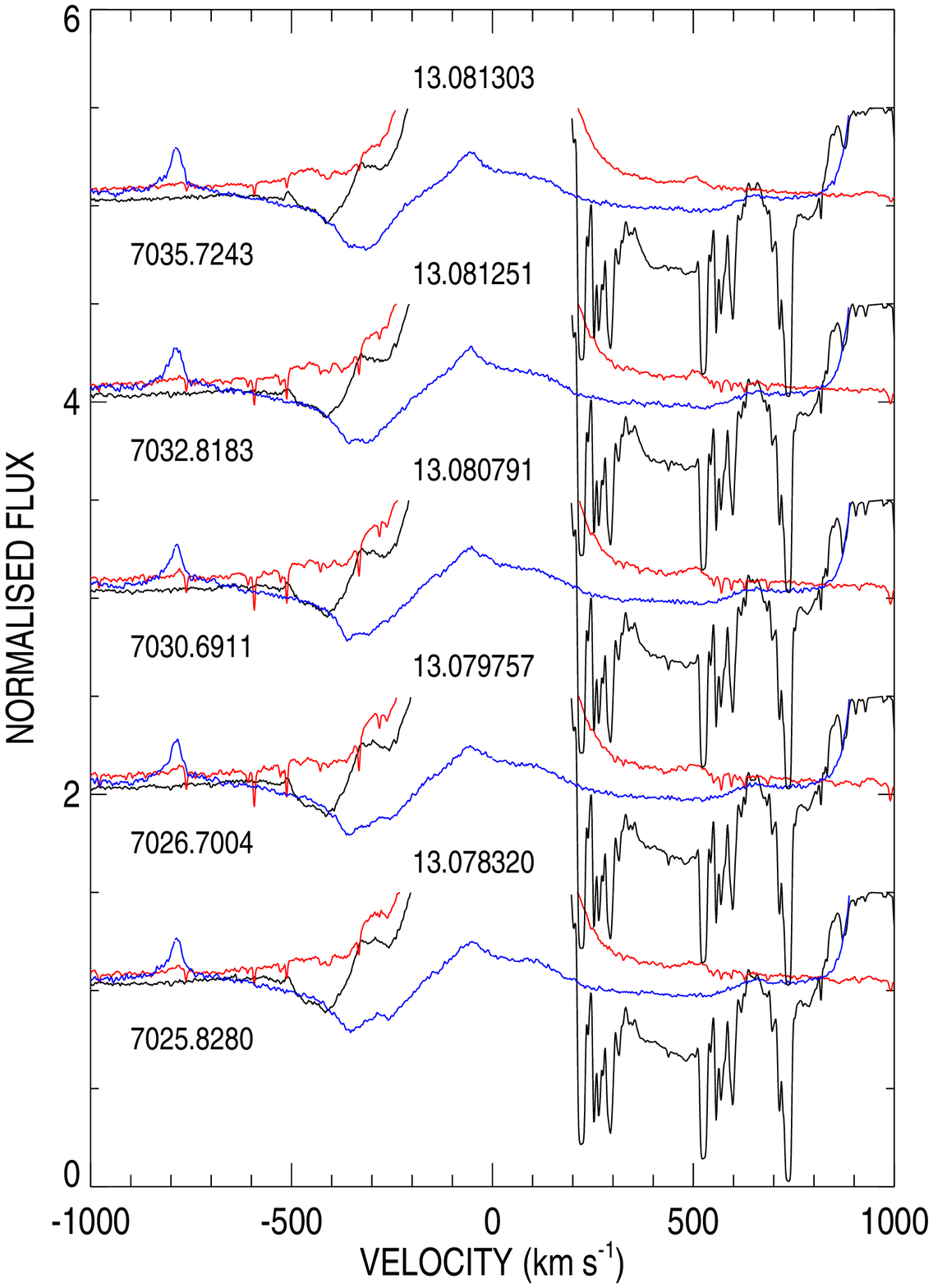}
\contcaption{\label{figprofiles_app}}
\end{figure*}

\begin{figure*}
 \includegraphics[width=50mm, angle=0]{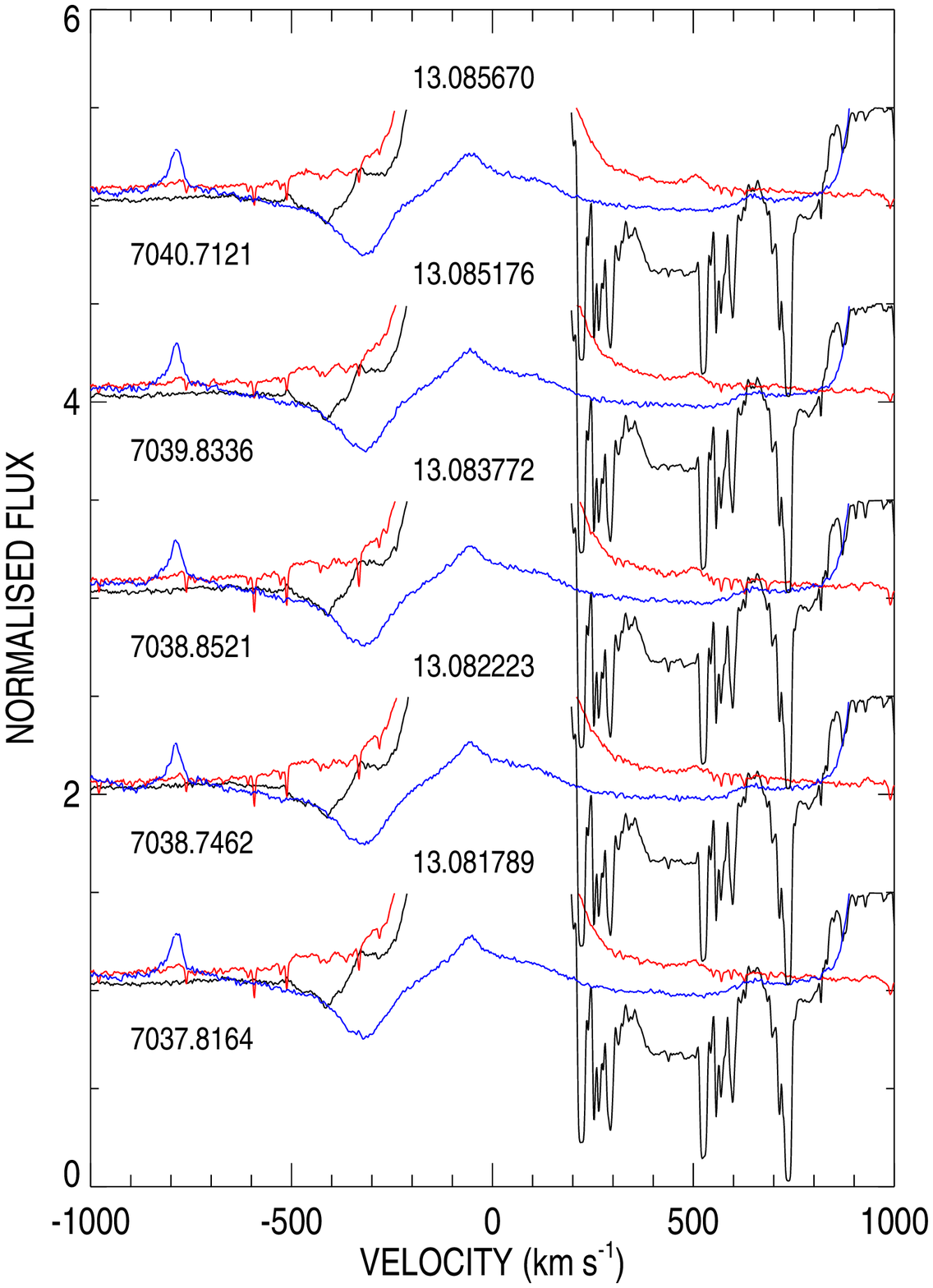}
 \includegraphics[width=50mm, angle=0]{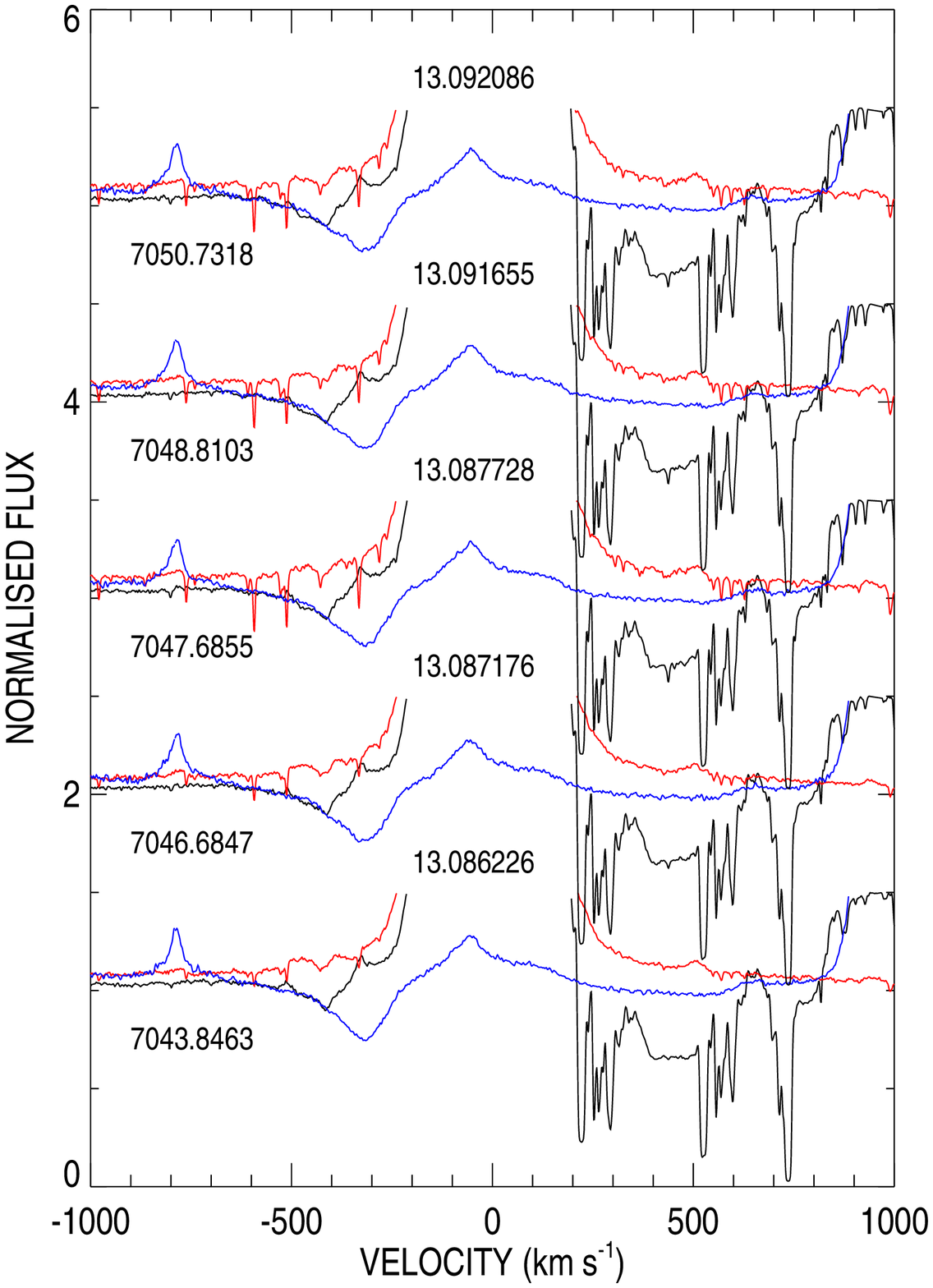}
 \includegraphics[width=50mm, angle=0]{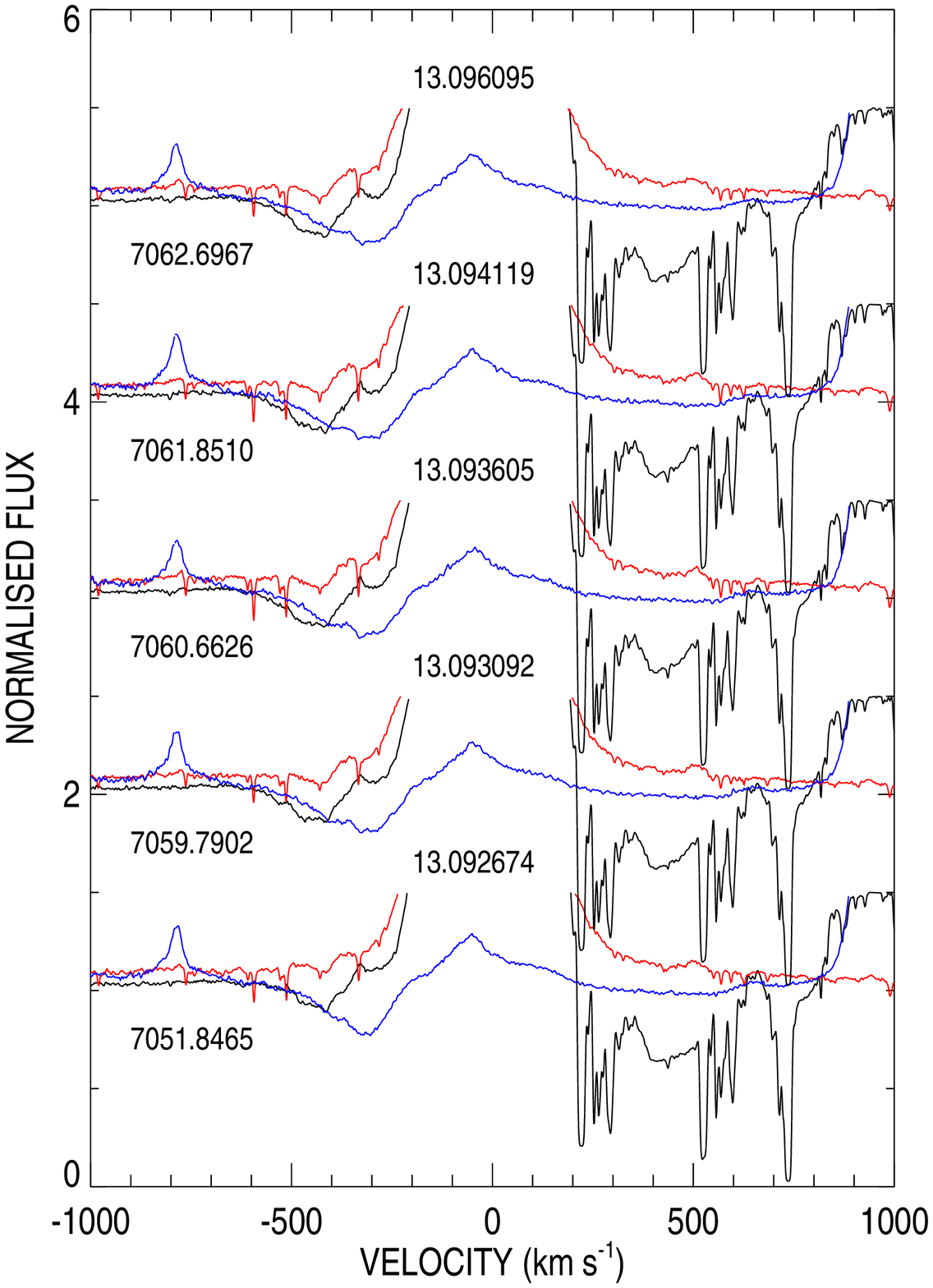}
 \includegraphics[width=50mm, angle=0]{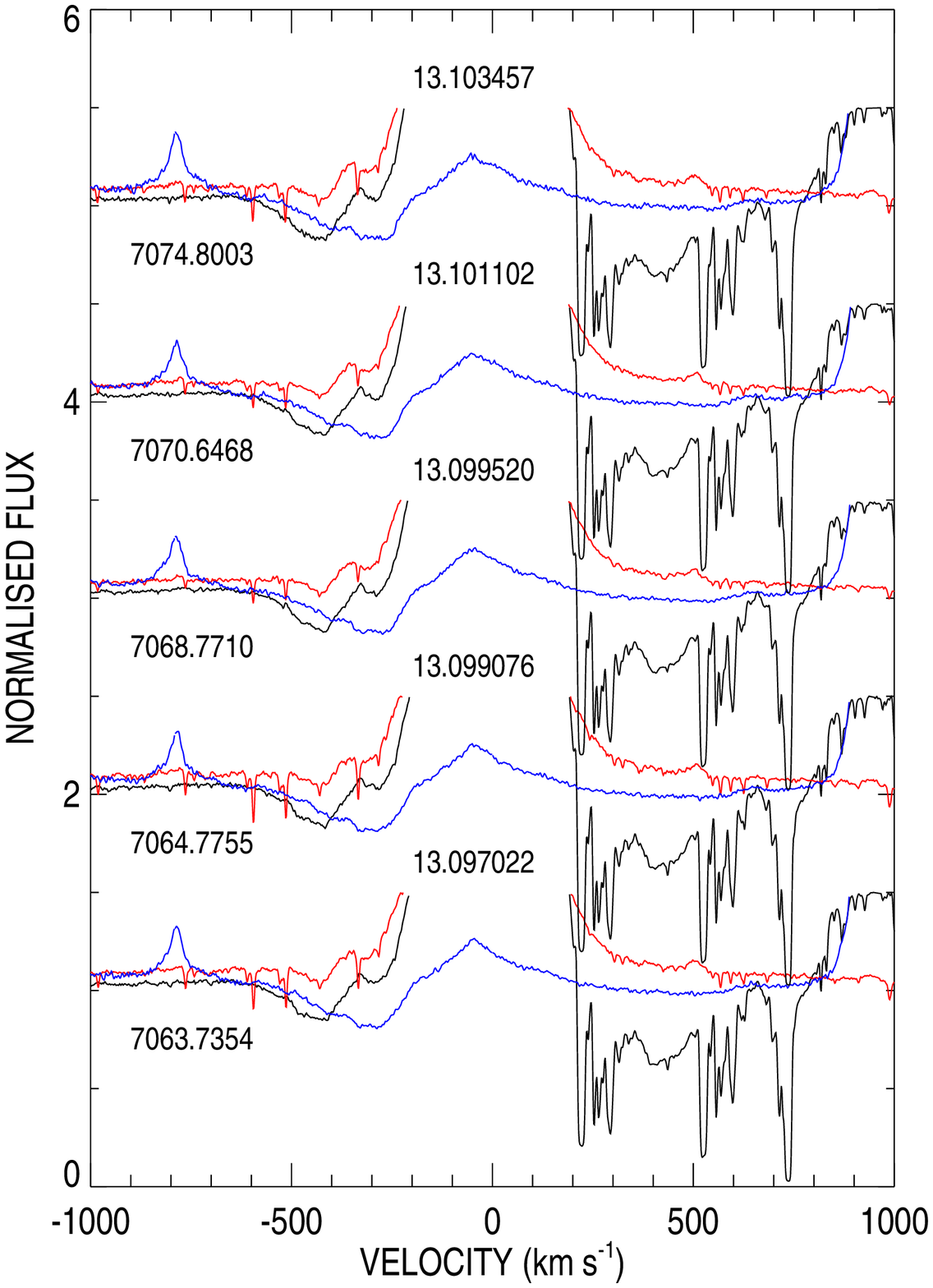}
 \includegraphics[width=50mm, angle=0]{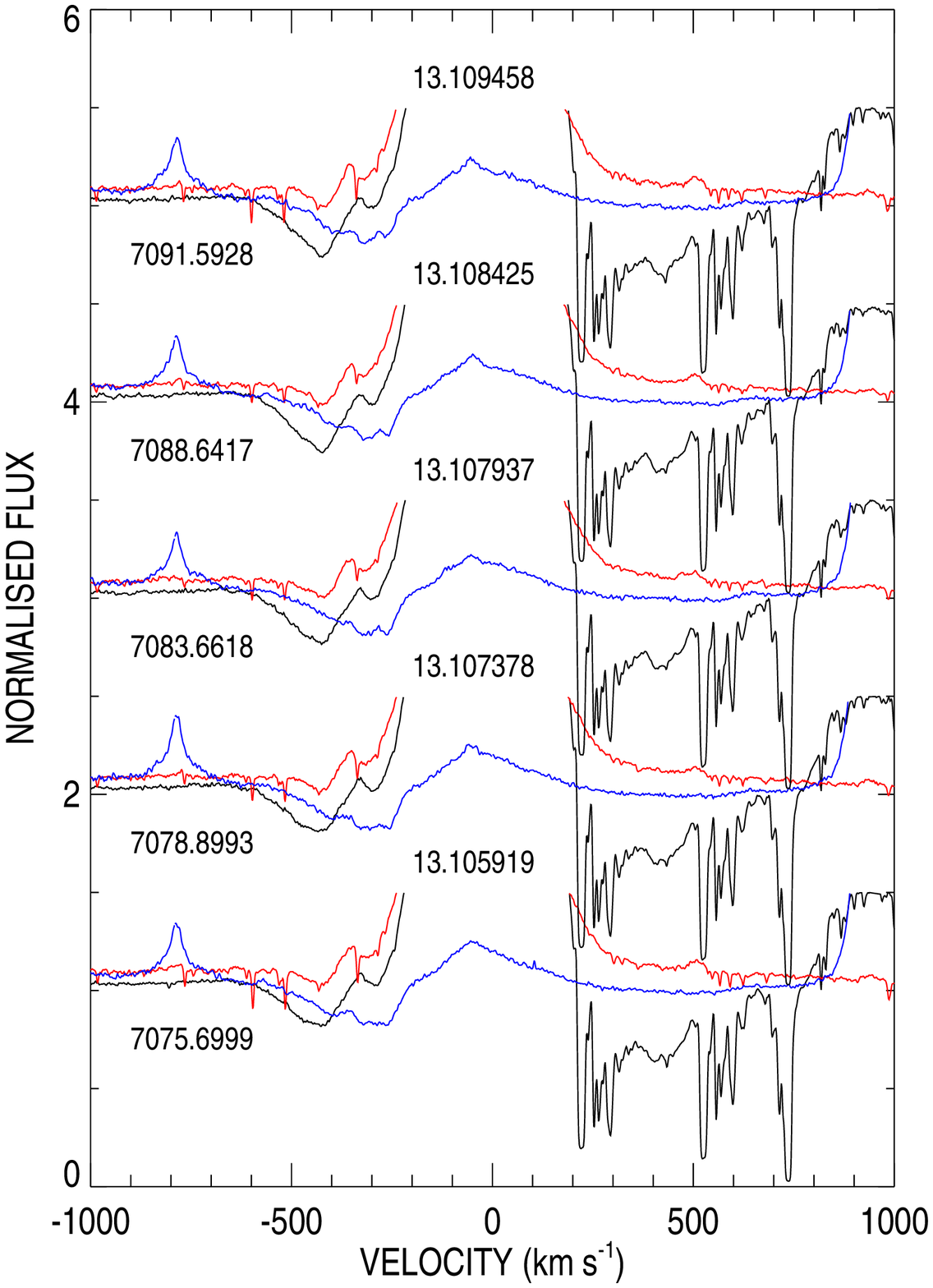}
 \includegraphics[width=50mm, angle=0]{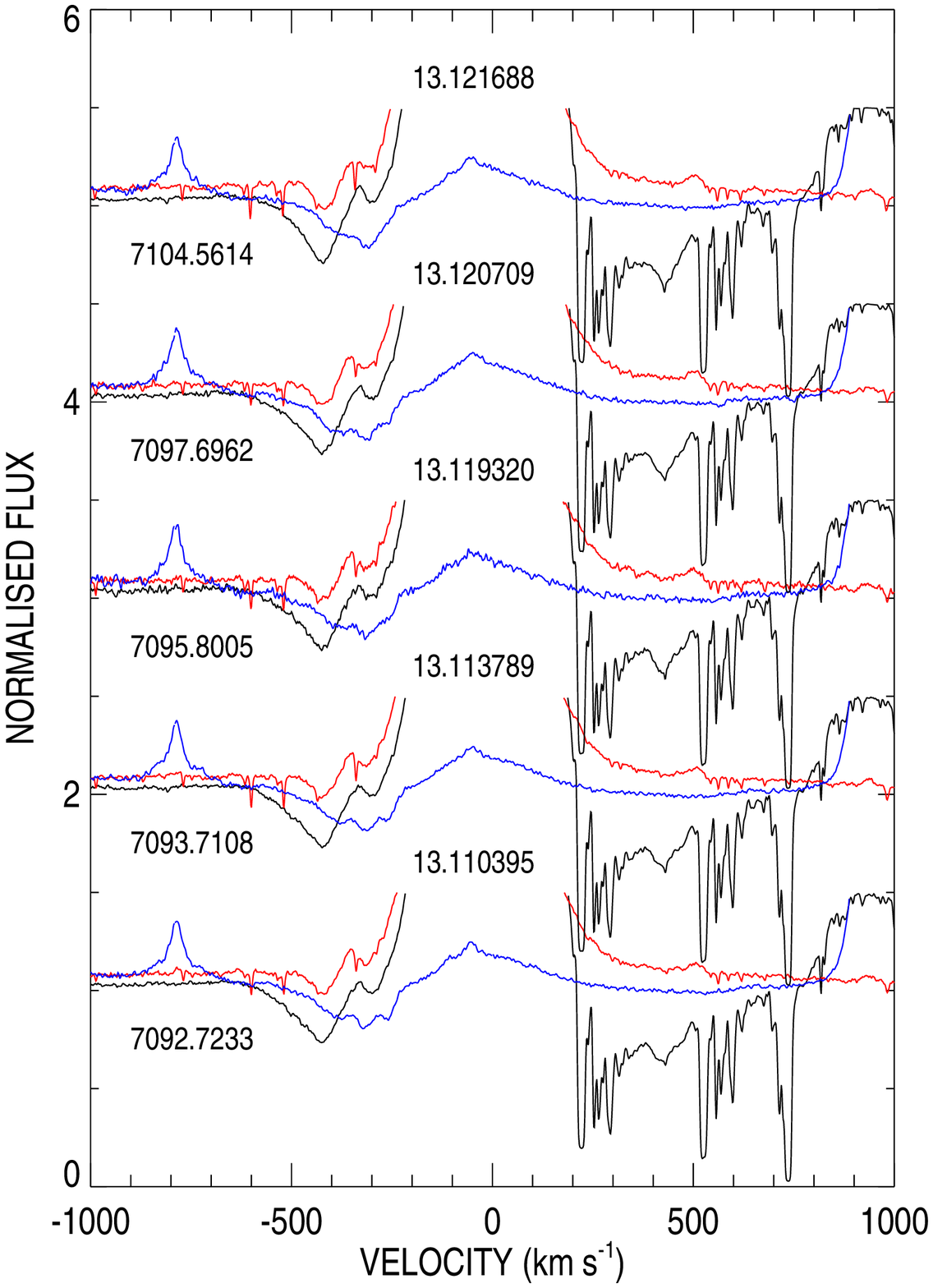}
 \includegraphics[width=50mm, angle=0]{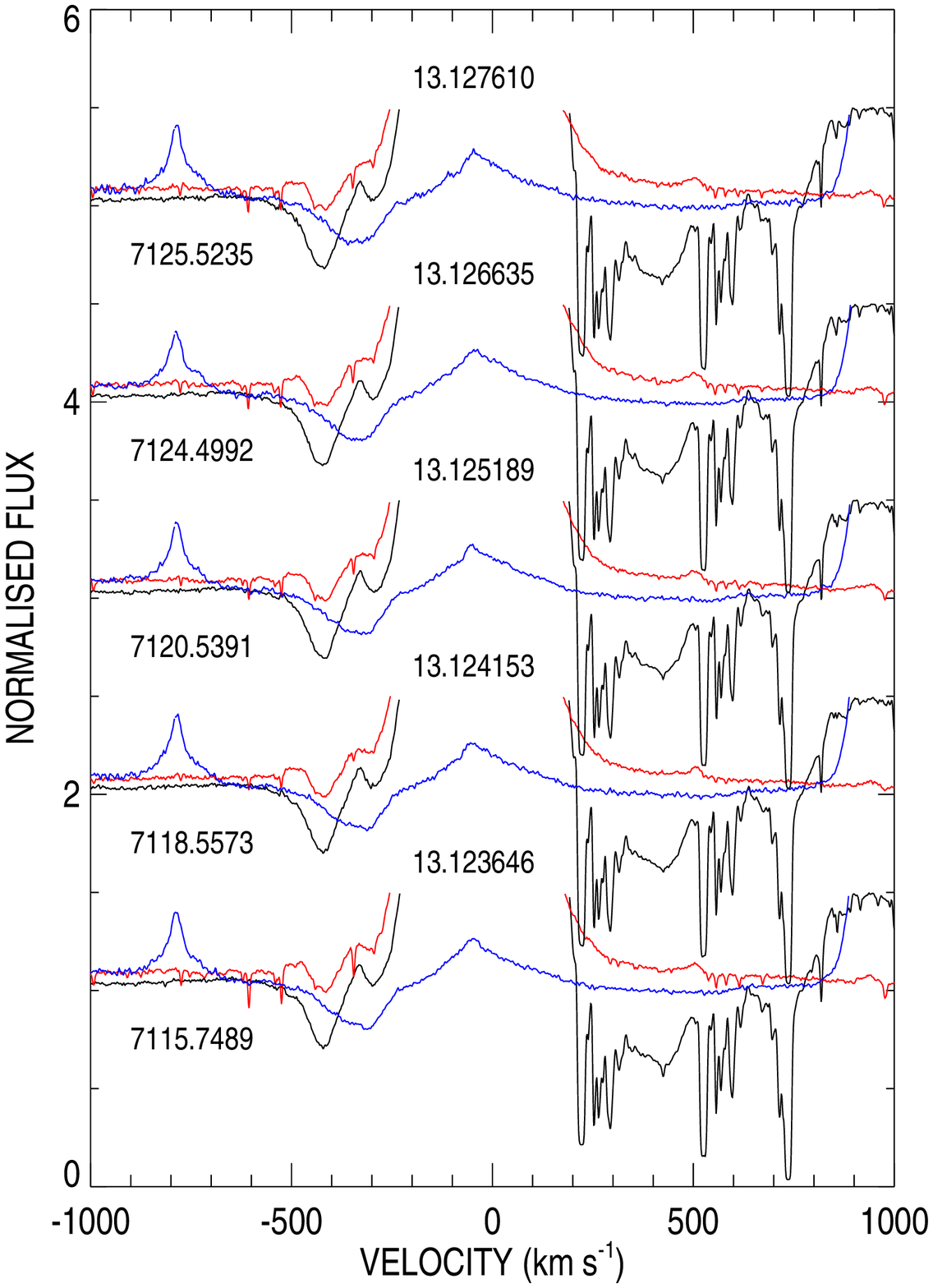}
 \includegraphics[width=50mm, angle=0]{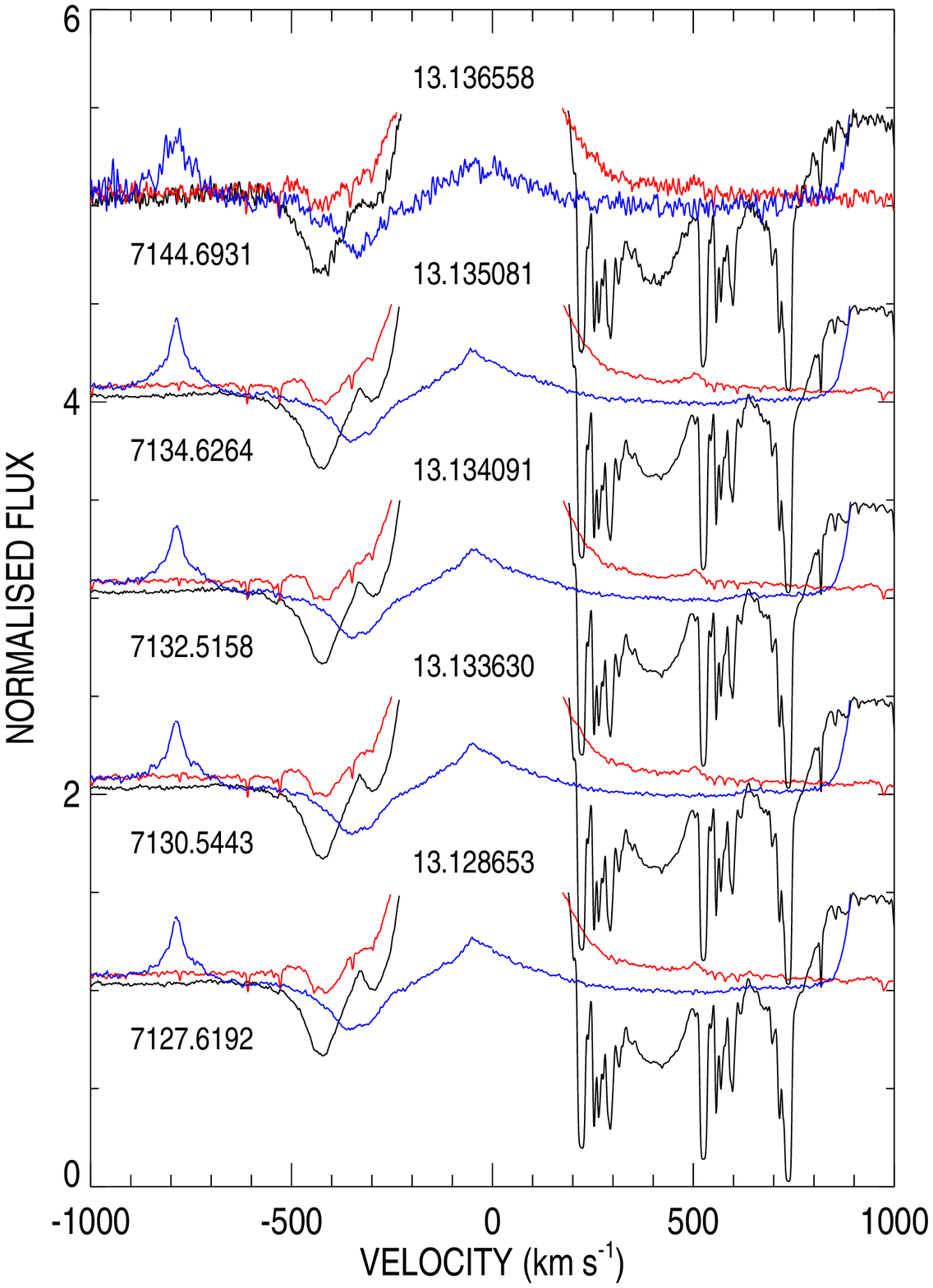}
 \includegraphics[width=50mm, angle=0]{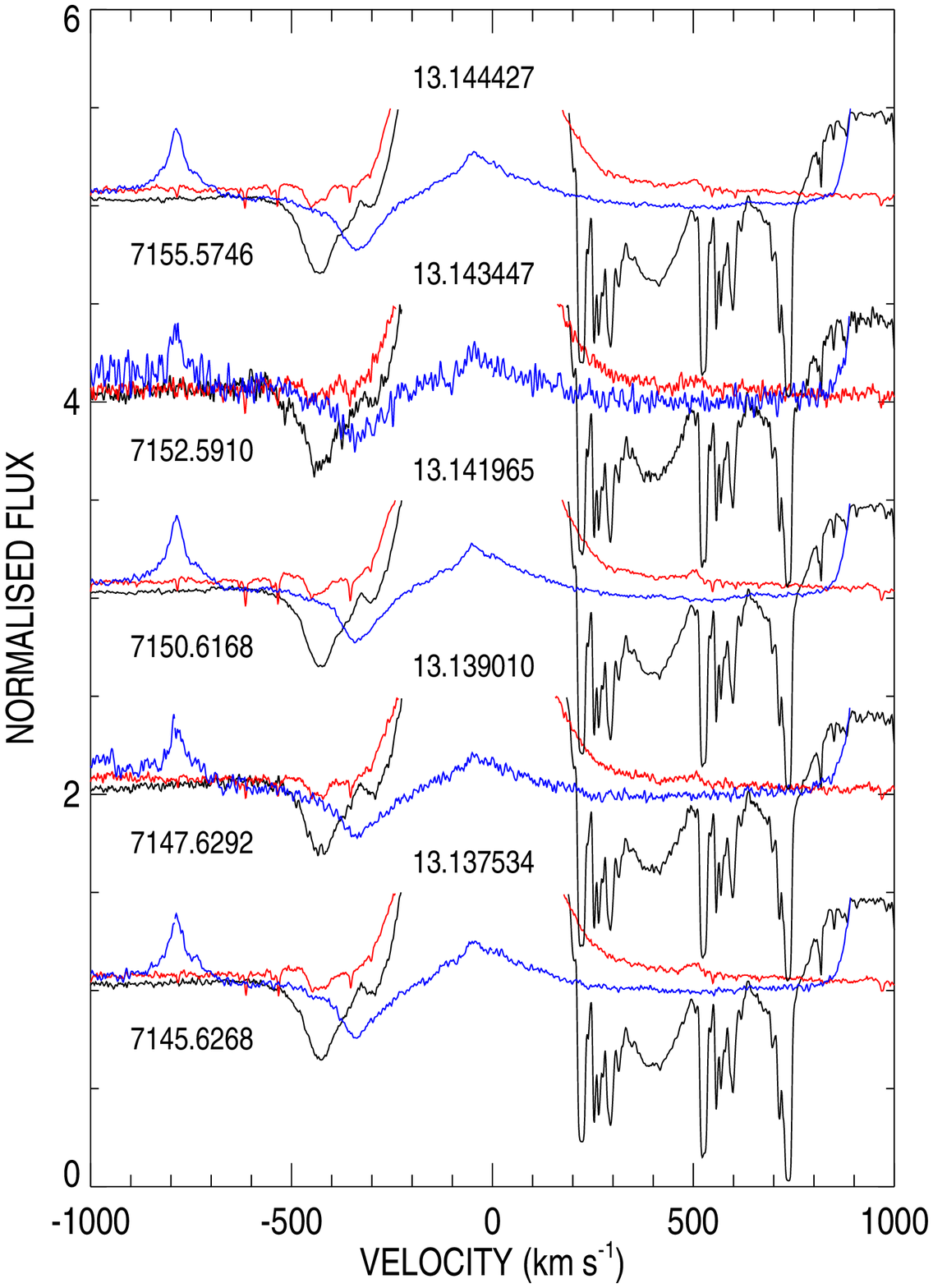}
\contcaption{\label{figprofiles_app}} 
\end{figure*}

\begin{figure*}
 \includegraphics[width=50mm, angle=0]{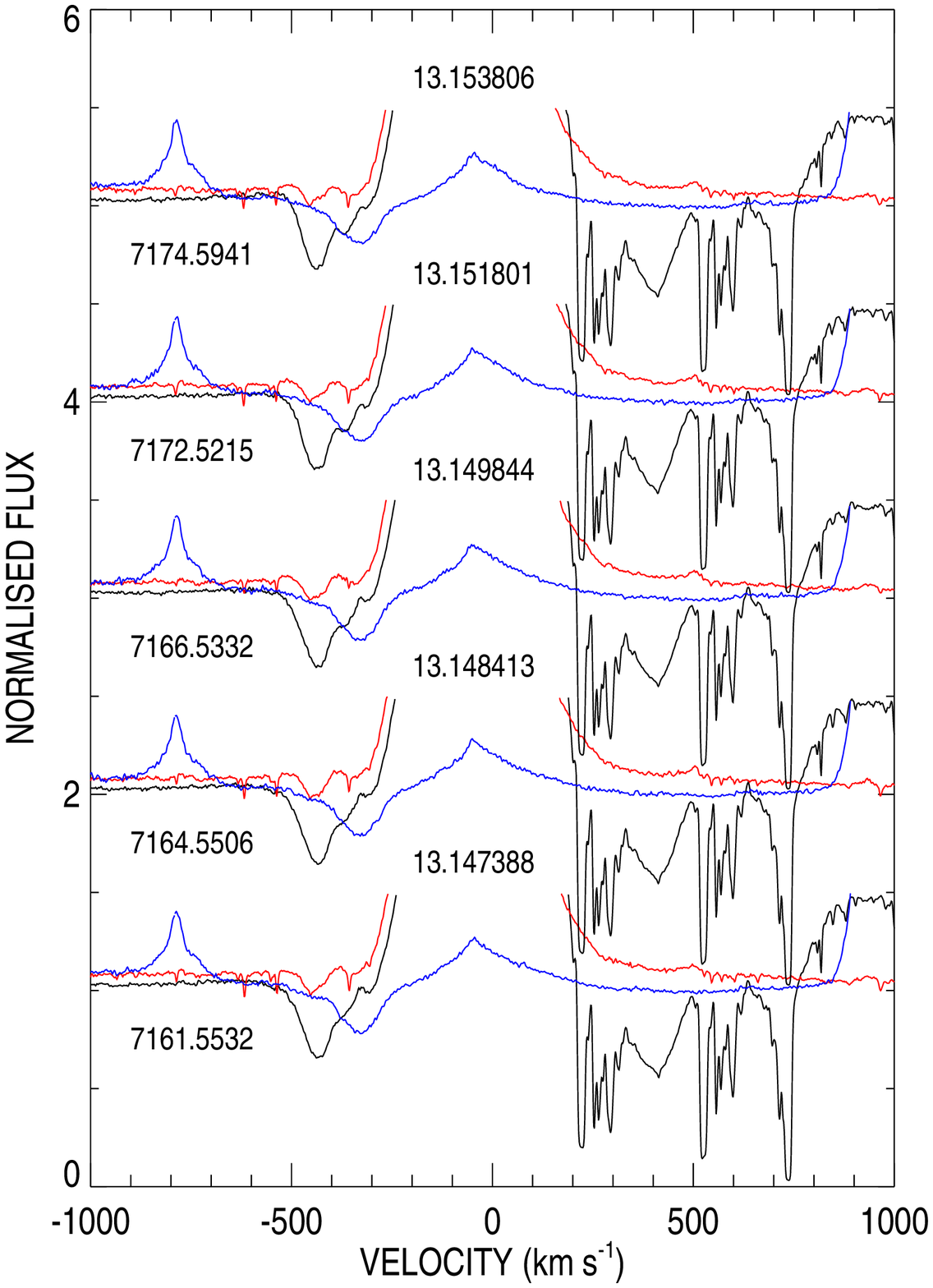}
\contcaption{\label{figprofiles_app}} 
\end{figure*}

\clearpage
\bsp \label{lastpage}

\end{document}